\documentclass[reprint,aps,prd,twocolumn,superscriptaddress,nofootinbib,preprintnumbers,floatfix]{revtex4-2}

\usepackage[utf8]{inputenc}
\usepackage[T1]{fontenc}
\DeclareUnicodeCharacter{2212}{\textendash}
\usepackage{graphicx}
\graphicspath{ {figures/} }
\usepackage[export]{adjustbox}
\usepackage{dcolumn}
\usepackage{bm}
\usepackage{amsmath}	
\usepackage{amssymb}
\usepackage{natbib}
\usepackage{listings}
\usepackage{hyperref}

\usepackage{placeins}
\usepackage{slashed}
\usepackage[normalem]{ulem}
\usepackage[dvipsnames]{xcolor}
\usepackage{xspace}
\usepackage{multirow}
\usepackage{booktabs}
\usepackage{aas_macros}
\usepackage{array}
\usepackage{subfigure}
\usepackage{rotating}
\usepackage{adjustbox}
\usepackage{orcidlink}
\hypersetup{colorlinks,linkcolor={red},citecolor={blue},urlcolor={blue}}

\newcommand{\code}[1]{\texttt{#1}}
\allowdisplaybreaks

\begin{document}
\include{notations}
%\preprint{APS/123-QED}

\title{The Indian Pulsar Timing Array Data Release 2: III. Search for a Stochastic Gravitational Wave Background}% Force line breaks with \\

\author{Hemanga Tahbildar \orcidlink{0009-0002-1036-9306}}
\email{hemanga22@iiserb.ac.in}
\affiliation{Department of Physics, IISER Bhopal, Bhopal Bypass Road, Bhauri, Bhopal 462066, Madhya Pradesh, India}

\author{Kunjal Vara \orcidlink{0009-0004-5501-1441}}
\affiliation{Department of Physics, IISER Bhopal, Bhopal Bypass Road, Bhauri, Bhopal 462066, Madhya Pradesh, India}

\author{Mayuresh Surnis \orcidlink{0000-0002-9507-6985}}
\affiliation{Department of Physics, IISER Bhopal, Bhopal Bypass Road, Bhauri, Bhopal 462066, Madhya Pradesh, India}

\author {Churchil Dwivedi \orcidlink{0000-0002-8804-650X}}
\affiliation{Astronomy and Astrophysics Division, Physical Research Laboratory, Thaltej Campus, Thaltej, Ahmedabad 380059, Gujarat, India}

\author{Bhal Chandra Joshi \orcidlink{0000-0002-0863-7781}}
\affiliation{National Centre for Radio Astrophysics, SP Pune University 
Campus, Pune, Maharashtra, 411007, India}
\affiliation{Department of Physics, Indian Institute of Technology Roorkee, 
Roorkee, Uttarakhand, 247667, India}

\author{Sharika Dhakappa \orcidlink{0009-0005-6671-9194}}
\affiliation{Department of Physics, IIT Hyderabad, Kandi, Telangana 502284, India}

\author{Aman Srivastava \orcidlink{0000-0003-3531-7887}}
\affiliation{Department of Physics, IIT Hyderabad, Kandi, Telangana 502284, India}

\author{Shantanu Desai \orcidlink{0000-0002-0466-3288}}
\affiliation{Department of Physics, IIT Hyderabad, Kandi, Telangana 502284, India}

\author{Abhimanyu Susobhanan  \orcidlink{0000-0002-2820-0931}}
\affiliation{School of Physics, Indian Institute of Science Education and Research Thiruvananthapuram, Marthamala PO, Thiruvananthapuram, Kerala 695551, India}

\author{Adya Shukla \orcidlink{0009-0005-7058-5539}}
\affiliation{Department of Physics, Indian Institute of Technology Roorkee, 
Roorkee, Uttarakhand, 247667, India}

\author{Himanshu Grover \orcidlink{0009-0004-1150-6151}}
\affiliation{National Centre for Radio Astrophysics, SP Pune University 
Campus, Pune, Maharashtra, 411007, India}
\affiliation{Department of Physics, Indian Institute of Technology Roorkee, 
Roorkee, Uttarakhand, 247667, India}

\author{P. Arumugam \orcidlink{0000-0001-9624-8024}}
\affiliation{Department of Physics, Indian Institute of Technology Roorkee, 
Roorkee, Uttarakhand, 247667, India}

\author{Manjari Bagchi \orcidlink{0000-0001-8640-8186}}
\affiliation{The Institute of Mathematical Sciences, C. I. T. Campus, Taramani, Chennai 600113, India}
\affiliation{Homi Bhabha National Institute, Training School Complex, Anushakti Nagar, Mumbai 400094, India}

\author{Neelam Dhanda Batra \orcidlink{0000-0003-0266-0195}}
\affiliation{Department of Physics, IIT Delhi, Hauz Khas, New Delhi 110016, India}

\author{Manoneeta Chakraborty \orcidlink{0000-0002-9736-9538}}
\affiliation{Department of Astronomy, Astrophysics, and Space Engineering, Indian Institute of Technology Indore, Indore 453552, India}

\author{Shaswata Chowdhury \orcidlink{0000-0001-5701-4014}}
\affiliation{Radio Astronomy Centre, Tata Institute of Fundamental Research, National Centre for Radio Astrophysics, Ooty, Tamil Nadu 643001, India}

\author{Debabrata Deb \orcidlink{0000-0003-4067-5283}}
\affiliation{ Centre for Space Research, North-West University, Private Bag X6001, Potchefstroom 2520, South Africa }
\affiliation{ National Institute for Theoretical and Computational Sciences (NITheCS), Stellenbosch 7604, South Africa}

\author{A. Gopakumar \orcidlink{0000-0003-4274-4369}}
\affiliation{Department of Astronomy and Astrophysics, Tata Institute of Fundamental Research, Mumbai 400005, Maharashtra, India}

\author{Kuldeep Meena  \orcidlink{0009-0007-8522-9574}}
\affiliation{UM-DAE Centre for Excellence in Basic Sciences, University of Mumbai, Vidyanagari, Mumbai 400098, India}

\author{Sushovan Mondal \orcidlink{0000-0002-3657-8256}}
\affiliation{Department of Physics, Indian Institute of Technology Bombay, Mumbai 400076, India}

\author{K Nobleson  \orcidlink{0000-0003-2715-4504}}
\affiliation{Faculty of Advanced Science and Technology, Kumamoto University, 2-39-1 Kurokami, Kumamoto 860-8555, Japan}

\author{Avinash Kumar Paladi \orcidlink{0000-0002-8651-9510}}
\affiliation{Joint Astronomy Programme, Department of Physics, Indian Institute of Science, Bengaluru, Karnataka, 560012, India}

\author{Arul Pandian B \orcidlink{0000-0002-0417-6308}}
\affiliation{Raman Research Institute, Bengaluru - 560080, India}
\affiliation{Department of Physics and Electronics, CHRIST (Deemed to be University), Bengaluru - 560029, India}

\author{Kaustubh Rai  \orcidlink{0009-0002-2175-7013}}
\affiliation{Department of Physics, IISER Bhopal, Bhopal Bypass Road, Bhauri, Bhopal 462066, Madhya Pradesh, India}

\author{Prerna Rana \orcidlink{0000-0001-6184-5195}}
\affiliation{Laboratoire de Physique et Chimie de l’Environnement et de l’Espace, Orléans, France 45071}

\author{Shubhit Sardana \orcidlink{0009-0007-2913-7704}}
\affiliation{Department of Physics, IISER Bhopal, Bhopal Bypass Road, Bhauri, Bhopal 462066, Madhya Pradesh, India}

\author{Vidit Singh \orcidlink{0009-0001-2049-3039}}
\affiliation{Department of Astronomy and Astrophysics, Tata Institute of Fundamental Research, Mumbai 400005, Maharashtra, India}

\author{Jaikhomba Singha \orcidlink{0000-0002-1636-9414}}
\affiliation{High Energy Physics, Cosmology and Astrophysics Theory Group (HEPCAT), Department of Mathematics and Applied Mathematics, University of Cape Town, Cape Town 7700, South Africa}

\author{Keitaro Takahashi \orcidlink{0000-0002-3034-5769}}
\affiliation{Faculty of Advanced Science and Technology, Kumamoto University, 2-39-1 Kurokami, Kumamoto 860-8555, Japan}

\author{Pratik Tarafdar \orcidlink{0000-0001-6921-4195}}
\affiliation{INAF - Osservatorio Astronomico di Cagliari, via della Scienza 5, 09047 Selargius (CA), Italy}

\author{Zenia Zuraiq \orcidlink{0009-0000-6980-6334}}
\affiliation{Department of Physics, Indian Institute of Science, Bangalore 560012, India}
%\collaboration{InPTA Collaboration}%\noaffiliation


\begin{abstract}
We present the first independent search for an isotropic stochastic gravitational wave background in the second data release of the Indian Pulsar Timing Array, comprising 27 millisecond pulsars monitored simultaneously in two frequency bands with the upgraded Giant Metrewave Radio Telescope over a maximum 7.2 year baseline. Building on a comprehensive single pulsar noise analysis, we search for a common uncorrelated red noise process within a Bayesian inference framework and with the noise-marginalized optimal statistic, and we test the robustness of the result through per-pulsar dropout analyses and solar-wind exclusion cuts. Leaving the spectral index free, we recover a broad amplitude posterior, $\log_{10} A_{\rm CURN} = -13.71^{+1.06}_{-3.28}$, with an unconstrained spectral index $\gamma_{\rm CURN} = 2.98^{+3.62}_{-2.70}$ and a Savage-Dickey Bayes factor of $2.5$ for a common red process over the no signal model. The optimal-statistic signal to noise ratios for the monopole, dipole, and Hellings-Downs correlations are all consistent with zero. Fixing the spectral index to $\gamma = 13/3$, the value predicted by an idealized toy model in which the background is sourced by a population of supermassive black hole binaries in circular orbits evolving purely under leading-order gravitational radiation reaction, we place a $95\%$ upper limit on the common-process amplitude of $A_{\rm GWB} < 3.4\times10^{-14}$, stable across solar elongation cuts of $10^\circ$, $20^\circ$, and $30^\circ$. This limit lies approximately an order of magnitude above the amplitudes reported by other, longer-running pulsar timing array experiments. Simulations with injected achromatic red noise and dispersion measure noise, based on a single noise realization, suggest that a baseline of at least $\sim 10$ years may be needed before the common red-noise signal begins to be recovered.
\end{abstract}

\maketitle

%\begin{keyword}

%Pulsars \sep Neutron Stars \sep Gravitational Waves \sep Pulsar Timing Arrays

%\end{keyword}

%\tableofcontents

\section{Introduction}

The existence of gravitational waves (GWs) constitutes one of the most important predictions of Einstein's general theory of relativity \citep{E16}. The first indirect evidence for the presence of GWs was achieved through observing the orbital decay of the binary pulsar PSR B1913+16 \citep{HT75,TFM79,TW82}, where part of the orbital energy is released through gravitational radiation. On 14 September 2015, the Advanced Laser Interferometer Gravitational-wave Observatory \citep[Advanced LIGO;][]{AAA+16a} recorded the first direct detection of GWs, designated event GW150914, fundamentally transforming our understanding of the transient universe. 

Ground-based interferometric detectors such as LIGO function within the 10 to 1000 Hz range \citep{AA+15}. Space-based gravitational wave detectors, such as the Laser Interferometer Space Antenna (LISA), will operate primarily in the millihertz (mHz) frequency regime \citep{AAB+17}. Pulsar Timing Array \citep[PTA:][]{S78,FB90} experiments are sensitive to GWs in the nanohertz (nHz) frequency regime. A PTA experiment operates by synthesizing a galaxy-sized observatory by monitoring an ensemble of millisecond pulsars (MSPs) distributed across the sky over decades. Numerous PTA experiments are being conducted globally. Several of them have reported evidence, at varying levels of significance, for a stochastic gravitational wave background (SGWB) \citep{AAB+23, AAA+23b, RD+23, XH+23, MSR+25}, although the current datasets and analytical methods do not yet support a definitive detection.

Pulsars are rapidly rotating, highly magnetized neutron stars emitting beams of radiation from their magnetic poles \citep{HB+68}. When our line of sight crosses one of these beams, we detect the radiation as a pulse. There are many different types of pulsars based on their observational properties. MSPs \citep{BK+82} with spin periods less than $\sim$30 ms are particularly of interest as the arrival of their radio pulses exhibits a consistency rivaling that of atomic clocks \citep{HGC+20}. PTA experiments are based on the fundamental principle that GWs modulate the time of travel of the radio signals as they propagate across the line of sight between the Earth and pulsars \citep{D79}. The ionized interstellar medium (IISM) also influences the radio pulses, inducing dispersion and scattering delays. In addition to such propagation delays, the measured times of arrival (TOAs) of pulses are influenced by inaccuracies in the solar system ephemeris, observatory clock errors, variable solar wind, random variations in the pulse shape (pulse jitter), and instrumental noise \citep{AAA+23}. The effect of some of these noise sources on the measured TOAs can resemble the effect of GWs. Timing models characterize these perturbations by constructing timing residuals, defined as the difference between the observed pulse TOAs and those predicted by a deterministic timing ephemeris which explicitly accounts for an array of pulsar intrinsic parameters alongside external propagation effects \citep{EHM06}. Consequently, the remaining timing residuals contain all unmodelled stochastic noise processes alongside the signatures of the SGWB \citep{LT+15}. 

As mentioned earlier, the PTA experiments are sensitive to GWs in the nHz frequency range. Potential sources of GWs within this low-frequency band include supermassive black hole binary (SMBHB) systems \citep{PS00}, primordial quantum fluctuations amplified through inflation \citep{LM+16}, cosmic strings \citep{BS+22}, and cosmological phase transitions \citep{MMP+23}. The most prominent signal in this frequency range is expected to be the SGWB, resulting from the incoherent superposition of GWs emitted by a large number of inspiraling SMBHBs or by cosmological sources \citep{BT+19, DZ+24}. One of the principal objectives for contemporary PTA collaborations in gravitational wave detection is the identification of an SGWB.

The timing residuals from multiple MSPs can be analyzed to search for pulsar-correlated signals arising from an SGWB simultaneously with various noise processes. A defining characteristic of an SGWB generated by a cosmological population of SMBHBs is that, in contrast to most astrophysical and instrumental noise sources, it produces spatially correlated timing variations across different Earth–pulsar baselines \citep{HD83}. For an isotropic, stationary, and unpolarized SGWB composed of tensor GWs, general relativity predicts a unique angular correlation pattern between pairs of pulsars, known as the Hellings–Downs (HD) overlap reduction function \cite{HD83}. This correlation function describes the expected response of PTA baselines to both the plus ($+$) and the cross ($\times$) polarization modes of GWs. Consequently, PTAs can test for the presence of an SGWB by comparing the measured cross-correlations in pulsar timing residuals with the HD prediction \citep{JH06}. Alternative theories of gravity may permit additional gravitational wave polarization states, while anisotropic gravitational wave backgrounds can produce angular correlation patterns that differ from the standard HD curve. Therefore, searches for SGWBs must also account for these possibilities once it is detected conclusively, as they provide important tests of both gravitational physics and the astrophysical origin of the background \citep{GJ+15, MS+13}.

The timing residuals induced by an isotropic SGWB are expected to follow a steep red-noise power spectral density (PSD), given by $P(f)\propto f^{-\gamma}$, where $f$ is the gravitational wave frequency \citep{PS00, JH06}. This spectral behavior is related to the characteristic strain spectrum, $h_c(f)=A\left(\frac{f}{1 \mathrm{yr}^{-1}}\right)^{\alpha}$, where $A$ is the strain amplitude and $\alpha$ is the strain spectral index. The relationship between the PSD and the strain spectral indices is given by $\gamma = 3 - 2\alpha$ \citep{JH06}. For a gravitational wave background generated by a population of inspiraling SMBHBs evolving solely through gravitational wave emission, the expected strain spectral index is $\alpha=-2/3$, corresponding to a timing-residual spectral index of $\gamma=13/3$ \citep{PS00}.

The Indian Pulsar Timing Array \citep[InPTA:][]{JAB+18,Joshi2022} is an Indo-Japanese PTA collaboration which utilizes the unique simultaneous multi-band capabilities of the upgraded Giant Metrewave Radio Telescope \citep[uGMRT:][]{sak+1991,gak+2017}. While the InPTA Data Release 1 \citep[DR1;][]{TNR+22} was used together with the European Pulsar Timing Array Data Release 2 \citep[EPTA DR2;][]{ABN+23} to search for the SGWB, the InPTA dataset has never been searched independently. With the publication of the InPTA data release 2 \citep[DR2;][]{RTN+25}, we have now, for the first time, probed this dataset for the presence of the SGWB signals. In Section~\ref{sec:dataset} we briefly describe the InPTA DR2 dataset. Section~\ref{sec:noiseanalysis} summarizes the single-pulsar noise analysis. Section~\ref{sec:Method} details our Bayesian and frequentist methodology. Section~\ref{sec:results} presents the DR1 and DR2 SGWB results and solar-wind robustness checks. Section~\ref{subsec:data_extension} forecasts the constraints achievable with an extended time baseline. We discuss the implications of our findings in Section~\ref{sec:discussion} and conclude in Section~\ref{sec:conclusion}.

\section{Dataset}
\label{sec:dataset}

The InPTA has released two datasets to date. The first data release \citep{TNR+22} contained TOAs and dispersion measures (DMs) from 14 pulsars with a roughly biweekly cadence. The data spanned around 3.5 years between 2018 and 2021. The second (and the latest) data release \citep{RTN+25} contains TOAs and DMs from 27 pulsars with a time span of around 7.2 years between 2016 and 2024. The TOAs and DMs themselves were obtained from observations carried out with the uGMRT at two distinct frequency bands from 300$-$500 MHz (Band 3) and 1260$-$1460 MHz (Band 5). Please refer to the respective publications for more details on the datasets. The InPTA dataset constitutes the widest frequency coverage with simultaneous observation at the two frequency bands, allowing a very precise estimation of the effects of the IISM on the TOAs. This, in turn, provides an excellent handle on the characterization of the noise properties of the PTA experiment, as is described briefly in the next section.

\vspace{0.3cm}

\section{Single Pulsar Noise analysis}
\label{sec:noiseanalysis}
Accurate characterization of noise processes in the PTA observations is essential for detecting and characterizing the SGWB. An extensive single pulsar noise analysis (SPNA) of the 27 MSPs included in the InPTA DR2 was carried out by \cite{ND+26} (See \cite{SDK+23} for InPTA DR1 noise analysis). The analysis employs rigorous Bayesian inference methodology with stationary Gaussian process models in the Fourier domain to systematically characterize and quantify multiple stochastic noise sources present in the timing residuals. These noise processes are classified into time-uncorrelated (radiometer noise, template matching errors, pulse jitter) and time-correlated processes, with the latter further subdivided into achromatic (like spin noise) and chromatic processes (like DM variations). The noise analysis showed significant heterogeneity in the noise properties of the pulsars. Some of the pulsars had enough time baseline to be able to characterize many noise processes while some pulsars showed only white noise given the relatively short time baseline. The extensive analysis has enabled us to search for the SGWB with accurate SPNA providing decent constraints on the noise other than the SGWB in the InPTA DR2.

\section{SGWB Search Method}
\label{sec:Method}

We employ a Bayesian inference framework to search for the SGWB signal in our pulsar timing array dataset. This approach enables simultaneous parameter estimation and model comparison while properly accounting for all sources of measurement uncertainty and astrophysical noise. Our methodology follows the established procedures developed by Refs.\cite{VN+09,LT+15,TV+16} and implemented in the \code{enterprise} software package \citep{EVM20}.

The single pulsar noise processes modelled in our analysis comprise the white noise (EFAC, EQUAD, and ECORR), achromatic red noise (ARN), dispersion measure noise (DMN), free chromatic noise (FCN), and solar-wind induced noise. At the array level, we additionally incorporate a common but mutually uncorrelated red-noise process (CURN). In the following subsections, we provide a detailed description of each of these processes.

\subsection{Likelihood Function}
\label{sec:likelihood}

The fundamental assumption underlying our analysis is that pulsar timing residuals, the difference between observed and predicted pulse TOAs, are realizations of stochastic Gaussian processes. Under this assumption, the likelihood for a set of timing residuals $\boldsymbol{\delta t}$ is:

\begin{equation}
    \mathcal{L}(\boldsymbol{\theta}|\boldsymbol{\delta t}) = \frac{1}{\sqrt{\det\!\left[2\pi C(\boldsymbol{\theta})\right]}} \exp\left(-\frac{1}{2}\boldsymbol{\delta t}^T C(\boldsymbol{\theta})^{-1} \boldsymbol{\delta t}\right)
    \label{eq:likelihood}
\end{equation}

where $\boldsymbol{\theta}$ represents the full set of model parameters and $C(\boldsymbol{\theta})$ is the covariance matrix encoding the statistical properties of all noise sources and potential signals.

For computational efficiency, we analytically marginalize over timing model parameters (pulsar position, proper motion, spin frequency and derivatives, and binary orbital elements) following \cite{VV13}. 

\subsection{Noise Processes}
\label{sec:noise_processes}

Any PTA dataset has many sources of noise that result from various intrinsic and extrinsic effects. These effects include (but are not limited to) the spin noise of the pulsar, variations in the orbital parameters for a binary pulsar, the effects of the IISM on the radio signal. In order to separate the effect of the SGWB on the pulse TOAs, one needs to model these effects depending on whether they are independent of the radio frequency (achromatic) or change as a function of the radio frequency (chromatic).  

\subsubsection{White Noise}
\label{sec:white_noise}

White noise captures fluctuations that are statistically independent from one observation to the next. Three parameters are used to describe it for each backend/receiver combination: EFAC, which rescales the nominal ToA uncertainty; EQUAD, which adds an additional uncertainty term in quadrature; and ECORR, which accounts for noise correlated across sub-bands within a single observing epoch but uncorrelated between epochs. Together these combine to build the white noise covariance matrix, whose explicit form is given in Eq.~(2) of \citep{ND+26}.

\subsubsection{Achromatic Red Noise}
\label{sec:arn}

Achromatic red noise (ARN), sometimes referred to as spin noise, reflects long timescale, frequency independent variability in the pulsar's rotation. We represent it as a stationary Gaussian process in the Fourier domain, with a power-law spectrum set by an amplitude $A_{\rm arn}$ (normalized at $f_{\rm 1yr}$) and a spectral index $\gamma_{\rm arn}$. The spectral density, the induced time domain residuals, and the corresponding Fourier coefficient covariance are all defined in Eqs.~(3)-(5) of \citep{ND+26}.

\subsubsection{Dispersion Measure Noise}
\label{sec:dmn}

Time variable free electron content along the pulsar's line of sight produces frequency dependent delays scaling as $\nu^{-2}$, referred to as dispersion measure noise (DMN). As with ARN, this is captured by a power-law Gaussian process in the Fourier domain, here with an additional frequency scaling term evaluated relative to a fiducial observing frequency $\nu_{\rm ref} = 1400$~MHz, and parameterized by an amplitude $A_{\rm dmn}$ and spectral index $\gamma_{\rm dmn}$. The full spectral density expression appears in Eqs.~(6)-(7) of \citep{ND+26}.

\subsubsection{Free Chromatic Noise}
\label{sec:fcn}

Rather than assuming the fixed $\nu^{-4}$ scaling of a standard scattering noise model (defined in Eqs.~(8)-(9) of \citep{ND+26}), the single-pulsar analysis of \citep{ND+26} lets the chromatic index $\chi_{\rm fcn}$ vary freely, resulting in the free chromatic noise (FCN) model. This added flexibility accommodates pulsar-to-pulsar differences in the chromatic power spectrum that a fixed thin-screen scattering index would not capture, and FCN is the model we use in place of the scattering process throughout this work. Its spectral density is given in Eq.~(10) of \citep{ND+26}. In the array-level analyses, $\chi_{\rm fcn}$ is held at SPNA median, while the FCN amplitude $A_{\rm fcn}$ and spectral index $\gamma_{\rm fcn}$ are sampled (Section~\ref{sec:spn_treatment}).

\subsubsection{Solar Wind Noise}
\label{sec:solar_wind}

Ionized plasma flowing radially outward from the Sun modifies the total electron content along the line of sight within the Solar system, inducing a frequency dependent delay that varies with both the pulsar’s DM and its solar elongation as they change over the course of a year. We model this effect in a deterministic way by expanding the solar wind electron density at the location of the Earth, $n_{\rm earth}$ to first order in time (thereby introducing $\dot{n}_{\rm earth}$) around a fixed reference epoch $t_{\rm ref}$, chosen to coincide with the beginning of the observing baseline. The full form of this deterministic delay is provided in Eq.~(11) of \citep{ND+26}.

\subsubsection{Treatment of single-pulsar noise in the array analyses}
\label{sec:spn_treatment}

In all array-level analyses, the per-pulsar noise model, that is, the set of processes included for each pulsar and the number of Fourier components used for each, is taken from the corresponding single-pulsar noise analysis (SPNA): Ref.~\cite{ND+26} for DR2 and Ref.~\cite{SDK+23} for DR1. The white noise parameters (EFAC, EQUAD, and ECORR), as is standard practice in PTA analyses, and the FCN chromatic index ($\chi_{\rm fcn}$) are held fixed at the medians of their one-dimensional marginal posteriors from that analysis. All remaining time-correlated single-pulsar parameters are sampled simultaneously with the common-process parameters and marginalized over. These are the ARN amplitude and spectral index, the DMN amplitude and spectral index, the FCN amplitude and spectral index, and the solar-wind parameters $n_{\mathrm{earth}}$ and, for PSR~J1909$-$3744 only, $\dot{n}_{\mathrm{earth}}$, each with the priors of Ref.~\citep{ND+26}. Only six of the 27 pulsars include an ARN term in their SPNA model; in the other 21, unmodelled achromatic red power is absorbed mainly by the common process, so in this respect the common-process amplitude, and hence its upper limit, is conservative. The common process has a log-uniform prior on $A_{\rm CURN}$ over $[10^{-18}, 10^{-11}]$ in all array analyses and, where its spectral index is free, a uniform prior on $\gamma_{\rm CURN}$ over $[0, 7]$; all priors are summarized in Table~\ref{tab:priors}. The upper-limit runs use the same amplitude range. This configuration is used for the free-spectral-index CURN search, the dropout analysis and the HD reweighting (Section~\ref{sec:hd_reweight}), and, with the spectral index fixed at $\gamma = 13/3$, for the upper limit (Section~\ref{sec:upper_limit}). For the noise-marginalized optimal statistic (NMOS; Section~\ref{sec:os}), the white-noise parameters remain fixed at the SPNA medians, while the time-correlated single pulsar noise parameters are drawn from the posterior of the free spectral-index CURN analysis.

\begin{table}
\caption{Parameters and priors in the array-level analyses. ''Free-$\gamma$'' denotes the free-spectral-index CURN search, the dropout analysis, and the chains used for HD reweighting and the NMOS; ''UL'' denotes the fixed-$\gamma$ upper-limit runs, whose log-uniform amplitude posterior is reweighted to a prior uniform in $A$ over $[10^{-18}, 10^{-11}]$ (Section~\ref{sec:upper_limit}). Single-pulsar parameters are included only for pulsars whose SPNA model contains the corresponding process, with the numbers of Fourier components and the priors of Ref.~\citep{ND+26}. $\mathcal{U}(a,b)$ denotes a uniform distribution; $n_{\rm earth}$ is in cm$^{-3}$ and $\dot{n}_{\rm earth}$ in cm$^{-3}$\,yr$^{-1}$.}
\label{tab:priors}
\begin{tabular}{lll}
\hline\hline
Parameter & Prior & Analyses \\
\hline
\multicolumn{3}{l}{\textbf{Common process}} \\
$\log_{10} A_{\rm CURN}$ & $\mathcal{U}(-18, -11)$ & Free-$\gamma$, UL \\
$\gamma_{\rm CURN}$ & $\mathcal{U}(0, 7)$ & Free-$\gamma$ \\
 & Fixed at $13/3$ & UL \\
$k_a$ & $\{0, 1\}$, equal probability & Dropout \\
\hline
\multicolumn{3}{l}{\textbf{White noise}} \\
EFAC, EQUAD, ECORR & Fixed at SPNA medians & All \\
\hline
\multicolumn{3}{l}{\textbf{Achromatic red noise}} \\
$\log_{10} A_{\rm arn}$ & $\mathcal{U}(-20, -11)$ & All \\
$\gamma_{\rm arn}$ & $\mathcal{U}(0, 7)$ & All \\
\hline
\multicolumn{3}{l}{\textbf{Dispersion measure noise}} \\
$\log_{10} A_{\rm dmn}$ & $\mathcal{U}(-20, -10)$ & All \\
$\gamma_{\rm dmn}$ & $\mathcal{U}(0, 7)$ & All \\
\hline
\multicolumn{3}{l}{\textbf{Free chromatic noise}} \\
$\log_{10} A_{\rm fcn}$ & $\mathcal{U}(-20, -10)$ & All \\
$\gamma_{\rm fcn}$ & $\mathcal{U}(0, 7)$ & All \\
$\chi_{\rm fcn}$ & Fixed at SPNA medians & All \\
\hline
\multicolumn{3}{l}{\textbf{Deterministic solar wind}} \\
$n_{\rm earth}$ & $\mathcal{U}(0, 20)$ & All \\
$\dot{n}_{\rm earth}$ & $\mathcal{U}(-20, 20)$ & All \\
\hline\hline
\end{tabular}
\end{table}
\subsubsection{Common Uncorrelated Red Noise}

As an intermediate step, we first search for a common spectrum process affecting all pulsars with identical power spectral density but independent phase realizations. This common uncorrelated red noise model has covariance:

\begin{equation}
    C_{\text{CURN}} = \text{block-diag}(C_1^{\text{CURN}}, C_2^{\text{CURN}}, \ldots, C_M^{\text{CURN}}), 
    \label{eq:curn}
\end{equation}

where $M$ is the total number of pulsars in the array ($M = 27$ for DR2) and each block has the same functional form as red noise with common amplitude $A_{\text{CURN}}$ and spectral index $\gamma_{\text{CURN}}$, but independent Fourier coefficients for each pulsar. The block-diagonal structure reflects that phases are uncorrelated between pulsars.

The CURN model is physically motivated as a potential signature of errors in the Solar System ephemeris, errors in the terrestrial time standard, or inadequacies in the noise models that affect all pulsars similarly. It could also indicate the presence of an unmodelled excess noise signal in the data indicating that it could be the SGWB. Hence, detecting CURN is a necessary but not sufficient condition for claiming an SGWB detection.

\subsubsection{Hellings-Downs Correlations}

The true SGWB signature is distinguished from CURN by the presence of inter-pulsar correlations following the Hellings-Downs curve \citep{HD83}. For a stochastic background of gravitational waves with an isotropic and unpolarized distribution, the expected correlation between timing residuals from pulsars $a$ and $b$ separated by angle $\zeta$ is:

\begin{equation}
    \Gamma_{\text{HD}}(\zeta_{ab}) = \frac{1}{2} - \frac{1}{4}x_{ab} + \frac{3}{2}x_{ab} \ln x_{ab} + \frac{1}{2}\delta_{ab}
    \label{eq:hellings_downs}
\end{equation}

where $x_{ab} = (1 - \cos\zeta_{ab})/2$. The Kronecker delta $\delta_{ab}$ ensures that the autocorrelation ($a = b$, or $\zeta = 0$) is normalized to unity.

The SGWB covariance matrix includes both autocorrelation terms (affecting individual pulsars) and cross-correlation terms (coupling different pulsars):

\begin{multline}
    C_{\text{GWB}}[a,b][t_i, t_j] = \Gamma_{\text{HD}}(\zeta_{ab}) \\
    \times \sum_{k=1}^{N_{\text{GWB}}} S_{\text{GWB}}(f_k)\,\Delta f\, \cos\!\left[2\pi f_k (t_i - t_j)\right]
    \label{eq:gwb_covariance}
\end{multline}

where $a$ and $b$ index pulsars, $i$ and $j$ index observation epochs, the Fourier frequencies are $f_k = k/T_{\rm span}$ with spacing $\Delta f = 1/T_{\rm span}$, $T_{\rm span}$ is the total time span of the observations, and $S_{\text{GWB}}(f)$ follows the power law introduced above. Unless stated otherwise, the common process is modelled with $N_{\text{GWB}} = 7$ components (corresponding to $f_k \leq 7/T_{\rm span} \approx 30.8$~nHz, i.e., all harmonics below $1\,\mathrm{yr}^{-1}$, for the 7.2-yr DR2 span, and applied likewise to DR1 and the simulated datasets).

Equation~\eqref{eq:gwb_covariance}, like the expressions for the single-pulsar red processes, assumes a covariance that is diagonal in frequency. For a finite observing window this is an approximation: power leaks between Fourier bins, and power below $1/T_{\rm span}$ is partly absorbed by the timing-model fit and partly redistributed into the lowest bins. Neglecting these correlations can bias the common-process parameters toward flatter spectra and higher amplitudes, particularly for short baselines such as ours~\citep{CVM+26,FFF+26}. We do not expect this to affect our non-detection or upper limit significantly, but the free-spectral-index posteriors of Section~\ref{sec:dr2_results} should be read with this caveat in mind.

For an SGWB from an ensemble of circular SMBHBs in the gravitational wave driven regime, the expected spectral index is:

\begin{equation}
    \gamma_{\text{GWB}} = \frac{13}{3} \approx 4.33
    \label{eq:smbhb_index}
\end{equation}

corresponding to a characteristic strain spectrum $h_c(f) \propto f^{-2/3}$. We perform analyses both with $\gamma_{\text{GWB}}$ fixed at this value and with $\gamma_{\text{GWB}}$ as a free parameter to constrain the GWB signal and to find the upper limit.

\subsection{Dropout Factor}
\label{sec:dropout}

The detection of CURN is only meaningful if it is supported by the array as a whole rather than driven by one or a few pulsars. To assess the contribution of each pulsar to the common signal, we employ the \emph{dropout} method \citep{ABB+20}, in which the participation of each pulsar in the common process is itself made a free parameter of the model.

For each pulsar $a$ we introduce a binary indicator parameter $k_a \in \{0,1\}$ that toggles whether that particular pulsar's timing residual includes the common red process. When $k_a = 1$, the common term contributes to the covariance of pulsar $a$; when $k_a = 0$ it is removed, leaving the pulsar's intrinsic (white, achromatic, and chromatic) noise model unchanged. The common process covariance for pulsar $a$ is thus scaled as

\begin{equation}
    C^{\text{CP}}_{a} \;\longrightarrow\; k_a\, C^{\text{CP}}_{a},
    \qquad k_a \in \{0,1\}
    \label{eq:dropout_switch}
\end{equation}

with the amplitude and spectral index of the common process shared across all pulsars. The indicator parameters $\{k_a\}$ are sampled jointly with the common process and noise parameters in a single Bayesian run, so that one analysis yields a participation posterior for every pulsar simultaneously.

The quantity of interest is the per pulsar \emph{dropout factor}, defined as the posterior odds that pulsar $a$ participates in the common process, normalized by the corresponding prior odds:

\begin{equation}
{\rm DF}_a = \left.\frac{p(k_a = 1 \mid \delta t)}{p(k_a = 0 \mid \delta t)}
             \right/ \frac{\pi(k_a = 1)}{\pi(k_a = 0)}
\label{eq:dropout_factor}             
\end{equation}

where the posterior probabilities are estimated from the fraction of samples in which $k_a = 1$ and $k_a = 0$, respectively. With equal prior probability for the two states, $\pi(k_a=1) = \pi(k_a=0)$, the dropout factor reduces to the posterior odds and acts as a Bayes factor for the inclusion of pulsar $a$ in the common signal. A value $\mathrm{DF}_a > 1$ indicates that the data favor that pulsar's participation, $\mathrm{DF}_a < 1$ disfavors it, and $\mathrm{DF}_a \approx 1$ signifies indifference. In the absence of a genuine common signal, the dropout factors of all pulsars are expected to cluster about unity, whereas a true common process recovered consistently across the array produces dropout factors systematically above one. The dropout factors reported in Section~\ref{sec:results} are computed in this way.

\subsection{Savage-Dickey Bayes Factor}
\label{sec:savage_dickey}

As a measure of the evidence for CURN, we compute the Savage-Dickey Bayes factor \citep{DM71}. When the no signal hypothesis is nested at the lower edge of the amplitude prior, the Bayes factor in favor of the common process is the ratio of the prior to the posterior density of $\log_{10} A_{\rm CURN}$ evaluated at that no signal value,

\begin{equation}
    \mathrm{BF}^{\rm CURN}_{10}
    = \frac{\pi(\log_{10} A_{\rm CURN} = \log_{10} A_{0})}
           {p(\log_{10} A_{\rm CURN} = \log_{10} A_{0} \mid \boldsymbol{\delta t})},
    \label{eq:savage_dickey}
\end{equation}

where the posterior density is estimated from the CURN chain with a reflection-corrected kernel density estimate, and $\log_{10} A_0 = -18$ is the lower bound of the log-uniform prior, which spans $[-18, -11]$. For a fixed prior, the flat low-amplitude posterior plateau means the prior-to-posterior density ratio does not depend on where within the plateau it is evaluated. The Bayes factor itself, however, depends on the prior width $\Delta\log_{10} A = \log_{10} A_{\max} - \log_{10} A_0$, scaling toward unity as the lower bound is extended because the added prior volume has the no-signal likelihood and therefore dilutes the evidence for the common process. Lowering the bound to $\log_{10} A_0 = -20$, for example, reduces the Bayes factor from $2.5$ to $\approx 2.2$, and $\mathrm{BF}^{\mathrm{CURN}}_{10} \to 1$ as $\log_{10} A_0 \to -\infty$. The reported values are therefore conditional on the conventional $[-18, -11]$ prior range.

\subsection{Upper Limit on the Common-Process Amplitude}
\label{sec:upper_limit}

In the absence of a common-process detection, we place an upper limit on the common red-noise amplitude. The spectral index is fixed at the SMBHB value $\gamma_{\rm GWB}=13/3$ (Equation~\ref{eq:smbhb_index}), and the common process is modelled as a spatially uncorrelated power law shared across all pulsars. As described in Section~\ref{sec:spn_treatment}, the white-noise parameters are held fixed at the SPNA medians, while the time-correlated single-pulsar noise parameters are sampled simultaneously with the common-process amplitude and marginalized over.

Upper limits on a positive amplitude are prior-dependent. We therefore sample efficiently under a log-uniform prior on $A_{\rm GWB}$ and then obtain the conventional uniform in amplitude (LinearExp) posterior by importance reweighting, assigning each post burn-in sample the prior ratio

\begin{equation}
    w_i \propto \exp\!\big[\ln\pi_{\rm LinearExp}(\boldsymbol{\theta}_i)
                          - \ln\pi_{\rm log\text{-}uniform}(\boldsymbol{\theta}_i)\big].
    \label{eq:importance_weight}
\end{equation}

Because the two models differ only in this prior, the likelihood cancels and the reweighting is exact up to Monte Carlo error, with the Kish effective sample size \citep{Kish95} confirming a stable reweighted posterior in each case. The $95\%$ upper limit is then the $0.95$ weighted quantile of the marginal amplitude posterior.

\subsection{Likelihood Reweighting from CURN to HD}
\label{sec:hd_reweight}

A full HD search requires evaluating the cross-correlation terms of the array covariance matrix at every sample, which is substantially more computationally expensive than the block-diagonal CURN model. Following \citep{HM+23}, we instead sample under the CURN model and reweight the resulting chain to the HD-correlated model by importance sampling. Each post burn-in sample is assigned the weight

\begin{equation}
    w_i = \frac{\mathcal{L}_{\rm HD}(\boldsymbol{\delta t}\mid\boldsymbol{\theta}_i)}
               {\mathcal{L}_{\rm CURN}(\boldsymbol{\delta t}\mid\boldsymbol{\theta}_i)},
    \label{eq:hd_weight}
\end{equation}

the ratio of the HD to the CURN likelihood evaluated at the same parameters $\boldsymbol{\theta}_i$, comprising the common-process parameters $(\gamma_{\rm CURN}, \log_{10} A_{\rm CURN})$ and the single-pulsar red-noise parameters of sample $i$ (Section~\ref{sec:spn_treatment}). Because the two models share an identical prior and differ only in the spatial-correlation structure of the covariance, this reweighting is exact up to Monte Carlo error, along with the Kish effective sample size \citep{Kish95} quantifying its reliability. The reweighted chain approximates the posterior that a direct HD search would have produced, albeit at a small fraction of the computational cost.

\subsection{Optimal Statistic}
\label{sec:os}

In addition to the Bayesian analysis, we use the hybrid frequentist optimal statistic \citep{AB+09,DF+13,CC+15}, whose amplitude estimator is

\begin{equation}
\hat{A}^2_{\rm GWB} =
\frac{\sum_{I=1}^{M-1}\sum_{J=I+1}^{M} r_I^{T} P_I^{-1}\tilde{S}_{IJ}P_J^{-1} r_J}
     {\sum_{I=1}^{M-1}\sum_{J=I+1}^{M}\mathrm{Tr}\!\left[P_I^{-1}\tilde{S}_{IJ}P_J^{-1}\tilde{S}_{JI}\right]},
\label{eq:os_amp}
\end{equation}
where $P_I$ is the auto-covariance matrix of pulsar $I$, $\tilde{S}_{IJ}$ is the
cross-covariance template for the pulsar pair $(I,J)$, $r_I$ is the vector of timing
residuals of pulsar $I$, and the sums run over the $M(M-1)/2$ distinct pulsar pairs.
The corresponding signal-to-noise ratio (S/N) is
\begin{equation}
\frac{S}{N} = \frac{\hat{A}^2_{\rm GWB}}{\sigma_0} =
\frac{\sum_{I=1}^{M-1}\sum_{J=I+1}^{M} r_I^{T} P_I^{-1}\tilde{S}_{IJ}P_J^{-1} r_J}
     {\left(\sum_{I=1}^{M-1}\sum_{J=I+1}^{M}\mathrm{Tr}\!\left[P_I^{-1}\tilde{S}_{IJ}P_J^{-1}\tilde{S}_{JI}\right]\right)^{1/2}}.
\label{eq:os_snr}
\end{equation}

When pulsars have significant red noise, the pure optimal statistic gives biased results due to the strong covariance between the individual red noise parameters and the SGWB amplitude. Hence, we use the Noise Marginalized Optimal Statistic (NMOS) method developed by \cite{VI+18}. In this method, instead of fixing the noise parameter, we marginalize the optimal statistic over the posterior distribution of noise parameters.

\begin{equation}
    \label{eq:nmos_amp}
    \left\langle \hat{A}_{\mathrm{GWB}}^{2} \right\rangle = \int \hat{A}_{\mathrm{GWB}}^{2}(\phi)\, p\!\left(\phi \mid \delta t\right)\, d\phi
\end{equation}

where $\phi$ denotes a single sample from the posterior of the free spectral-index CURN analysis, comprising the time-correlated single-pulsar noise parameters and the common-process parameters $(\gamma_{\rm CURN}, \log_{10} A_{\rm CURN})$; the white noise is fixed at the SPNA medians (Section~\ref{sec:spn_treatment}). For each draw, $\mathbf{P}_I$ includes the intrinsic noise and the common process at the drawn values, and $\tilde{\mathbf{S}}_{IJ}$ assumes a $\gamma = 13/3$ power law; repeating this over many draws gives distributions of $\hat{A}^2_{\rm GWB}$ and of the S/N.

\section{Results}
\label{sec:results}

We apply our search to two InPTA datasets where we use the DR1 as a consistency check. We apply the dropout method (Sec.~\ref{sec:dropout}), confirm the non-detection is robust to removing the pulsar with the highest dropout factor and set an upper limit that depends noticeably on the amplitude prior (Sec.~\ref{sec:upper_limit}). DR2 provides our headline constraint, analyzed with the full battery of four diagnostics, namely the Bayesian common-spectrum posterior, its Savage-Dickey Bayes factor (Sec.~\ref{sec:savage_dickey}), the per-pulsar dropout factors (Sec.~\ref{sec:dropout}), and the noise marginalized optimal statistic (Sec.~\ref{sec:os}), together with CURN to Hellings-Downs reweighting (Sec.~\ref{sec:hd_reweight}), an amplitude upper limit, and robustness checks against solar-wind cuts and Band 3 removal. Unless stated otherwise, quoted uncertainties are 90\% credible intervals (5th-95th percentiles) about the median.

\subsection{InPTA DR1}
\label{sec:dr1_dropout}

To assess whether any single pulsar drives the evidence for the CURN in the InPTA DR1 dataset, we applied the dropout method.

\begin{figure}[htbp]
    \centering
    \includegraphics[width=\linewidth]{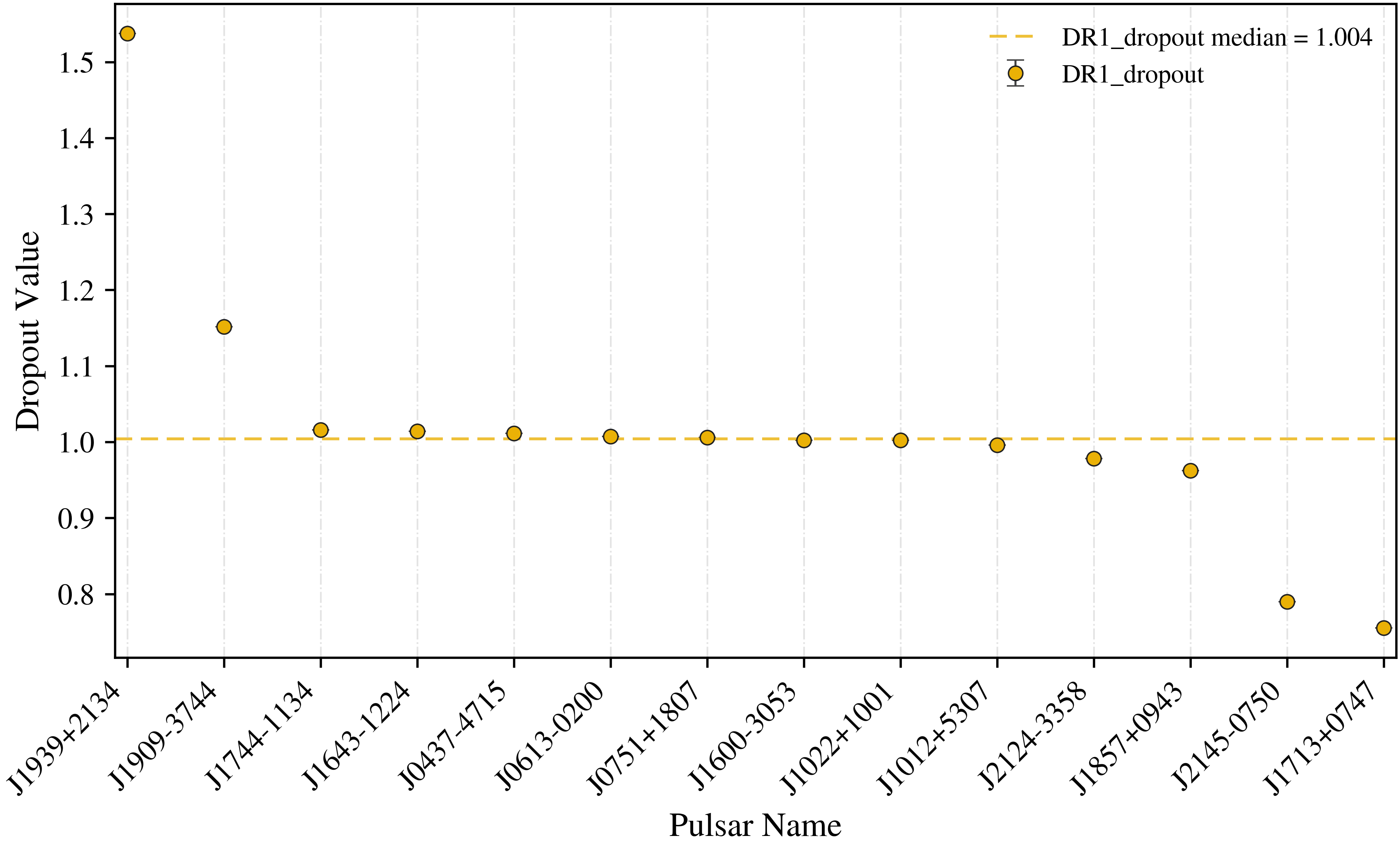}
    \caption{Per-pulsar dropout factors for the common process in InPTA DR1, sorted by value. Values above unity indicate support for that pulsar's participation in the common process. The dashed line marks the median, $1.004$; all pulsars lie below $1.2$ except PSR~J1939$+$2134.}
    \label{fig:DR1_dropout}
\end{figure}

The recovered dropout values (Figure~\ref{fig:DR1_dropout}) cluster about unity, with a median of $1.004$ and all but one pulsar below $1.2$. The exception is PSR~J1939$+$2134 at $\simeq 1.55$. As a robustness check, we removed J1939$+$2134 and repeated the analysis; the recovered amplitude spectral index posterior is essentially unchanged, confirming that the non-detection is not driven by any single pulsar.

\begin{figure}[htbp]
    \centering
    \includegraphics[width=\linewidth]{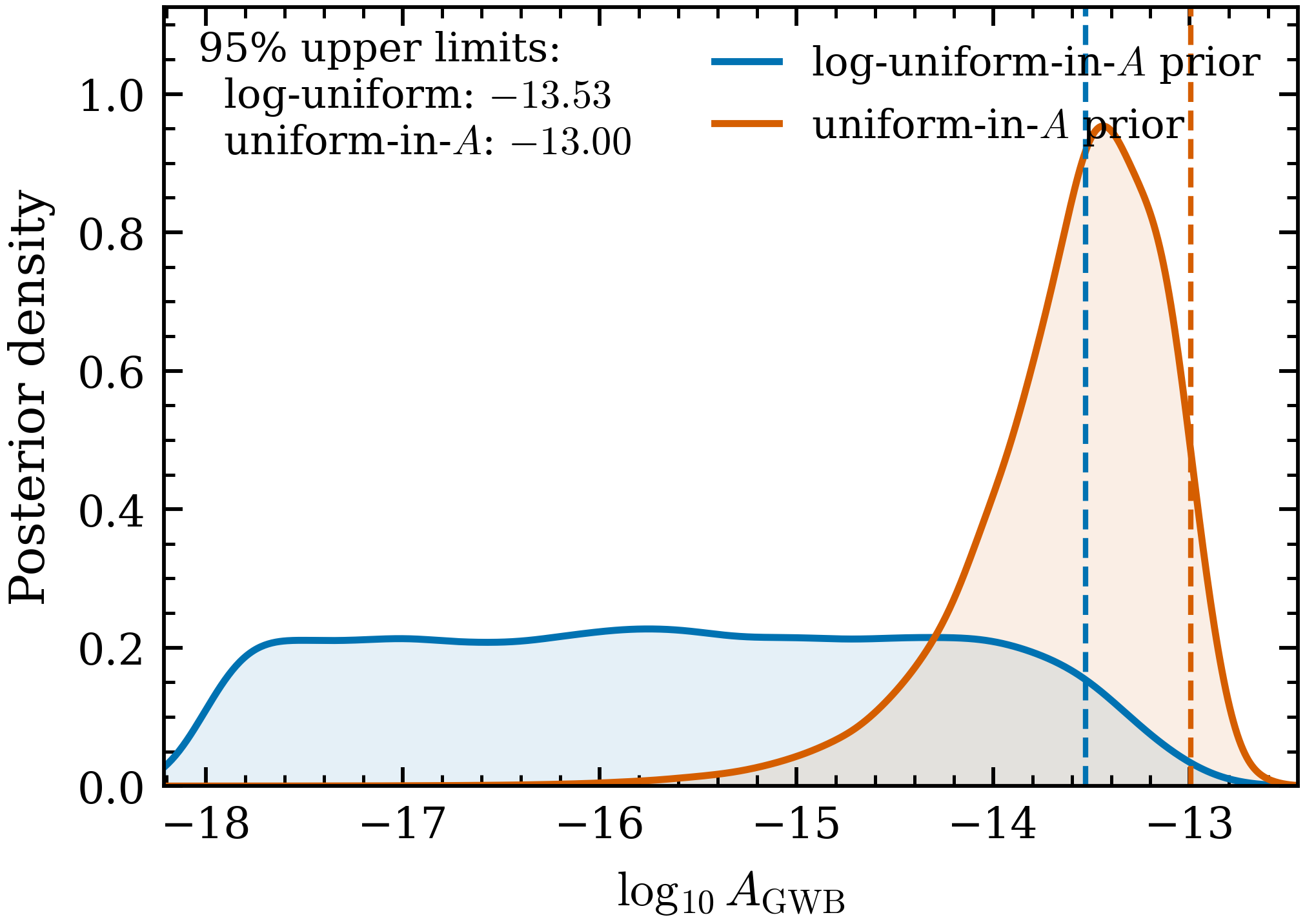}
    \caption{Marginal posterior on $\log_{10} A_{\text{GWB}}$ at fixed
    $\gamma_{\text{GWB}} = 13/3$ for the \textbf{InPTA DR1} dataset: raw
    log-uniform-prior chain (blue) and LinearExp-reweighted posterior (orange).
    Dashed lines mark the $95\%$ upper limits ($-13.53$ and $-13.00$); the
    latter is adopted as fiducial.}
    \label{fig:upper_limit_crn}
\end{figure}

Given the absence of a common-process detection in DR1, we apply the upper-limit procedure of Section~\ref{sec:upper_limit}. The log-uniform prior gives $\log_{10} A_{\rm GWB} < -13.53$, while the fiducial LinearExp prior yields $\log_{10} A_{\rm GWB} < -13.00$ ($95\%$ C.L.; Figure~\ref{fig:upper_limit_crn}). This limit lies above the amplitudes reported by the longer baseline PTAs \citep{AAB+23,AAA+23b,RD+23,XH+23}; for the short DR1 baseline the limit is prior-sensitive, shifting by $0.53$ in $\log_{10} A$ between the two priors, and a short span is known to weaken the constraint and bias amplitudes high \citep{PTV+21}.

\subsection{InPTA DR2}
\label{sec:dr2_results}

We now present the principal results of this study, obtained from the complete DR2 dataset comprising 27 pulsars observed over a $\sim$7.2-year timespan. We evaluate the presence of a common red-noise process using four complementary diagnostics: (i) the Bayesian common-spectrum amplitude posterior, (ii) its Savage-Dickey Bayes factor against the no-signal hypothesis, (iii) the per-pulsar dropout factors, and (iv) the noise marginalized optimal statistic. In the absence of a statistically significant detection across these diagnostics, we then place an upper limit on the common-process amplitude.

\begin{figure*}[htbp]
    \centering

    \subfigure{
        \includegraphics[width=0.48\textwidth]{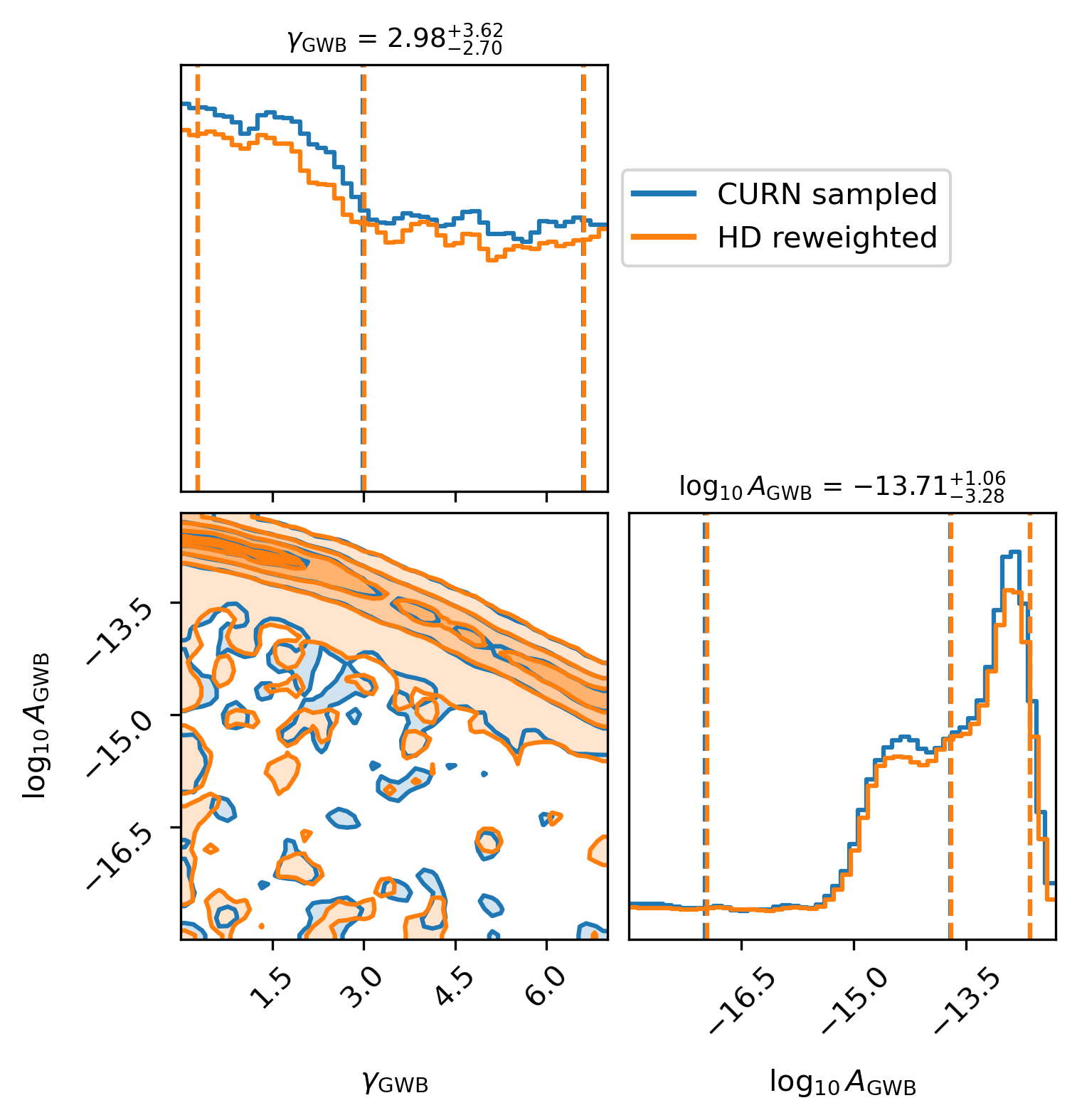}
        \label{fig:corner_dr2}
    }
    \hfill
    \subfigure{
        \includegraphics[width=0.48\textwidth]{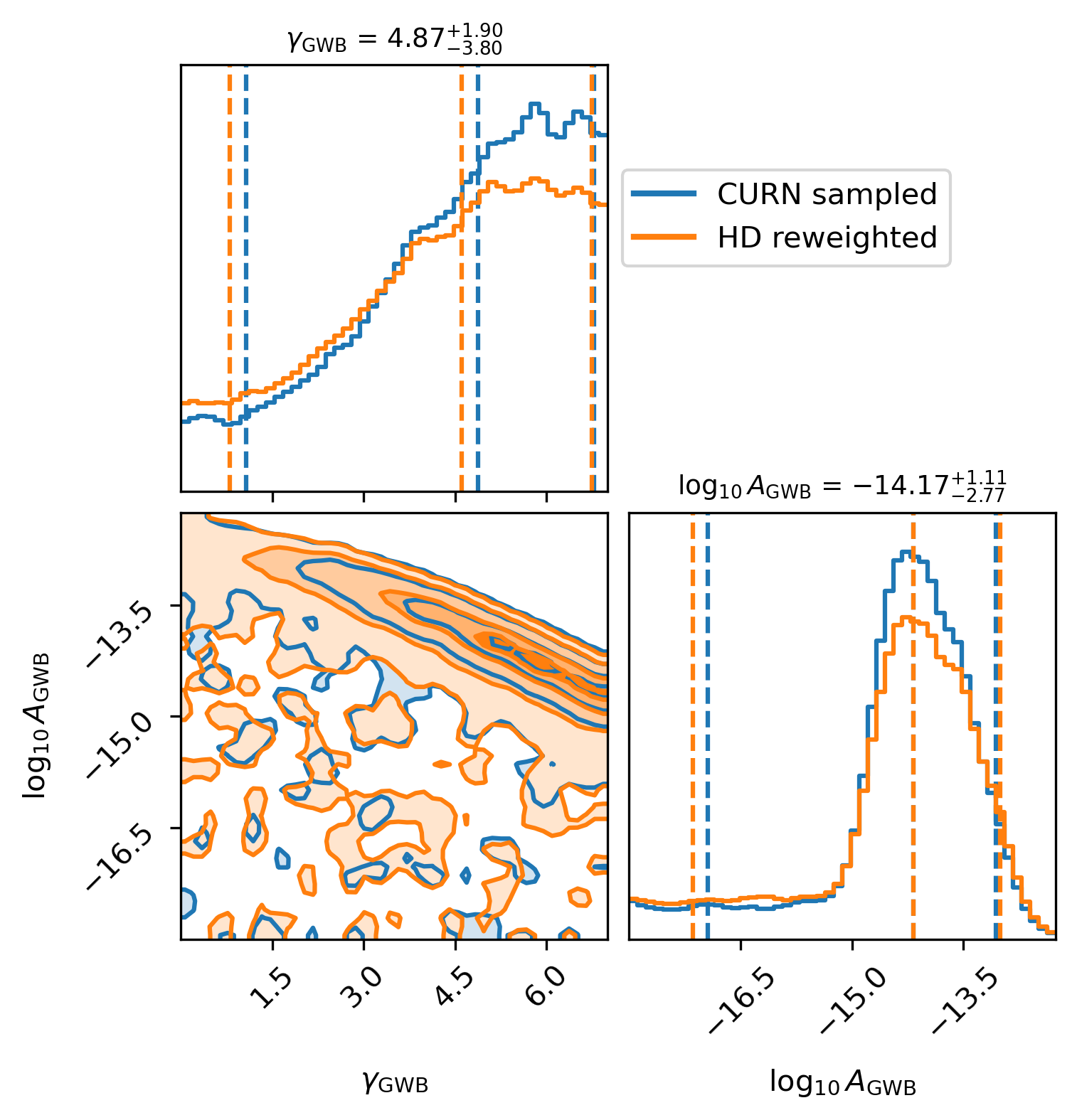}
        \label{fig:corner_band5}
    }

        \caption{Joint posterior on the common-process amplitude $\log_{10} A_{\rm CURN}$ and spectral index $\gamma_{\rm CURN}$ for the full DR2 (left) and Band 5-only (right) datasets, showing the CURN-sampled chain (blue) and the same chain reweighted to the Hellings-Downs correlated model (orange). Titles give the median and 90\% credible interval (5th-95th percentiles), marked by the dashed lines. The two posteriors are nearly identical, indicating that the data do not prefer the HD spatial correlation over a spatially uncorrelated common process.}
    \label{fig:DR2_corner_baselines}
\end{figure*}

\begin{figure*}[t]
    \centering
    \includegraphics[width=\textwidth]{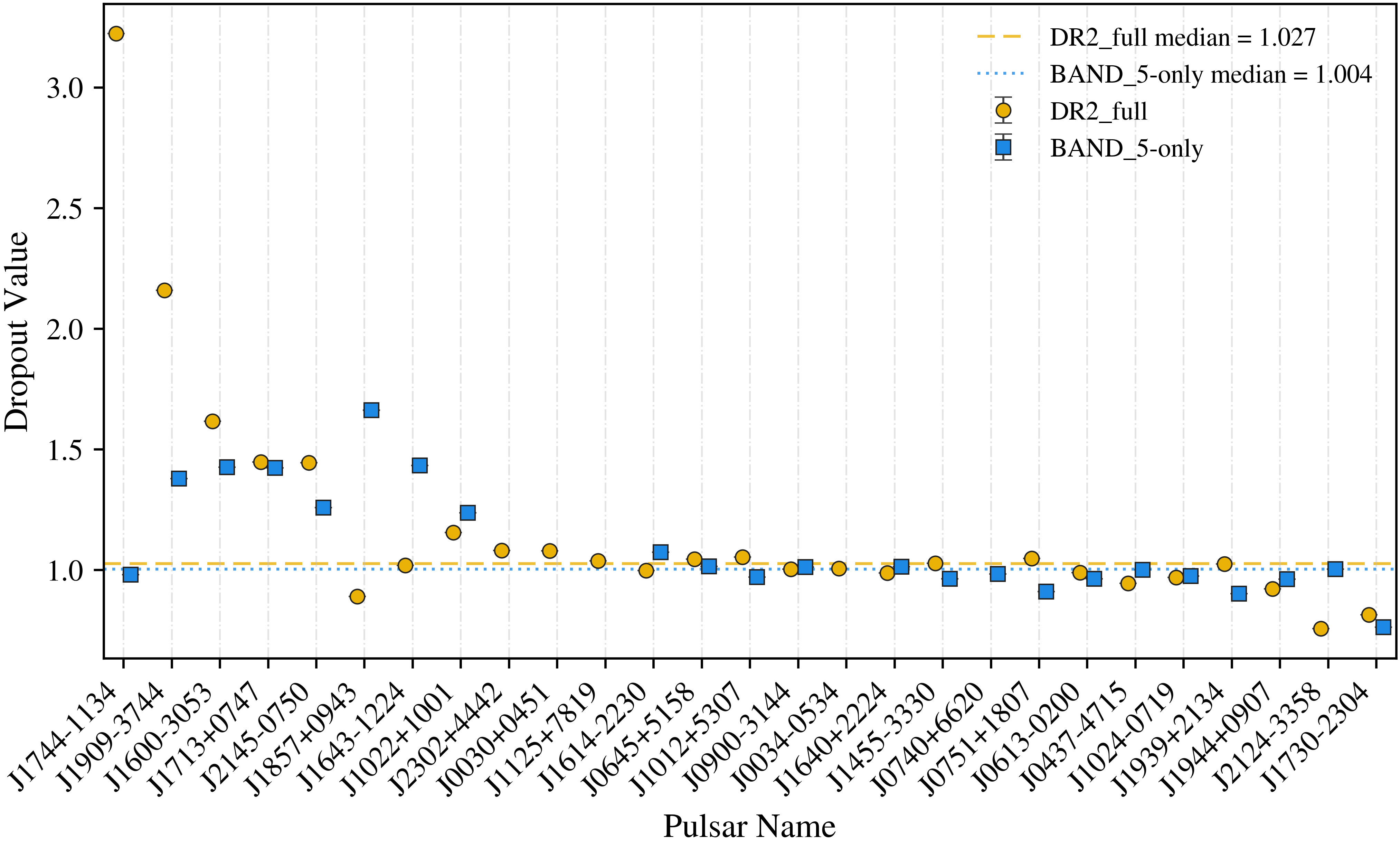}
    \caption{Per-pulsar dropout factors for the common process for the InPTA full DR2 (yellow) and Band 5-only (blue) datasets. Median values (dashed) are $1.027$ and $1.004$, respectively, both consistent with unity. Values above one indicate support for participation in the common process; when Band~3 is removed, PSR~J1744$-$1134 drops from $3.22$ to $0.98$ and PSR~J1909$-$3744 from $2.16$ to $1.38$, comparable to the scatter among several pulsars in the Band~5-only analysis.}
    \label{fig:dropout}
\end{figure*}

\subsubsection{Common-process posterior and Bayes factor}
\label{sec:dr2_cp}

We first search for a CURN process across the array using the Bayesian framework of Section~\ref{sec:Method}, with the spectral index left as a free parameter. The recovered marginal posterior on the amplitude is broad, with $\log_{10} A_{\rm CURN} = -13.71^{+1.06}_{-3.28}$ and a spectral index $\gamma_{\rm CURN} = 2.98^{+3.62}_{-2.70}$ consistent with a wide range of values, indicating the absence of a well constrained common-process signal (Figure~\ref{fig:DR2_corner_baselines}).

The full DR2 posterior is not a single compact peak. Besides the region consistent with the Band~5-only posterior, it places substantial weight at low spectral index and correspondingly high amplitude, along the amplitude spectral-index degeneracy. The marginal $\gamma_{\rm CURN}$ posterior declines smoothly from low $\gamma$ rather than separating into distinct peaks, so we do not regard this as clear evidence for two separate modes; for brevity, we refer to this structure as the low-$\gamma$ mode.

The low-$\gamma$ mode appears to depend on the high-frequency end of the common-process basis. Restricting the search to $N_{\text{GWB}} = 3$ components ($f \leq 3/T_{\rm span} \approx 13$~nHz) removes the preference for shallow spectra, whereas the Band~5-only posterior changes little between three and seven components (Table~\ref{tab:ngwb}). Of these four combinations, only the full DR2 data with seven components favour low $\gamma$. This suggests that the mode is supported mainly by the power in the bins $4/T_{\rm span}$-$7/T_{\rm span}$ ($\approx 18$-$31$~nHz). A power-law common process can accommodate additional power in these bins only by flattening its spectrum, and because the amplitude is referenced to $f = 1\,\mathrm{yr}^{-1}$, close to the highest of these bins, this also raises $\log_{10} A_{\rm CURN}$. The Band~5-only data, which cover the same observing span, show no such preference, so the excess power appears only when the Band~3 data are included. Because the Band~5-only data are less sensitive, this comparison alone does not establish where the power originates. We examine the contribution of individual pulsars in Section~\ref{sec:dr2_dropout} and test a solar-wind origin in Section~\ref{sec:solar_robust}.

\begin{table}
    \caption{Free spectral-index CURN posteriors (median and 90\% credible interval, as in Figure~\ref{fig:DR2_corner_baselines}) for the full DR2 and Band 5-only data, with the common process modelled by $N_{\text{GWB}} = 3$ and $7$ Fourier components, and for the full DR2 data without PSR~J1744$-$1134. The $N_{\text{GWB}} = 7$ entries here are from dedicated CURN-only runs, whereas Figure~\ref{fig:DR2_corner_baselines} displays the separate chains used for HD reweighting; the two sets of runs agree to within $0.03$ for the full DR2 dataset and well within the broad credible intervals for the less-constrained Band 5-only dataset.}
    \label{tab:ngwb}
    \begin{tabular}{llcc}
        \hline\hline
        Data & $N_{\text{GWB}}$ & $\gamma_{\rm CURN}$ & $\log_{10} A_{\rm CURN}$ \\
        \hline
        Full DR2 & 3 & $4.05^{+2.69}_{-3.52}$ & $-14.15^{+1.39}_{-3.08}$ \\
        Full DR2 & 7 & $3.01^{+3.60}_{-2.73}$ & $-13.70^{+1.05}_{-3.25}$ \\
        Full DR2 without J1744$-$1134 & 7 & $4.00^{+2.73}_{-3.53}$ & $-14.21^{+1.37}_{-3.11}$ \\
        Band 5 only & 3 & $4.61^{+2.14}_{-3.80}$ & $-14.18^{+1.17}_{-2.96}$ \\
        Band 5 only & 7 & $4.54^{+2.22}_{-3.74}$ & $-14.20^{+1.13}_{-3.10}$ \\
        \hline\hline
    \end{tabular}
\end{table}

As a direct test for Hellings-Downs spatial correlations, we reweight the CURN chain to the HD-correlated model following Section~\ref{sec:hd_reweight}. The reweighted HD posterior is essentially indistinguishable from the CURN posterior in both amplitude and spectral index (Figure~\ref{fig:DR2_corner_baselines}).

To quantify the evidence for the common-process, we compute the Savage-Dickey Bayes factor (Section~\ref{sec:savage_dickey}), which yields a value of 2.5 in favor of CURN over the no-signal hypothesis. On the Jeffreys scale~\cite{Trotta08} this does not exceed the level of a bare mention, confirming that the data provide no statistically significant support for a common red signal.

\begin{figure}[htbp]
    \centering
    \includegraphics[width=0.95\linewidth]{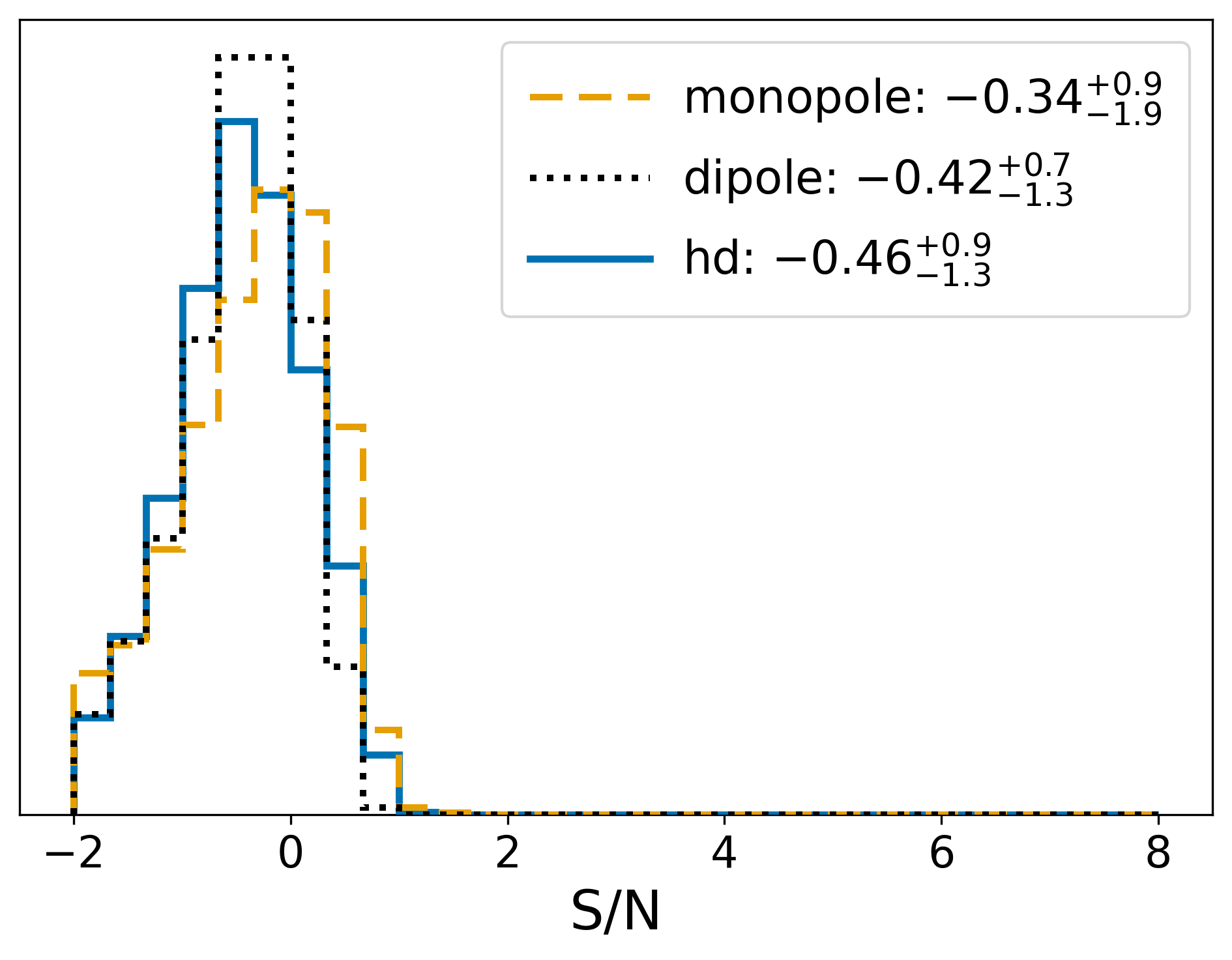}
    \caption{Noise marginalized distributions of the optimal-statistic signal to noise ratio ($\mathrm{S/N}$) for the three overlap reduction functions: the Hellings-Downs quadrupolar correlation expected for an isotropic SGWB (blue, solid), a monopole (orange, dashed), and a dipole (black, dotted). The colours are taken from a colour-blind-safe palette, and the line styles distinguish the curves independently of colour. All three distributions peak near zero and are consistent with non-detection; in particular, the Hellings-Downs $\mathrm{S/N}$ shows no significant excess, consistent with the absence of a detectable isotropic SGWB. Legend values give the median and 90\% interval (5th-95th percentiles).}
    \label{fig:os_sn}
\end{figure}

\subsubsection{Dropout factors}
\label{sec:dr2_dropout}

To test whether the common-process is supported by the array as a whole rather than driven by a few pulsars, we compute the per-pulsar dropout factors of Section~\ref{sec:dropout} for two configurations (Figure~\ref{fig:dropout}). For the fiducial combined full DR2 data, the dropout factors have a median of $1.027$, and for the Band 5 only data, a median of $1.004$; both are consistent with unity, so the array as a whole shows no coherent preference for a common process. The median is, however, insensitive to individual outliers, which we examine next.

The dual-band data, however, reveal a structure that would be difficult to identify with a single-band array. Two of the most precisely timed pulsars, PSRs~J1744$-$1134 and J1909$-$3744, show dropout factors above unity in the combined full DR2 analysis (3.22 and 2.16), but fall to 0.98 and 1.38 once the Band~3 data are removed. One possible explanation is red power that the per-pulsar noise model does not fully capture and that the common process absorbs in the most precisely timed pulsars. We examine this possibility below.

This possibility was anticipated by the companion noise analysis \citep{ND+26}, which identifies improved chromatic noise modelling as a priority to prevent such leakage into the achromatic component; the dropout excess may be its array-level manifestation.

Because the dropout factors are calculated over the full common-process posterior, they may depend on the low-$\gamma$ mode. With $N_{\text{GWB}} = 3$, which removes that mode without an arbitrary split in $\gamma$, the dropout factor of PSR~J1744$-$1134 decreases from 3.22 to 1.15, whereas that of PSR~J1909$-$3744 changes little (2.16 to 1.98). Given the sampling uncertainty, we draw conclusions only from the large change for PSR~J1744$-$1134, which suggests that its preference is associated with the high-frequency power supporting the low-$\gamma$ mode. Repeating the $N_{\text{GWB}} = 7$ search without PSR~J1744$-$1134 also removes the low-$\gamma$ preference ($\gamma_{\rm CURN} = 4.00^{+2.73}_{-3.53}$, $\log_{10} A_{\rm CURN} = -14.21^{+1.37}_{-3.11}$; Table~\ref{tab:ngwb}). Since this pulsar is among the most precisely timed in the array, its removal also reduces the sensitivity at these frequencies, so we regard this result as indicative rather than conclusive.

These tests do not establish the physical origin of the mode. The chromatic models of these pulsars are not truncated within the common-process band: their DMN bases, and for PSR~J1909$-$3744 also the FCN basis, extend to at least $\approx 150$~nHz. Nor does the common process trade power with the modelled noise of these pulsars. In the free spectral-index chain, the Spearman rank correlation coefficients between $(\gamma_{\rm CURN}, \log_{10} A_{\rm CURN})$ and the single-pulsar noise parameters are small, with $|\rho_s| \leq 0.15$ for PSR~J1744$-$1134 and $|\rho_s| \leq 0.17$ for PSR~J1909$-$3744. The SPNA model of PSR~J1744$-$1134 contains DM noise ($\propto \nu^{-2}$) and a deterministic solar-wind term, but neither an achromatic red-noise nor a free chromatic term. In the array analysis the common process is therefore the only achromatic red-noise term for this pulsar. The same is true of all 21 pulsars without an ARN term (Section~\ref{sec:spn_treatment}), but PSR~J1744$-$1134 is among the most precisely timed of them, so unmodelled power in its residuals carries particular weight in the common-process likelihood. Red power in this pulsar that is achromatic, or chromatic with a steeper frequency dependence than $\nu^{-2}$ such as scattering variations, is not fully described by any other term in its model, and unmodelled red power of either kind is a plausible origin of the mode. We have not tested this directly. Because the Band~5-only data are less sensitive and a common chromatic signal would also appear mainly in Band~3, the present data also do not exclude a common chromatic process.

\subsubsection{Optimal Statistic}

As an independent hybrid frequentist check, we apply the NMOS (Section~\ref{sec:os}), computing the S/N for three overlap-reduction functions (Figure~\ref{fig:os_sn}). The noise marginalized S/N distributions peak near zero for the monopole ($-0.34$), dipole ($-0.42$), and HD correlations ($-0.46$), showing no preference for the quadrupolar spatial correlation that would signify a gravitational wave origin and corroborating the non-detection from the Bayesian analysis.

\begin{table}
    \caption{$95\%$ upper limits on the common-process amplitude $\log_{10} A_{\rm GWB}$ at fixed $\gamma_{\rm GWB}=13/3$. The first row is the fiducial no-cut result; the remaining rows apply solar-wind exclusion cuts at three elongation angles. The LinearExp column uses the fiducial uniform-in-amplitude prior; the log-uniform $95\%$ quantile is listed for reference.}
    \label{tab:solar_cuts}
    \begin{tabular}{lcc}
    \hline\hline
        Cut & $\log_{10} A$ (LinearExp) & $\log_{10} A$ (log-unif) \\
        \hline
        No cut      & $-13.47$ & $-13.55$ \\
        $10^\circ$  & $-13.46$ & $-13.55$ \\
        $20^\circ$  & $-13.46$ & $-13.56$ \\
        $30^\circ$  & $-13.44$ & $-13.59$ \\
        \hline\hline
    \end{tabular}
\end{table}

\subsubsection{Upper limit on the common-process amplitude}

\begin{figure*}[htbp]
    \centering

    \subfigure[Full DR2 Upper Limit plot]{
        \includegraphics[width=0.48\textwidth]{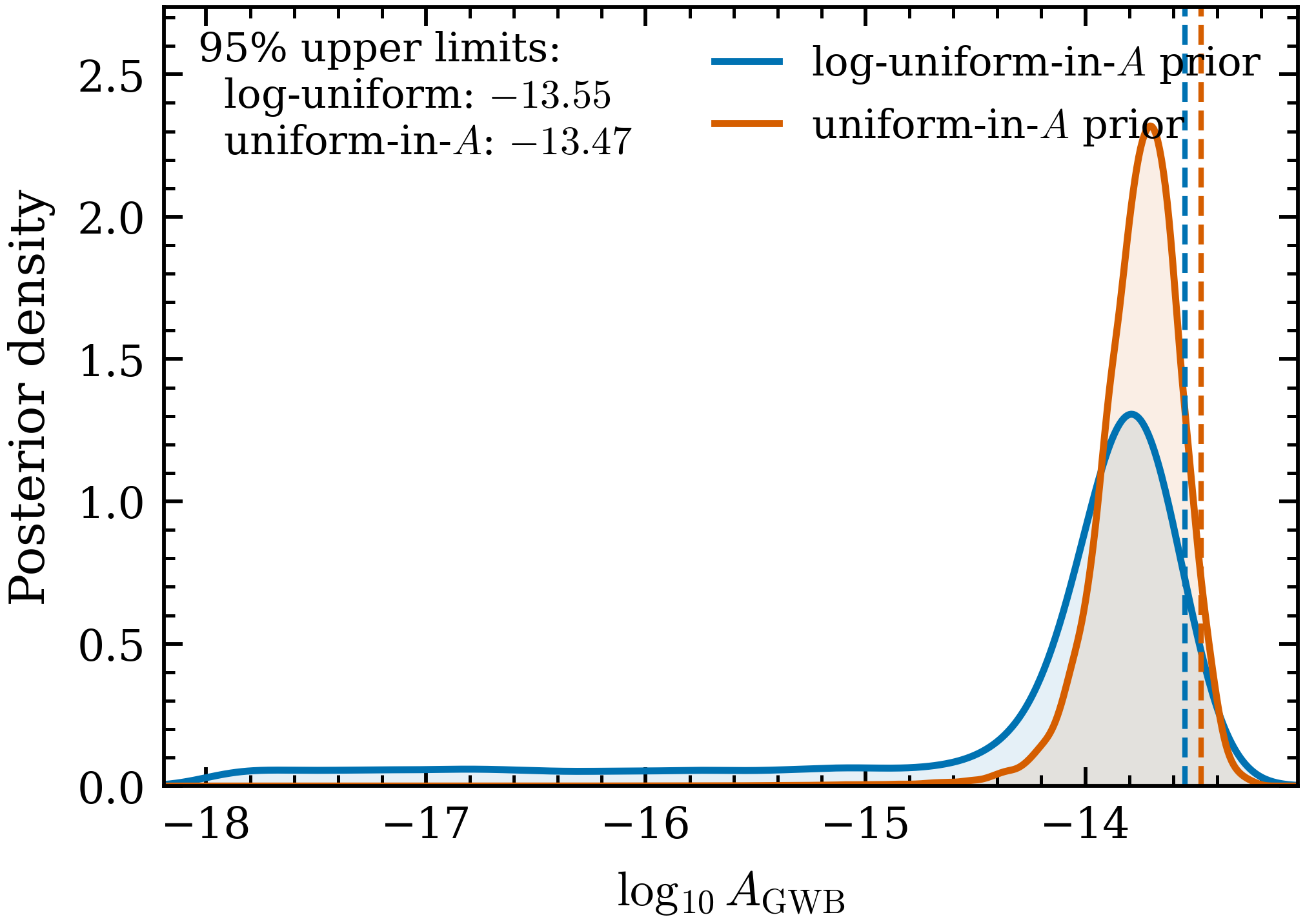}
        \label{fig:corner_dr2_ul}
    }
    \hfill
    \subfigure[DR2 Band 5 Upper Limit plot]{
        \includegraphics[width=0.48\textwidth]{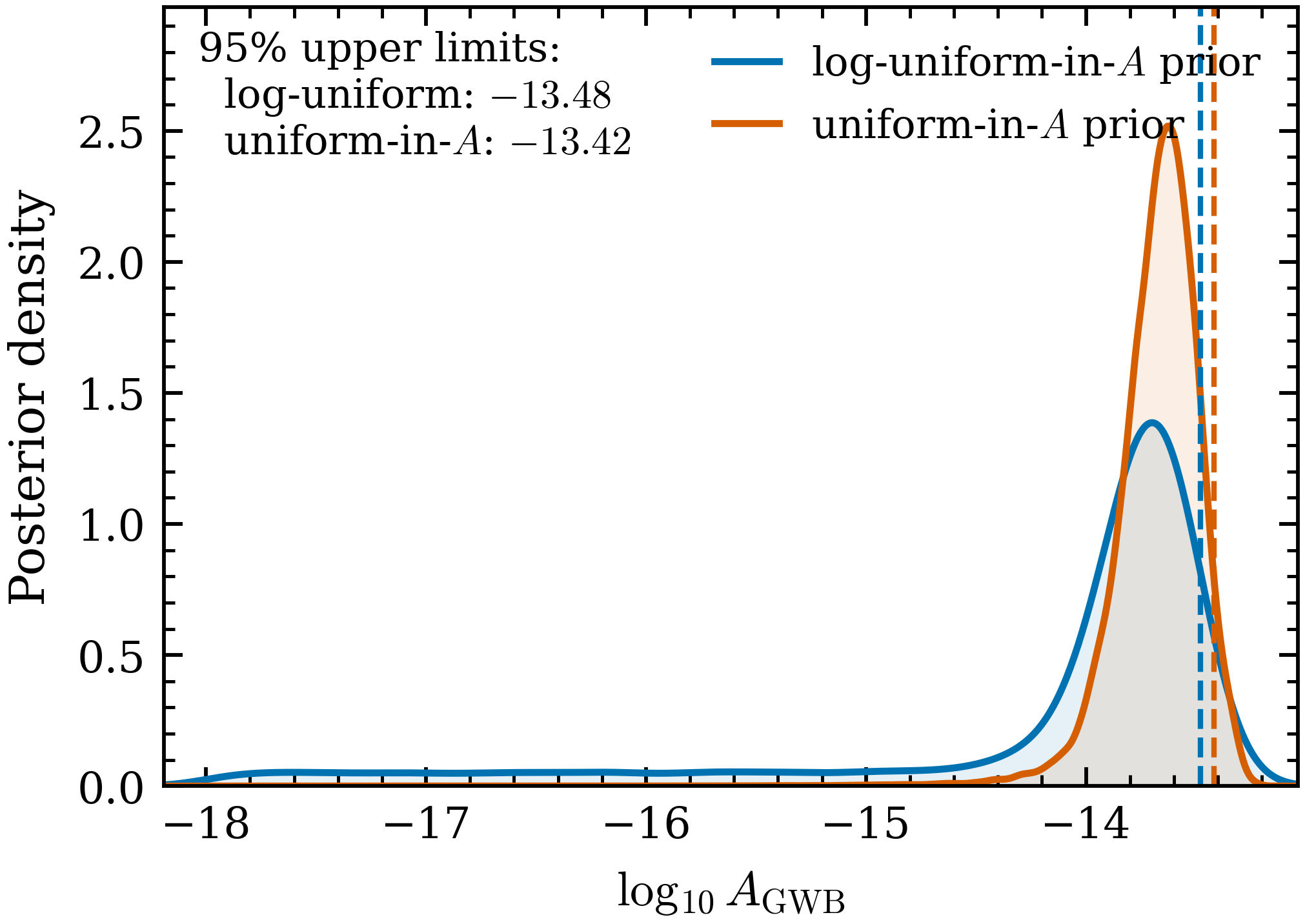}
        \label{fig:corner_band5_ul}
    }

    \caption{Marginal posterior distributions for $\log_{10} A_{\rm GWB}$ at fixed $\gamma_{\rm GWB} = 13/3$ are shown for the combined full DR2 data (left panel) and for the Band 5 only data (right panel). In each panel, the posterior derived from the original log-uniform prior chain (blue) and the LinearExp-reweighted posterior (orange) are displayed, with dashed vertical lines indicating the $95\%$ upper limits. The two data configurations yield mutually consistent upper limits ($-13.47$ and $-13.42$ under a prior uniform in $A$), demonstrating that the inferred constraint on the gravitational-wave background amplitude is robust against the inclusion of the lower-frequency Band~3 data.}
    \label{fig:ul_dr2}

\end{figure*}

Given the absence of a common-process detection, we place an upper limit on the common-spectrum amplitude following the procedure of Section~\ref{sec:upper_limit}, with the spectral index fixed at the SMBHB value expected from a population of inspiraling SMBHBs $\gamma_{\rm GWB} = 13/3$. We compute the limit for both dataset configurations, the combined full DR2 and the Band 5 only data, with the results shown in Figure~\ref{fig:ul_dr2}. The two configurations give consistent limits: the full DR2 yields

\begin{equation}
    \log_{10} A_{\rm GWB} < -13.47 \quad (95\%\ \text{C.L.}),
    \label{eq:ul_dr2}
\end{equation}

corresponding to $A_{\rm GWB} < 3.4\times10^{-14}$, with a log-uniform reference quantile of $-13.55$, while the Band~5-only data give $-13.42$ (LinearExp) and $-13.48$ (log-uniform). The amplitude upper limit is therefore robust to the inclusion of the Band~3 data, differing by $0.05$ in $\log_{10} A$ between the two configurations. We adopt the full DR2 result, Equation~\eqref{eq:ul_dr2}, as our fiducial DR2 constraint.

The dropout excess (Section~\ref{sec:dr2_dropout}) does not appear to bias the array-level limit. A $\gamma_{\rm GWB} = 13/3$ spectrum places most of its power in the lowest Fourier bins, where the full DR2 and Band~5-only posteriors agree (Table~\ref{tab:ngwb}), so the fixed-index limit is largely insensitive to the low-$\gamma$ mode of the free spectral-index posterior (Section~\ref{sec:dr2_cp}). With $N_{\text{GWB}} = 3$ the limits, $-13.48$ (full DR2) and $-13.41$ (Band~5 only), are essentially unchanged. The per-pulsar dropout factors and the fixed-index limit are thus complementary diagnostics.

\subsubsection{Robustness to Solar-wind contamination}
\label{sec:solar_robust}

As a final robustness check, we assess the sensitivity of the constraint to residual Solar-wind contamination by recomputing the $95\%$ upper limit after excluding all TOAs within solar elongations of $10^\circ$, $20^\circ$, and $30^\circ$. As in Section~\ref{sec:upper_limit}, the spectral index is fixed to $\gamma_{\rm GWB} = 13/3$, the limit is obtained by importance reweighting a log-uniform chain to the LinearExp prior, and the Kish effective sample size \citep{Kish95} confirms a stable reweighted posterior in each case. The resulting limits are summarised in Table~\ref{tab:solar_cuts}. The fiducial LinearExp limit is stable across all three cuts, ranging from $-13.44$ to $-13.46$ and differing from the no-cut fiducial value of $-13.47$ by at most $0.03$ in $\log_{10} A$, with the log-uniform reference quantile likewise consistent (ranging from $-13.55$ to $-13.59$), indicating that the inferred upper limit on the gravitational-wave background amplitude is not sensitive to residual solar-wind contamination near solar conjunction.

This array-level stability is consistent with, and complements, the per-pulsar findings of the companion noise analysis \citep{ND+26}. There, repeating the single pulsar analysis on solar wind excised sub-datasets reveals that unmodelled solar-wind power biases the recovered red-noise parameters of the most affected pulsars, with the dispersion-measure noise of PSR~J1022$+$1001 shifting by as much as $3\sigma$ in tension and the posteriors of PSRs~J1744$-$1134 and J1909$-$3744 also moving appreciably between the full and cut datasets. That PSRs J1744$-$1134 and J1909$-$3744, both precisely timed and at low ecliptic latitude, are prominent both in the solar-wind sensitivity of the noise analysis and in the Band 3 dropout excess of Section~\ref{sec:dr2_dropout} is consistent with a common origin in residual chromatic power only partly captured at the single-pulsar level, although only the excess of PSR J1744$-$1134 is associated with the low-$\gamma$ mode.

Because the fixed index limits are largely insensitive to the low-$\gamma$ mode, we also repeated the free-spectral-index search ($N_{\text{GWB}} = 7$) on the cut datasets; the highest common-process bin, $7/T_{\rm span} \approx 0.97\,\mathrm{yr}^{-1}$, lies close to the annual frequency at which solar-wind delays are concentrated. The low-$\gamma$ preference remains under all three cuts ($\gamma_{\rm CURN} = 2.79^{+3.68}_{-2.50}$, $2.73^{+3.76}_{-2.48}$ and $2.54^{+3.92}_{-2.31}$ for the $10^\circ$, $20^\circ$ and $30^\circ$ cuts, compared with $3.01^{+3.60}_{-2.73}$ without cuts), and the upper end of the amplitude credible interval stays near $-12.6$ in every case. This suggests that the mode is not dominated by solar-wind delays near conjunction, although mismodeling at larger elongations is not tested.

\section{Projection for Future InPTA Data Releases}
\label{subsec:data_extension}

Since in the current observational baseline there is no concrete evidence for CURN, to forecast how constraints on the common-process amplitude evolve with increased observational baseline, we employ a purpose built simulation pipeline following the methodology of Ref.\citep{PTV+21}. The script extends the timing data for each pulsar from the current $\sim$7.2 year baseline to target baselines of 10 and 15 years by generating synthetic epochs whose cadence and TOA uncertainties are drawn empirically from the final year of real observations, preserving the statistical character of each pulsar's observing history without assuming a fixed cadence model. Crucially, the same SGWB signal with fixed spectral index $\gamma = 13/3$ and a single injected amplitude  $\log_{10} A_{\rm inj} = -14$ are used consistently across all three baselines, so that differences between the recovered posteriors reflect only the increase in observational sensitivity and not any change in the injected signal. While the overall framework is inherited from \cite{PTV+21}, several significant modifications have been introduced to reflect the specific noise properties, observing strategy, and data format of the InPTA dataset, as detailed below.

%\subsubsection{Differences with Respect to \citet{PTV+21}}
\subsection{Differences with Respect to \texorpdfstring{\citet{PTV+21}}{PTV+21}}
\label{subsubsec:pol_diff}

The original pipeline~\cite{PTV+21} was developed for the NANOGrav 12.5-year dataset and consequently made a number of assumptions that are either inapplicable or sub-optimal for the InPTA data. The following modifications have been introduced.

\paragraph{Noisefile driven injection with per-pulsar component detection:}
The \citet{PTV+21} implementation assumes a largely homogeneous noise model across the array, with a fixed set of noise components applied to all pulsars. The InPTA noise model is considerably more heterogeneous. Different pulsars require different combinations of red noise, DM noise, free chromatic noise (e.g. scattering), and Solar-wind corrections, as determined by their individual single pulsar analyses. The pipeline detects the noise components present for each pulsar from the noise-file keys, but injects only achromatic red and DM noise, with amplitudes and spectral indices fixed to their posterior median values (Section~\ref{subsec:noise_injection}). This avoids injecting spurious noise into pulsars whose noise analyses do not support a given term, which would artificially degrade the sensitivity and produce a biased validation.

\paragraph{Anchoring the common-process filter to the original baseline:}
The per-pulsar red-noise parameters were measured on the Fourier basis $k/T_{\rm orig}$ of the real data and therefore include any common-process power at those frequencies. The common-process (CP) filter subtracts the injected SGWB spectrum from this measured spectrum and refits a power law to the remainder, which is injected as intrinsic red noise (Section~\ref{subsec:noise_injection}), so that common power is not counted twice. \citet{PTV+21} define the filter on the extended basis $k/T_{\rm ext}$, but the single-pulsar analysis contains no information on how power between $1/T_{\rm ext}$ and $1/T_{\rm orig}$ divides into intrinsic and common parts. We therefore anchor the filter to $k/T_{\rm orig}$, where the double counting occurs. This affects only the intrinsic-noise injection: the SGWB, injected separately with \texttt{createGWB} on a frequency grid of spacing  $1/(10\,T_{\rm ext})$ over the full extended span (Section~\ref{subsec:sgwb_injection}), retains its power below $1/T_{\rm orig}$.

\paragraph{Multi-backend ECORR selection:}
\citet{PTV+21} treat ECORR as a single per-pulsar parameter. The InPTA data contain observations in two frequency bands (e.g. Band 3 and Band 5 of the uGMRT), each with distinct jitter properties. The pipeline implements per-group ECORR selection using the \texttt{-group} flag in the \texttt{.tim} files, computing the epoch-averaged ECORR contribution as a weighted mean over backends present within each bin. Because each bin combines all backends into a single epoch (Section~\ref{subsec:epoch_avg}), the simulated epochs carry a TOA-weighted jitter level rather than separate per-backend values.

\subsection{Epoch Averaging and ECORR Treatment}
\label{subsec:epoch_avg}

The pipeline operates on epoch-averaged TOAs in which, multiple observations within a 14-day bin are combined via inverse-variance weighting. The effective uncertainty on the averaged TOA is:

\begin{equation}
    \sigma_\mathrm{epoch}^2 = \frac{1}{\displaystyle\sum_i w_i} + \sigma_\mathrm{ECORR}^2,
    \label{eq:epoch_avg}
\end{equation}

\noindent where $w_i = 1/\sigma_i^2$ are per-TOA weights derived from EFAC and EQUAD-scaled uncertainties, and $\sigma_\mathrm{ECORR}$ is the TOA weight-averaged ECORR value across all backends contributing to the bin. This treatment is critical. Unlike EFAC and EQUAD noise, which average down as $\sim 1/\sqrt{N_\mathrm{TOA}}$, ECORR arises from pulse-phase jitter that is correlated within an epoch and therefore enters the epoch-averaged uncertainty as an irreducible additive floor rather than being suppressed by averaging. A limitation of this epoch-averaging scheme is that all TOAs within each 14-day bin are combined into a single TOA assigned the observing frequency of the most precise TOA in that bin. For the simulated full DR2 configuration, this merges Band 3 and Band 5 observations into single-frequency epochs. Consequently, the simulated full DR2 datasets lack the simultaneous dual-band coverage of the real observations, making the injected DM noise nearly degenerate with achromatic red noise. We therefore do not draw conclusions regarding dual-band chromatic noise mitigation from the simulations. The Band 5-only simulations are unaffected.

\begin{figure*}[htbp]
    \centering

    %--- Top row ---
    \subfigure[7 year baseline (validation)]{
        \includegraphics[width=0.31\textwidth]{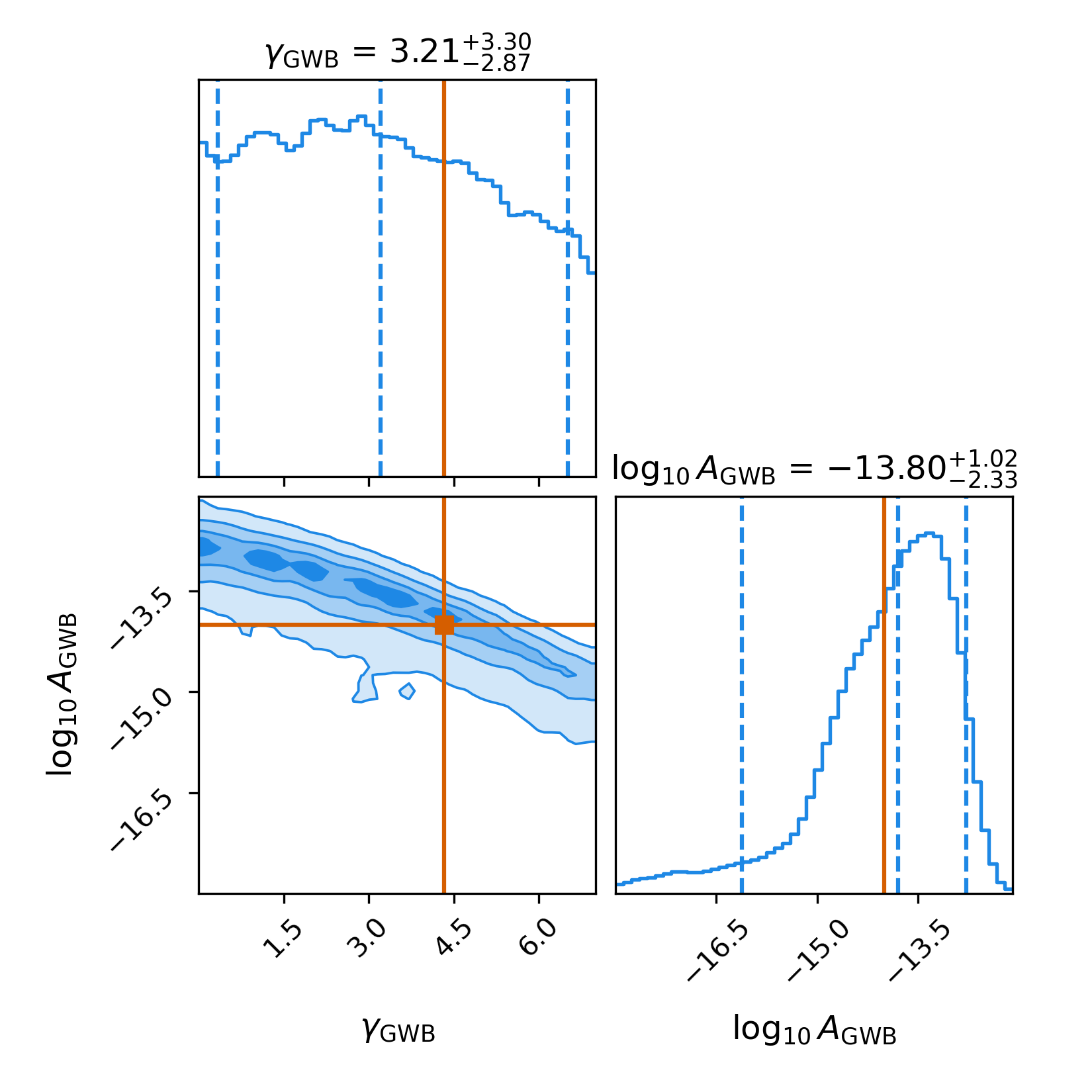}
        \label{fig:corner_7yr}
    }
    \hfill
    \subfigure[10 year baseline (forecast)]{
        \includegraphics[width=0.31\textwidth]{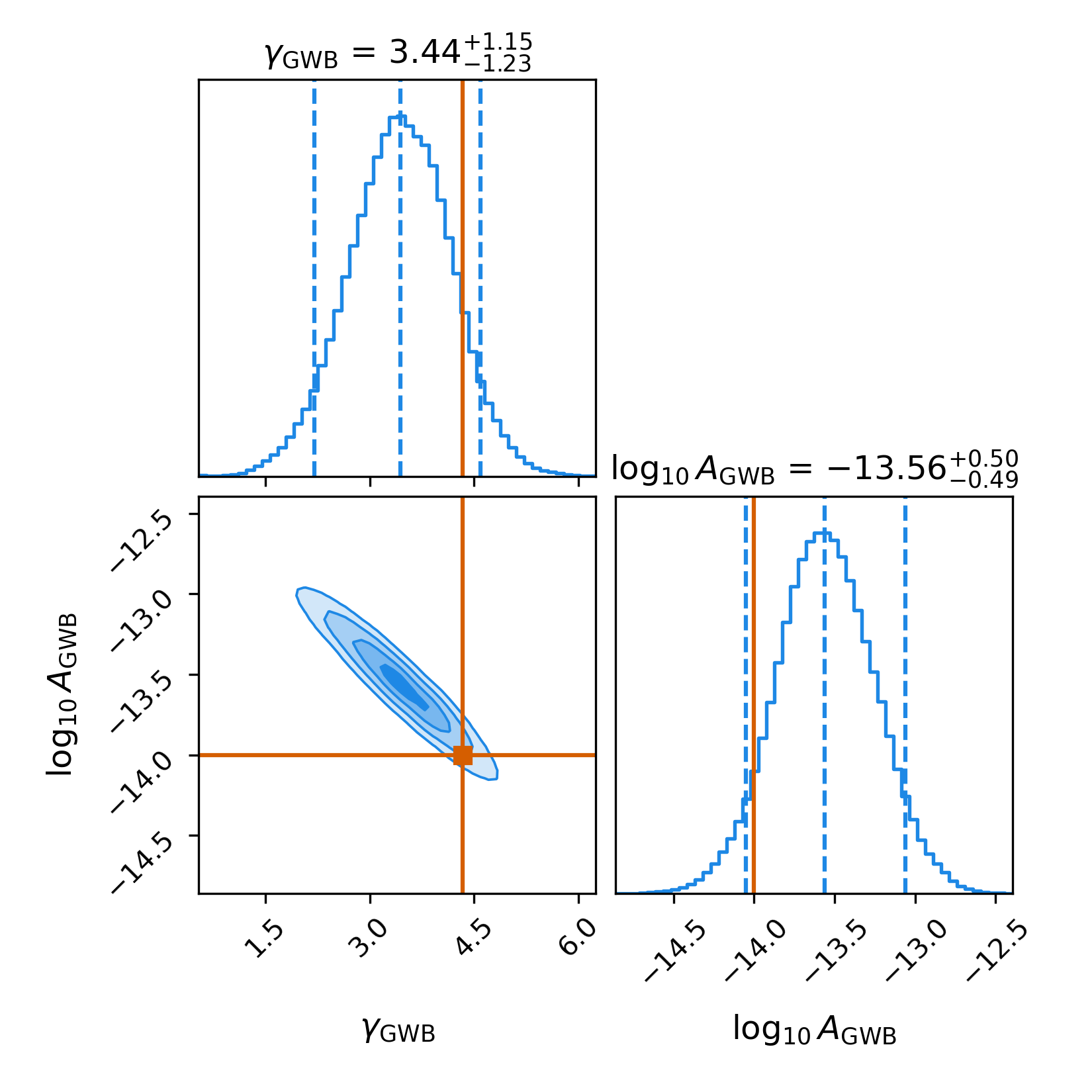}
        \label{fig:corner_10yr}
    }
    \hfill
    \subfigure[15 year baseline (forecast)]{
        \includegraphics[width=0.31\textwidth]{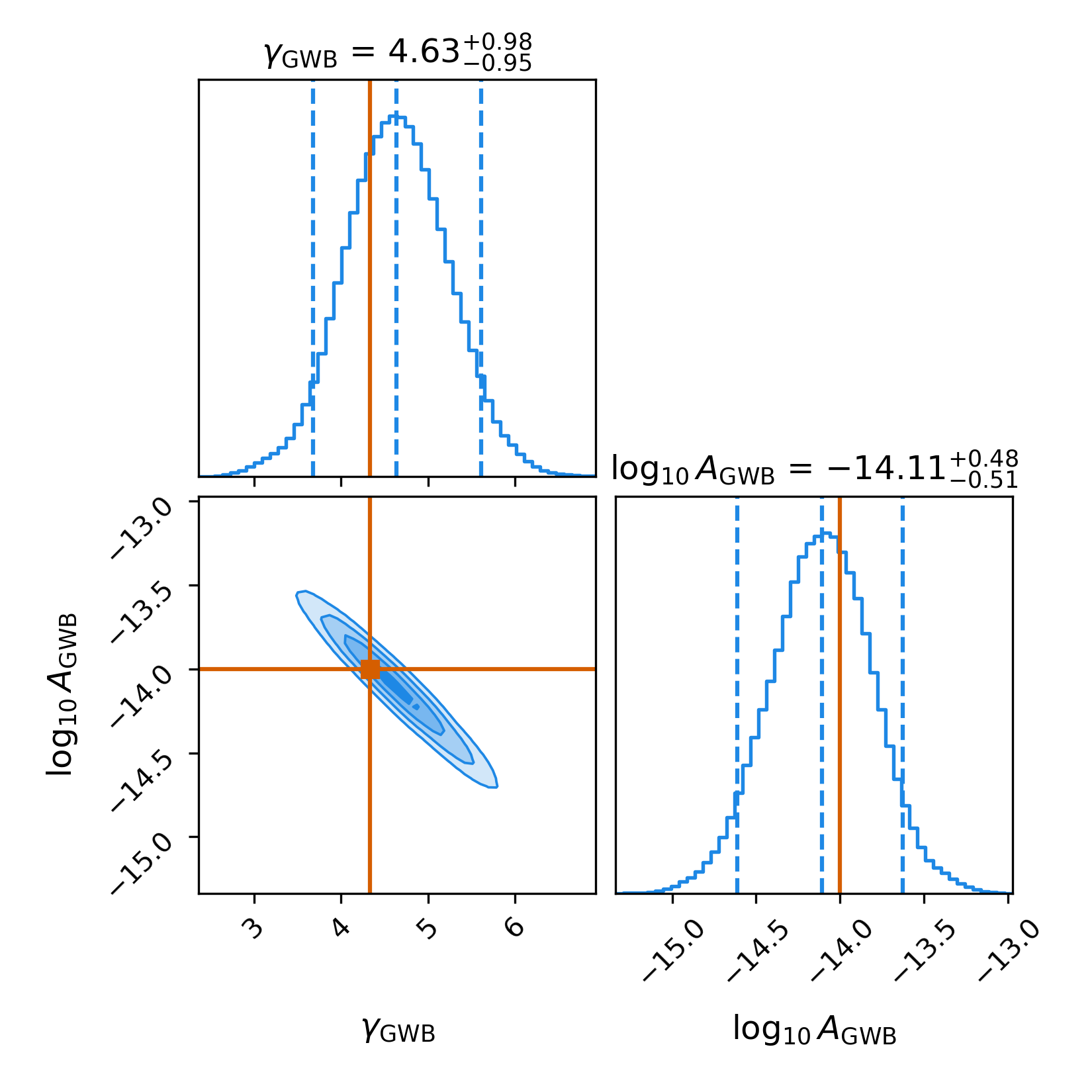}
        \label{fig:corner_15yr}
    }

    \vspace{0.5cm}

    %--- Bottom row ---
    \subfigure[7 year Band 5 baseline (validation)]{
        \includegraphics[width=0.31\textwidth]{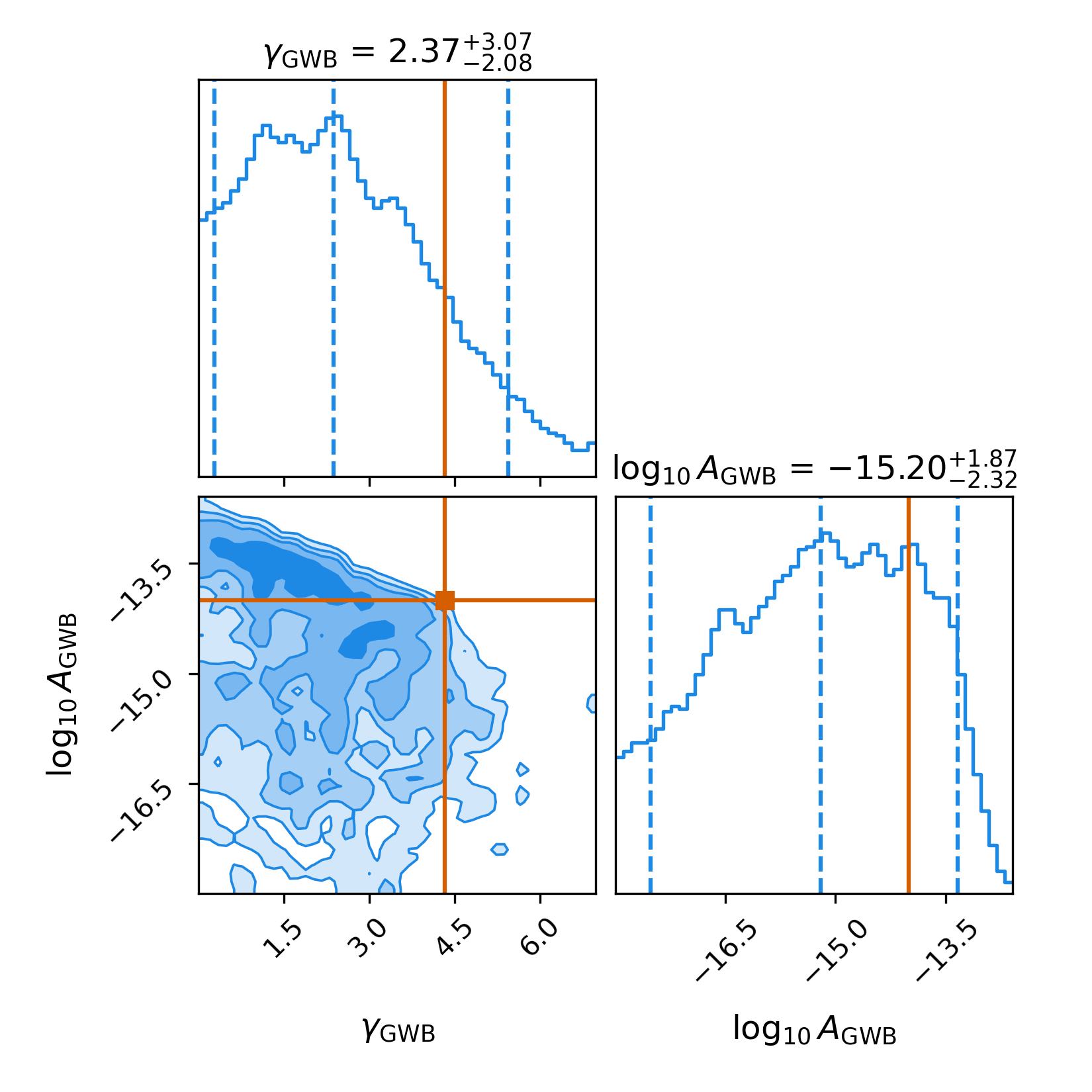}
        \label{fig:corner_20yr}
    }
    \hfill
    \subfigure[10 year Band 5 baseline (forecast)]{
        \includegraphics[width=0.31\textwidth]{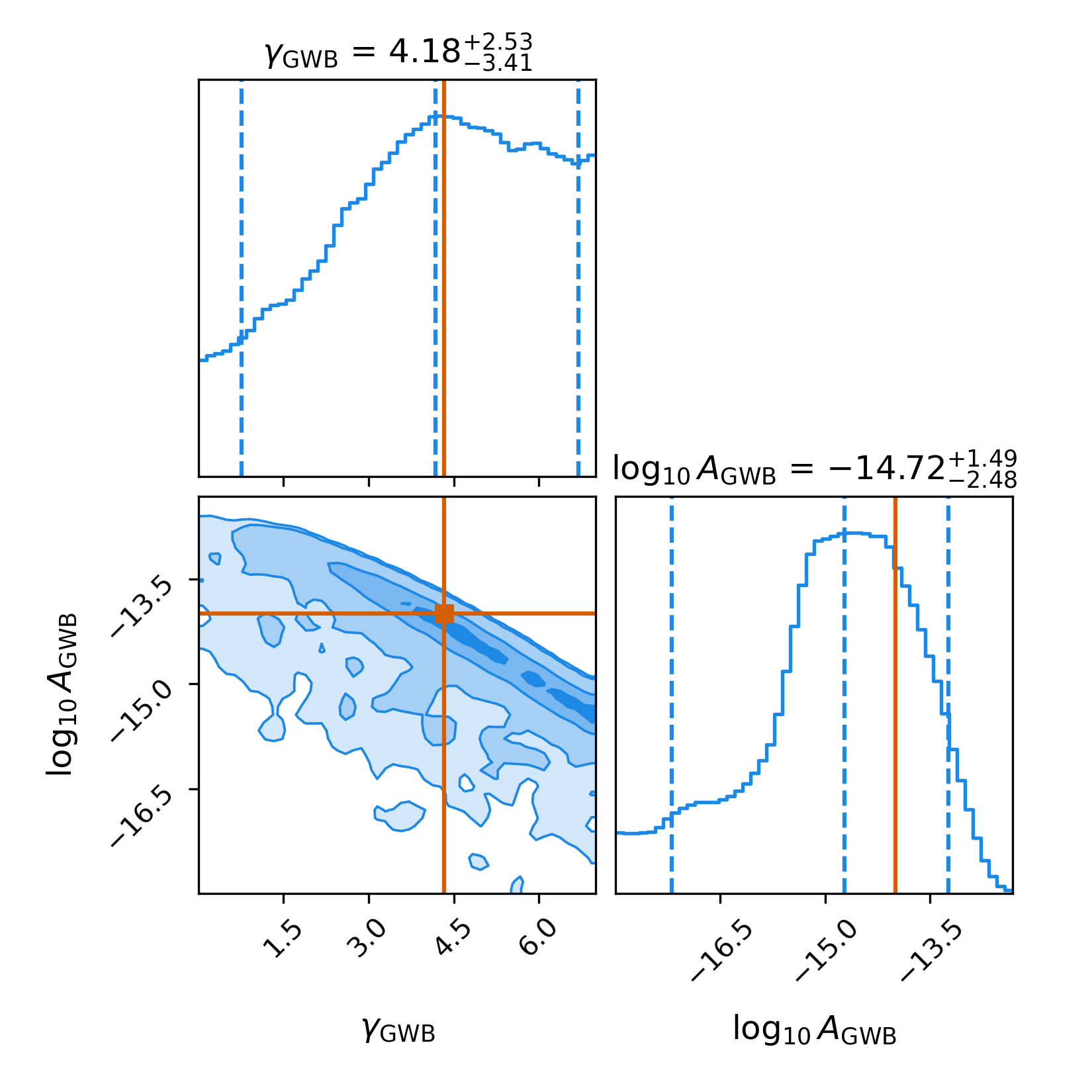}
        \label{fig:corner_25yr}
    }
    \hfill
    \subfigure[15 year Band 5 baseline (forecast)]{
        \includegraphics[width=0.31\textwidth]{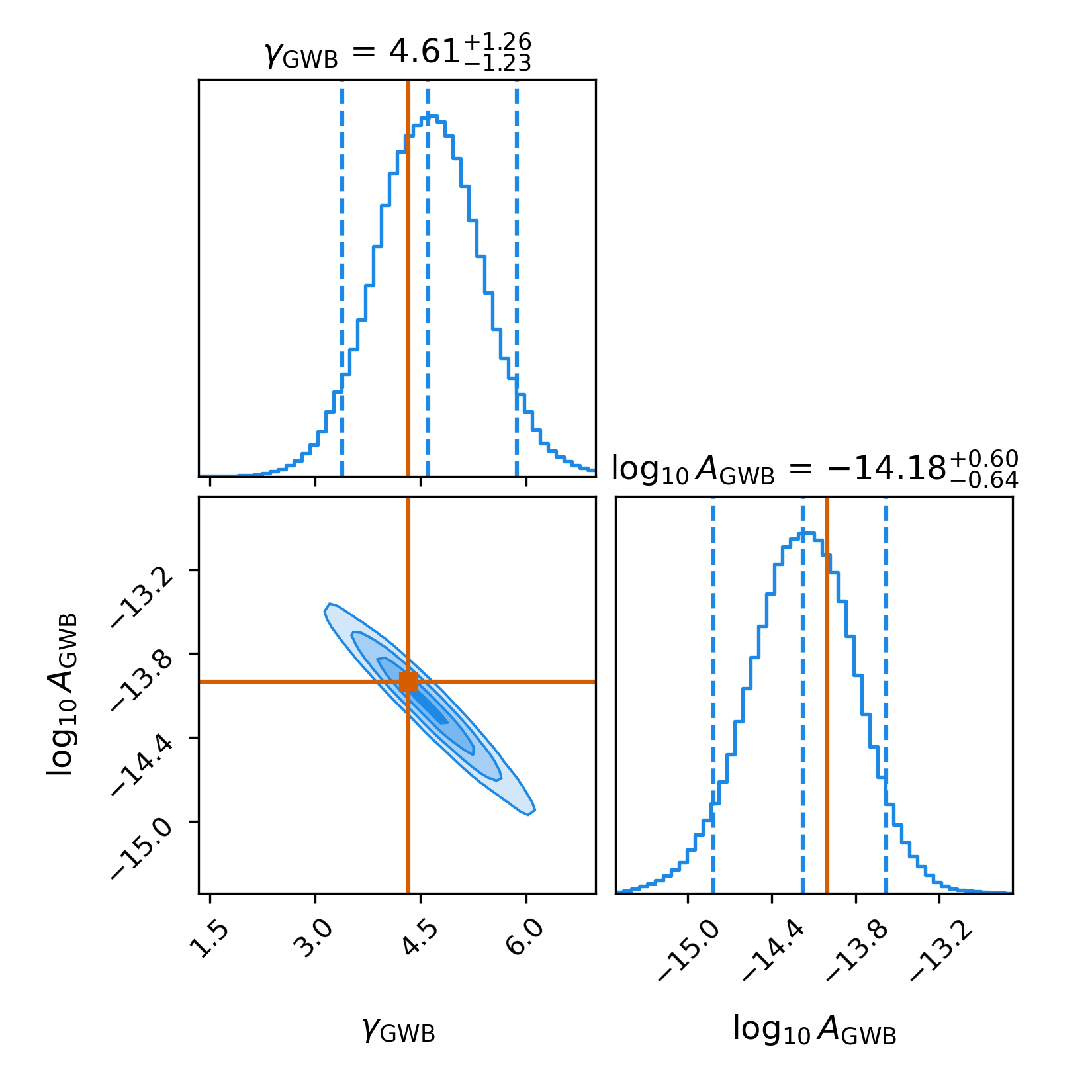}
        \label{fig:corner_30yr}
    }

    \caption{Recovered posterior distributions for the common-process amplitude, $\log_{10} A$, and spectral index, $\gamma$, for the simulated datasets. The top row (a–c) corresponds to the combined full DR2 configuration, and the bottom row (d–f) to the Band~5-only configuration, each evaluated at observational baselines of 7, 10, and 15 years, with an identical GWB signal injected in all cases and a single noise realization per baseline. Orange crosshairs indicate the injected parameter values ($\gamma = 13/3$, $\log_{10} A_{\rm inj} = -14 $). For both configurations, the posterior credible regions are considerably tighter around the injected values at 15 years than at 7 years.}
    \label{fig:corner_baselines}
\end{figure*}

\subsection{Noise Injection}
\label{subsec:noise_injection}

Following epoch averaging, the noise model is injected in stages. All noise parameters used in the injection are taken directly from the single-pulsar noise files produced by the InPTA DR2 noise analysis pipeline \citep{ND+26}. Apart from the power-law refit performed by the CP filter described below, no parameters are refit or drawn from priors at this stage. White noise (EFAC, EQUAD) is reapplied per backend using the noise-file values, and ECORR enters the injected white noise as an epoch-level term through the averaged uncertainty of Eq.~\eqref{eq:epoch_avg}. Intrinsic achromatic red noise is injected per pulsar in the Fourier domain using both the amplitude $A_{\rm arn}$ and spectral index $\gamma_{\rm arn}$, which are fixed to their posterior median values from the per-pulsar noise files. Where a pulsar's noise file includes DM noise parameters, DM noise is additionally injected as a chromatic Fourier process with the same frequency resolution, with amplitude $A_{\rm dmn}$ and spectral index $\gamma_{\rm dmn}$ likewise fixed to their noise-file values. Free chromatic and solar-wind noise are not injected, so the simulations do not reproduce the full chromatic structure of the real data.

To avoid double-counting the SGWB-correlated component at the array level, a CP filter is applied during per-pulsar red noise injection. The filter evaluates the measured red-noise spectrum and the injected SGWB spectrum on the grid $f_k = k/T_{\rm orig}$, $k \leq \min(N_{\rm RN}, 30)$, where $N_{\rm RN}$ is the number of red-noise components, subtracts the latter, and fits a power law to the frequencies with positive remainder; this power law is injected over the full simulated span. If fewer than two frequencies remain positive, no intrinsic red noise is injected. The filter does not affect the injected SGWB (Section~\ref{subsubsec:pol_diff}). Because the subtracted spectrum is the injected one, with $A = 10^{-14}$, the filter removes more intrinsic red power than double counting requires whenever the real data contain less common power than this, as the amplitudes reported by the longer-baseline arrays suggest; in this respect, too, the forecast is optimistic.

\subsection{SGWB Injection}
\label{subsec:sgwb_injection}

The SGWB is injected coherently across the full array using the \texttt{createGWB} function from \textsc{libstempo} \citep{VM20}. The injected signal follows the power-law PSD defined earlier, with the spectral index fixed to the SMBHB prediction $\gamma = 13/3$ and a single amplitude $\log_{10} A_{\mathrm{inj}} = -14$ ($A_{\mathrm{GW}} = 10^{-14}$) held identical across all baselines and both band configurations, so that differences between the recovered posteriors reflect only the change in observing span rather than any change in the injected signal. This amplitude lies below our upper limit but is several times larger than the amplitudes reported by the longer-baseline arrays~\citep{AAB+23,AAA+23b,RD+23,XH+23}, so the forecast is optimistic for a background at the reported level. Although the injected background is spatially correlated, the recovery in every dataset uses the spatially uncorrelated CURN model of Section~\ref{sec:Method}, exactly as for the real data; the posteriors therefore track the CP amplitude constraint, and we assess the Hellings-Downs correlation separately with the optimal statistic (Section~\ref{sec:forecast_results}).

Each configuration was simulated with a single realization of the noise and the SGWB, and the 7 and 10 yr datasets are truncations of the 15 yr simulation. The realization to realization scatter, which at these baselines can be comparable to the differences between baselines~\citep[e.g.,][]{FFF+26}, is therefore not captured, and the trends in Section~\ref{sec:forecast_results} may not generalize.

\subsection{Pipeline Validation}

As a validation of the simulation pipeline, the Bayesian analysis described in Section~\ref{sec:Method} is applied to a simulated dataset constructed to replicate the noise properties, cadence, and 7.2-year baseline of the real observations. For the combined full DR2 configuration the recovered posterior, $\log_{10} A = -13.80^{+1.02}_{-2.33}$ with $\gamma = 3.21^{+3.30}_{-2.87}$, overlaps that obtained from the real data. We caution that this is a weak test. At a 7.2-year baseline, the posterior is broad and largely uninformative, so a simulation containing an injected signal at $\log_{10} A_{\rm inj} = -14$ and one containing no common signal at all would both be expected to reproduce it. The comparison therefore establishes that the noise injection and epoch-averaging machinery introduces no gross distortion of the recovered posterior; it does not, on its own, demonstrate that the simulated array reproduces the statistical sensitivity of the real one. The Band~5 only validation run returns $\log_{10} A = -15.20^{+1.87}_{-2.32}$, a lower median than the corresponding real data posterior though consistent with it within the credible intervals, reflecting the prior-dominated character of that configuration at short time baselines.

\subsection{Results: Evolution of Posteriors with Baseline}
\label{sec:forecast_results}

The Bayesian analysis described above is applied to simulated datasets at baselines of 7, 10, and 15\, years, for both the full DR2 and the Band~5 only configurations, with the same SGWB injected in all cases. The injected values ($\gamma = 13/3 \approx 4.33$, $\log_{10} A_{\rm inj} = -14 $) are indicated by the orange cross-hairs in Figure~\ref{fig:corner_baselines}. The recovered medians and $90\%$ credible intervals are summarized in Table \ref{tab:forecast}.

\begin{table}
        \caption{Recovered common-process parameters (median and $90\%$ credible interval, 5th-95th percentiles) from the simulated datasets, for both configurations at three baselines. The same signal ($\gamma = 13/3$, $\log_{10} A_{\rm inj} = -14$) is injected in all cases. Each entry comes from a single realization of the noise and the SGWB, so realization to realization scatter is not included.}
    \vspace{0.2cm}
    \label{tab:forecast}
    \begin{tabular}{llcc}
        \hline\hline
        Config. & Baseline & $\gamma$ & $\log_{10} A$ \\
        \hline
        \multirow{3}{*}{Full DR2}
        & 7\,yr  & $3.21^{+3.30}_{-2.87}$ & $-13.80^{+1.02}_{-2.33}$ \\
        & 10\,yr & $3.44^{+1.15}_{-1.23}$ & $-13.56^{+0.50}_{-0.49}$ \\
        & 15\,yr & $4.63^{+0.98}_{-0.95}$ & $-14.11^{+0.48}_{-0.51}$ \\
        \hline
        \multirow{3}{*}{Band 5 only}
         & 7\,yr  & $2.37^{+3.07}_{-2.08}$ & $-15.20^{+1.87}_{-2.32}$ \\
        & 10\,yr & $4.18^{+2.53}_{-3.41}$ & $-14.72^{+1.49}_{-2.48}$ \\
        & 15\,yr & $4.61^{+1.26}_{-1.23}$ & $-14.18^{+0.60}_{-0.64}$ \\
        \hline\hline
    \end{tabular}
\end{table}

In both configurations, the joint posteriors exhibit the characteristic amplitude spectral-index degeneracy intrinsic to power-law SGWB searches at finite S/N, producing the elongated, correlated contours of Figure~\ref{fig:corner_baselines}; and in both, extending the baseline from 7 to 15\,years substantially narrows the credible regions around the injected signal.

\begin{figure*}[htbp]
    \centering
    \subfigure[10 year baseline]{
        \includegraphics[width=0.45\textwidth]{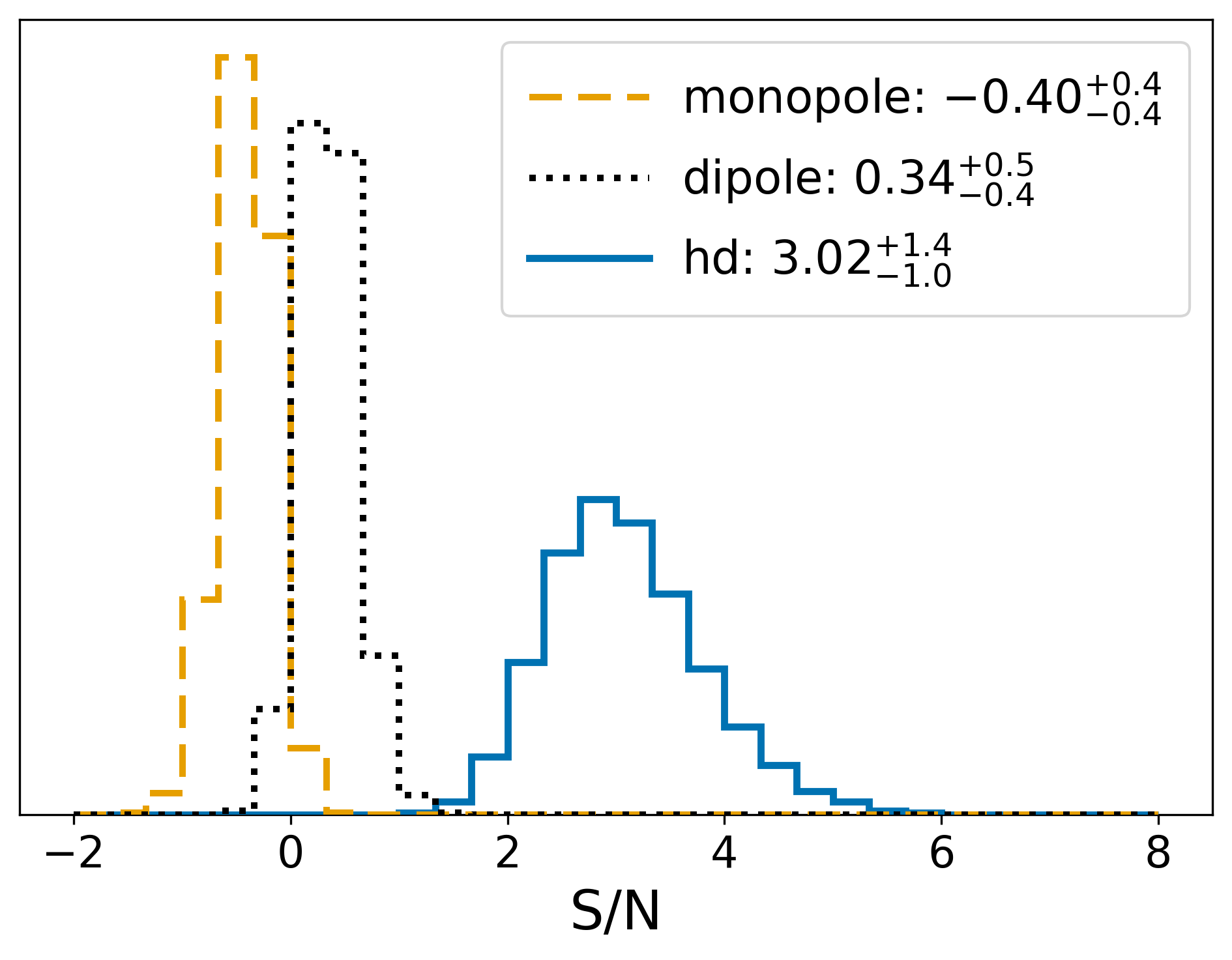}
        \label{fig:os_sim_10yr}
    }
    \hfill
    \subfigure[15 year baseline]{
        \includegraphics[width=0.45\textwidth]{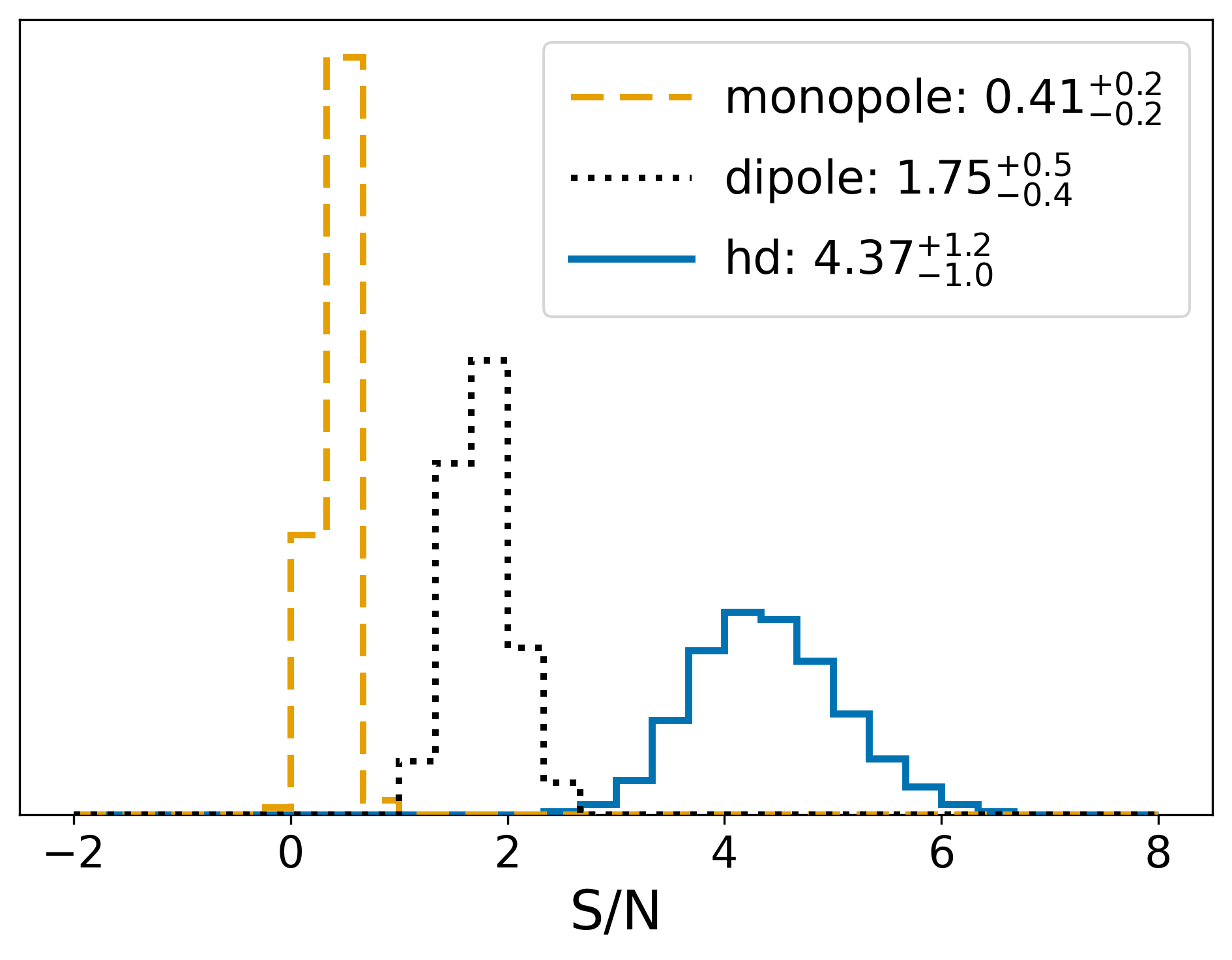}
        \label{fig:os_sim_15yr}
    }
    \caption{ Noise-marginalized optimal-statistic S/N for the simulated full DR2 datasets at (a) 10 and (b) 15 years, for the Hellings-Downs (blue, solid), monopole (orange, dashed), and dipole (black, dotted) overlap-reduction functions. The injected signal ($\gamma = 13/3$, $\log_{10} A_{\rm inj} = -14$) is Hellings-Downs correlated, and each panel comes from a single noise realization. Legend values give the median and 90\% interval (5th-95th percentiles).}
    \label{fig:os_sim}
\end{figure*}

For a single realization, most differences between the two configurations lie within the credible intervals. The full DR2 amplitude median at 10\,yr, $\log_{10} A = -13.56^{+0.50}_{-0.49}$, lies above the injected value, which falls just inside the lower edge of its $90\%$ interval, with $\gamma = 3.44$. Three effects may contribute to this offset, and a single realization cannot separate them. First, because the simulated full DR2 epochs are single-frequency (Section~\ref{subsec:epoch_avg}), the injected DM noise is nearly degenerate with achromatic noise. Second, the Fourier-diagonal approximation of Eq.~\eqref{eq:gwb_covariance}, applied to an injected background that retains power below $1/T$ (Section~\ref{subsec:sgwb_injection}), biases recoveries toward flatter spectra and higher amplitudes~\citep{CVM+26,FFF+26}. Third, there is realization scatter. The Band~5-only posterior is prior-dominated in amplitude at 7\,yr ($\log_{10} A = -15.20$), and at 10\,yr its median $\gamma$ moves closer to the injected value while its credible interval remains broad ($4.18^{+2.53}_{-3.41}$). By 15\,years both configurations recover the injected amplitude and index well within their credible intervals ($\gamma \approx 4.6$, $\log_{10} A \approx -14.1$), with the injected values lying within the inner posterior contours.

This improvement between 7 and 15\,years arises because extending the time baseline lowers the minimum sampled frequency $f_{\min} \simeq 1/T$, providing additional low-frequency modes that partially break the amplitude spectral-index degeneracy; for the full DR2 configuration, the $90\%$ interval on $\gamma$ narrows from $^{+3.30}_{-2.87}$ at 7\,yr to $^{+1.15}_{-1.23}$ at 10\,yr and $^{+0.98}_{-0.95}$ at 15\,yr, while the $\log_{10} A$ interval width drops from $3.35$~dex at 7\,yr to $\approx 1.0$~dex at 10 and 15\,yr. In this realization, the concentration of both 15-yr posteriors around the injected signal suggests that, for the current 27-pulsar array, the observing baseline is the dominant limitation on the common-process amplitude and spectral index, and that extending the timespan alone may suffice to recover the injected signal. We have not simulated larger pulsar arrays; since the cross-correlation S/N grows with the number of pulsar pairs \citep{SEJ+13}, expanding the pulsar sample constitutes a complementary route to sensitivity that our forecast does not address. Whether these trends hold across realizations, and for simulated datasets with simultaneous dual-band epochs, remains to be tested.

To follow the cross-correlation signature, we also computed the noise-marginalized optimal statistic (Section~\ref{sec:os}) for the simulated full DR2 datasets at 10 and 15 years (Figure~\ref{fig:os_sim}). The Hellings-Downs S/N rises from $3.02^{+1.4}_{-1.0}$ at 10\,yr to $4.37^{+1.2}_{-1.0}$ at 15\,yr, while the monopole S/N stays close to zero. The dipole S/N also grows, to $1.75^{+0.5}_{-0.4}$ at 15\,yr. This is expected for a purely Hellings-Downs correlated signal: for a finite array the overlap-reduction functions are not orthogonal over the sampled pulsar pairs, so part of the injected correlation projects onto the dipole template. These S/N values come from a single noise realization, and they are not by themselves detection significances, which would require a null distribution from, e.g., sky scrambles or phase shifts.

The approximate 1.45 times increase in the median Hellings-Downs S/N between 10 and 15\,yr can be compared with the scaling laws of Ref.~\citep{SEJ+13}. In the weak-signal regime, where white noise dominates at all sampled frequencies, the cross-correlation S/N of a $\gamma = 13/3$ background grows as $T^{13/3}$ ($\approx 5.8\times$ from 10 to 15\,yr). In the intermediate-signal regime, where the lowest sampled frequencies are gravitational-wave dominated while white noise still dominates at high frequencies, it grows only as $T^{1/2}$ ($\approx 1.22\times$). The simulated growth lies far below the weak-signal prediction and close to the $T^{1/2}$ scaling. This is consistent with the handful of sub-microsecond pulsars that dominate the cross-correlation sum already being in the intermediate-signal regime at this injected amplitude, where Ref.~\citep{SEJ+13} identifies the number of pulsars, rather than the timespan, as the primary lever on detectability. Because both values come from a single realization, we regard this comparison as indicative. The baseline therefore governs the common-process constraint of Figure~\ref{fig:corner_baselines}, while further growth of the cross-correlation S/N at this injected amplitude depends increasingly on the number of pulsars.
\section{Discussion}
\label{sec:discussion}

Our analysis yields no statistically significant evidence for a common red noise process across any of our four diagnostics. We compute the Savage-Dickey Bayes factor of 2.5 in favor of a common red-noise process, which on the Jeffreys scale is not worth more than a bare mention. The common-spectrum amplitude posterior is broad and prior-dominated, and the noise-marginalized optimal statistic returns S/N consistent with zero for the monopole, dipole, and Hellings-Downs correlations alike. From the fiducial full DR2 dataset we report a 95\% upper limit on the common-process amplitude of $\log_{10} A_{\rm GWB} < -13.47$ at the SMBHB index $\gamma_{\rm GWB} = 13/3$, robust to the exclusion of data near solar conjunction; the Band~5-only configuration gives a consistent limit of $-13.42$, the two differing by $0.05$ in $\log_{10} A$.

This limit lies above the amplitudes for which the longer-baseline arrays reported evidence in 2023 \citep{AAB+23,AAA+23b,RD+23,XH+23}, and is fully consistent with them. The difference reflects both the shorter baseline and the smaller array. In the weak-signal regime the S/N of a steep power-law background grows steeply with timespan \citep{SEJ+13}, and at the present $\sim$7.2-year span the lowest sampled Fourier frequency, $f_{\min} \approx 1/T$, is correspondingly higher than that of the longer-baseline arrays, offering little leverage on a $\gamma = 13/3$ spectrum. Baseline is not the only factor, however: the MeerKAT PTA reported evidence for a common signal from a dataset shorter than ours but with a substantially larger array \citep{MSR+25}, and the cross-correlation S/N grows with both the number of pulsar pairs and the per-pulsar timing precision \citep{SEJ+13}. The non-detection is therefore expected for an array of this size and age. The dense, simultaneous dual-band cadence of InPTA provides detailed per-pulsar chromatic noise characterization, and the constraint will sharpen as the baseline lengthens, $f_{\min}$ descends, and the pulsar sample grows.

\subsection{Simultaneous dual-band data as a probe of band-dependent systematics}

An important outcome of InPTA's dual-band coverage is the ability to identify band-dependent excess power that would be difficult to recognize with single-band arrays. By repeating the common-process dropout analysis with and without the low-frequency Band~3 data, we find that two of the most precisely timed pulsars, PSRs~J1744$-$1134 and J1909$-$3744, exhibit dropout factors of $3.22$ and $2.16$ in the combined full DR2 analysis, which decrease to $0.98$ and $1.38$ when Band~3 is excluded (within the typical Band~5-only scatter of $\lesssim 1.7$). Further tests (Section~\ref{sec:dr2_dropout}) associate the low-$\gamma$ structure of the full DR2 posterior largely with PSR~J1744$-$1134. This pulsar's noise model contains no achromatic red-noise or free chromatic term, and its modelled DM noise shows no trade-off with the common process. Unmodelled red power in this pulsar, absorbed by the common process, is therefore a plausible explanation, but we have not established it, and a common chromatic process cannot be excluded.

This effect appears in the per-pulsar diagnostics and the free-spectral-index posterior, but does not appear to bias the fixed-$\gamma = 13/3$ array-level amplitude limit, which is consistent between the full DR2 and Band~5-only configurations (Section~\ref{sec:dr2_results}). Comparing dual-band and single-band dropout factors may therefore provide a useful diagnostic of band-dependent mismodeling alongside the amplitude limit.

\subsection{Prospects for detection with simultaneous dual-band coverage}
\label{sec:prospects}

Because the statistical significance of a gravitational-wave background detection is constructed from cross-correlations between pulsar pairs, it increases sharply with the number of pulsars in a timing array, and the InPTA, with its 27 pulsars, is one of the smaller contributing datasets.

When a Hellings-Downs correlated background with fixed spectral index $\gamma = 13/3$ and amplitude $\log_{10} A_{\mathrm{inj}} = -14$ is injected and recovered with the CURN model at successively longer baselines in a single noise realization, the posteriors converge toward the injected values as the timespan increases. At 10 years, the full DR2 configuration recovers $\log_{10} A = -13.56^{+0.50}_{-0.49}$ and $\gamma = 3.44^{+1.15}_{-1.23}$, with the injected values inside its 90\% credible intervals (the amplitude just inside the lower edge), while the Band 5-only posterior remains broader ($\log_{10} A = -14.72^{+1.49}_{-2.48}$, $\gamma = 4.18^{+2.53}_{-3.41}$). By 15 years, both configurations recover the injected amplitude and spectral index well within their credible intervals ($\gamma \approx 4.6$, $\log_{10} A \approx -14.1$), and the noise-marginalized optimal-statistic $S/N$ for the Hellings-Downs correlation reaches $4.37^{+1.2}_{-1.0}$ for the full DR2 configuration. In this realization, the convergence of both configurations onto the injected signal suggests that, for the current 27-pulsar array, the observing baseline is the dominant limitation on the common-process constraint, and that extending the timespan alone may suffice to recover the signal. Because the simulated full DR2 epochs are single-frequency (Section~\ref{subsec:epoch_avg}) and come from a single noise realization, these simulations do not test dual-band chromatic mitigation or realization-to-realization scatter.

The extent to which sensitivity can be maintained with a relatively small number of pulsars in real observations is contingent upon the fidelity of the per-pulsar noise modelling. The statistical significance of cross-correlations is degraded both by the restricted number of available baselines and by unmodelled chromatic noise power, which can either imitate or conceal a common stochastic signal, a possibility highlighted by the Band~3 sensitivity of the dropout factors in Section~\ref{sec:dr2_dropout}. The InPTA's simultaneous dual-band observation strategy is designed to address this second limitation by enabling dispersion-measure and higher-order chromatic variations to be constrained via its extended low-frequency coverage, while also providing a direct cross-check between dual-band and high-frequency-only inferences. In this sense the InPTA's contribution is complementary to that of larger arrays: its enhanced control of chromatic effects compensates for its smaller pulsar sample and will be realized most effectively as the timing baseline increases and the data are incorporated into the IPTA, where the detailed chromatic characterization of the shared pulsars augments the raw sensitivity of the longer-baseline datasets.
\section{Conclusions}
\label{sec:conclusion}

We report the first independent search for an isotropic stochastic gravitational-wave background using data from the InPTA, based on the second data release comprising 27 pulsars over a temporal baseline of approximately 7.2 years. Building upon a comprehensive single-pulsar noise characterization, we investigated the presence of a common-spectrum red process employing a Bayesian inference framework in conjunction with the noise-marginalized optimal statistic, and we evaluated the robustness of our findings through per-pulsar dropout analyses and solar-wind exclusion cuts.

No statistically significant evidence for a common red process is found in any of our four diagnostic approaches. The corresponding Savage-Dickey Bayes factor is 2.5, which on the Jeffreys scale does not exceed the level of a bare mention. In agreement with this, the median per-pulsar dropout factors are consistent with unity, the common-process amplitude posterior is broad and prior-dominated, and the noise-marginalized optimal-statistic S/N for the monopole, dipole, and Hellings-Downs spatial correlations are all consistent with zero. From the fiducial full DR2 dataset we place a 95\% upper limit on the common-process amplitude of $\log_{10} A_{\rm GWB} < -13.47$ ($A_{\rm GWB} < 3.4 \times 10^{-14}$) at the fiducial SMBHB spectral index $\gamma_{\rm GWB} = 13/3$, stable to within $0.03$ in $\log_{10} A$ across solar-elongation cuts of $10^\circ$, $20^\circ$, and $30^\circ$. The amplitude limit is consistent between the full DR2 and Band~5-only configurations, differing by $0.05$ in $\log_{10} A$; the excess seen in the per-pulsar dropout factors does not appear to bias the array-level constraint. This constraint lies approximately an order of magnitude above, and is fully consistent with, the amplitudes inferred by longer-baseline pulsar-timing arrays \citep{AAB+23,AAA+23b,RD+23,XH+23}; the difference reflects primarily the shorter temporal extent and smaller size of our dataset rather than any fundamental limitation of the instrument.

An important capability of the InPTA configuration is that its simultaneous dual-band frequency coverage can reveal band-dependent systematic effects that would be difficult to identify with single-band arrays. By comparing the common-process inference with and without the low-frequency Band~3 data, we find that two of the most precisely timed pulsars display elevated dropout factors in the joint full DR2 analysis ($3.22$ for PSR~J1744$-$1134 and $2.16$ for PSR~J1909$-$3744), which decrease to $0.98$ and $1.38$ once Band~3 is excluded. Further tests associate the low-$\gamma$ structure of the full DR2 posterior largely with PSR~J1744$-$1134, plausibly, though not yet established, through red power that the noise model of this pulsar, which contains no achromatic red-noise or free chromatic term, does not capture and that the common process absorbs instead. The effect is visible in the per-pulsar dropout factors and the free-$\gamma$ posterior but not in the fixed $\gamma = 13/3$ array-level limits, which agree to within $0.05$ in $\log_{10} A$, so the dual-band dropout comparison may serve as a sensitive probe of band-dependent mismodeling alongside the amplitude limit. Such pulsar-by-pulsar diagnostic capability is likely to be valuable as the international pulsar-timing program advances toward a high-significance, spatially correlated detection.

Our simulation study, based on a single noise realization per configuration, further suggests that extending the observing timespan alone, at fixed array size, may suffice to recover the injected signal. We emphasize that these simulations hold the number of pulsars fixed at 27 and therefore do not address the sensitivity gain available from enlarging the array; since the cross-correlation signal-to-noise ratio grows with the number of pulsar pairs, and since Ref.~\citep{SEJ+13} shows that the number of pulsars overtakes the timespan as the dominant lever once an array enters the intermediate-signal regime, expanding the pulsar sample remains a complementary and potentially faster route to detection that our forecast does not evaluate. By 15 years both configurations recover the injected amplitude and index within their credible intervals in this realization (Section~\ref{sec:forecast_results}). Because the simulated full DR2 epochs lack simultaneous dual-band coverage, the simulations do not test dual-band chromatic mitigation. Owing to its dual-band capability, illustrated here through the dropout analysis, the InPTA is positioned as a distinctive and complementary component of the International Pulsar Timing Array. Its detailed characterization of the pulsars shared with other collaborations can enhance the effective sensitivity of longer-baseline data sets, supporting the transition of nanohertz gravitational-wave astronomy from the regime of marginal evidence to that of a robust and confident detection.

\section*{DATA AVAILABILITY}
All the scripts and datasets used in the analysis are available in the InPTA official GitHub repository.

\section*{Acknowledgements}

InPTA acknowledges the support of the GMRT staff in resolving technical difficulties and providing technical solutions for high-precision work. We acknowledge the GMRT telescope operators for the observations. The GMRT is run by the National Centre for Radio Astrophysics of the Tata Institute of Fundamental Research, India. Authors acknowledge the National Supercomputing Mission (NSM) for providing computing resources of 'PARAM Ganga' at the Indian Institute of Technology Roorkee, 'PARAM Smriti' at National Agri-Food Biotechnology Institute Mohali and 'PARAM Seva' at Indian Institute of Technology, Hyderabad. PARAM Ganga, PARAM Smriti and PARAM Seva are implemented by C-DAC and supported by the Ministry of Electronics and Information Technology (MeitY) and Department of Science and Technology (DST), Government of India. \\
We acknowledge the use of computational infrastructure built through the start-up grant awarded to MS by IISER Bhopal.\\
AG acknowledges support of the Department of Atomic Energy, Government of India, under Project Identification No. RTI 4002. \\
AKP is supported by CSIR fellowship Grant number 09/0079(15784)/2022-EMR-I. \\
AdS is supported by UGC-JRF fellowship.\\
AmS acknowledges the National Supercomputing Mission (NSM) for providing computing resources of ‘PARAM Seva’ at Indian Institute of Technology Hyderabad, which is implemented by C-DAC and supported by the Ministry of Electronics and Information Technology (MeitY) and Department of Science and Technology (DST), Government of India.\\
BCJ acknowledges the support from Raja Ramanna Chair fellowship of the Department of Atomic  Energy, Government of India (RRC – Track I Grant 3/3401 Atomic Energy Research 00 004 Research and Development 27 02 31 1002/2/2023/RRC/R\&D-II/13886 and 1002/2/2023/RRC/R\&D-II/14369). \\
CD acknowledges the Param Vikram-1000 High Performance Computing Cluster of the Physical Research Laboratory (PRL), a unit of the Department of Space, Government of India, for performing the intensive computations. The work of CD at the Physical Research Laboratory (PRL) was supported by the Department of Space, Government of India. \\
DD acknowledges the support from the Department of Atomic Energy, Government of India through ‘Apex-I Project - Advance Research and Education in Mathematical Sciences’ at The Institute of Mathematical Sciences. \\
HT is supported by DST INSPIRE Fellowship, INSPIRE code IF210656. \\
JS acknowledges the support from the University of Cape Town Vice Chancellor’s Future Leaders 2030 Awards programme and the South African Research Chairs Initiative of the Department of Science and Technology and the National Research Foundation.\\
KT is partially supported by JSPS KAKENHI grant Nos. 24H01813, 25K21670, 26H00838 and 26K21724.\\
KR is supported by UGC-JRF fellowship.\\
NDB acknowledges the support received from the Department of Science and Technology, Government of India vide the DST-WISE fellowship (DST/WISE-PDF/PM-17/2024).\\
PA acknowledges the support from SERB-DST, Govt. of India, via project code No. CRG/2022/009359.\\
PR acknowledges the support from Centre national de la recherche scientifique (CNRS) for this work.\\
SD acknowledges support from ANRF MTR/2023/000384.\\
VS acknowledges the support of the Department of Atomic Energy, Government of India, under project identification No. RTI 4002.\\
ZZ is supported by the Prime Minister’s Research Fellows (PMRF) scheme, Ref. No. TF/PMRF22-7307.\\
The authors acknowledge Siyuan Chen for providing constructive comments that contributed to improving the manuscript.

%\begin{acknowledgments}

%\end{acknowledgments}

\bibliographystyle{apsrev4-2}
\bibliography{InPTA_SGWB}% Produces the bibliography via BibTeX.

@ARTICLE{E16,
       author = {{Einstein}, Albert},
        title = "{N{\"a}herungsweise Integration der Feldgleichungen der Gravitation}",
      journal = {Sitzungsberichte der K{\"o}niglich Preussischen Akademie der Wissenschaften},
         year = 1916,
        month = jan,
        pages = {688-696},
       adsurl = {https://ui.adsabs.harvard.edu/abs/1916SPAW.......688E}
}

@ARTICLE{Trotta08,
       author = {{Trotta}, Roberto},
        title = "{Bayes in the sky: Bayesian inference and model selection in cosmology}",
      journal = {Contemporary Physics},
         year = 2008,
        month = mar,
       volume = {49},
       number = {2},
        pages = {71-104},
          doi = {10.1080/00107510802066753},
archivePrefix = {arXiv},
       eprint = {0803.4089},
 primaryClass = {astro-ph},
       adsurl = {https://ui.adsabs.harvard.edu/abs/2008ConPh..49...71T}
}

@ARTICLE{SDK+23,
       author = {{Srivastava}, Aman and {Desai}, Shantanu and {Kolhe}, Neel and {Surnis}, Mayuresh and {Joshi}, Bhal Chandra and {Susobhanan}, Abhimanyu and {Chalumeau}, Aur{\'e}lien and {Hisano}, Shinnosuke and {Nobleson}, K. and {Arumugam}, Swetha and {Kharbanda}, Divyansh and {Singha}, Jaikhomba and {Tarafdar}, Pratik and {Arumugam}, P. and {Bagchi}, Manjari and {Bathula}, Adarsh and {Dandapat}, Subhajit and {Dey}, Lankeswar and {Dwivedi}, Churchil and {Girgaonkar}, Raghav and {Gopakumar}, A. and {Gupta}, Yashwant and {Kikunaga}, Tomonosuke and {Krishnakumar}, M.~A. and {Liu}, Kuo and {Maan}, Yogesh and {Manoharan}, P.~K. and {Paladi}, Avinash Kumar and {Rana}, Prerna and {Shaifullah}, Golam M. and {Takahashi}, Keitaro},
        title = "{Noise analysis of the Indian Pulsar Timing Array data release I}",
      journal = {\prd},
         year = 2023,
        month = jul,
       volume = {108},
       number = {2},
          eid = {023008},
        pages = {023008},
          doi = {10.1103/PhysRevD.108.023008},
archivePrefix = {arXiv},
       eprint = {2303.12105},
 primaryClass = {astro-ph.HE},
       adsurl = {https://ui.adsabs.harvard.edu/abs/2023PhRvD.108b3008S}
}

@ARTICLE{HT75,
       author = {{Hulse}, R.~A. and {Taylor}, J.~H.},
        title = "{Discovery of a pulsar in a binary system.}",
      journal = {\apjl},
         year = 1975,
        month = jan,
       volume = {195},
        pages = {L51-L53},
          doi = {10.1086/181708},
       adsurl = {https://ui.adsabs.harvard.edu/abs/1975ApJ...195L..51H}
}

@ARTICLE{AAA+16a,
       author = {{Abbott}, B.~P. and {Abbott}, R. and {Abbott}, T.~D. and others},
        title = "{Observation of Gravitational Waves from a Binary Black Hole Merger}",
      journal = {\prl},
         year = 2016,
        month = feb,
       volume = {116},
       number = {6},
          eid = {061102},
        pages = {061102},
          doi = {10.1103/PhysRevLett.116.061102},
archivePrefix = {arXiv},
       eprint = {1602.03837},
 primaryClass = {gr-qc},
       adsurl = {https://ui.adsabs.harvard.edu/abs/2016PhRvL.116f1102A}
}

@BOOK{Kish95,
       author = {{Kish}, L.},
        title = "{Survey Sampling}",
        publisher = {Wiley \& Sons},
        year = {1995},
        isbn = {978-0-471-10949-5}
}

@ARTICLE{ABN+23,
       author = {{Antoniadis}, J. and {Babak}, S. and {Bak Nielsen}, A. -S. and others},
        title = "{The second data release from the European Pulsar Timing Array. I. The dataset and timing analysis}",
      journal = {\aap},
         year = 2023,
        month = oct,
       volume = {678},
          eid = {A48},
        pages = {A48},
          doi = {10.1051/0004-6361/202346841},
archivePrefix = {arXiv},
       eprint = {2306.16224},
 primaryClass = {astro-ph.HE},
       adsurl = {https://ui.adsabs.harvard.edu/abs/2023A&A...678A..48A}
}

@ARTICLE{TNR+22,
       author = {{Tarafdar}, Pratik and {Nobleson}, K. and {Rana}, Prerna and {Singha}, Jaikhomba and {Krishnakumar}, M.~A. and {Joshi}, Bhal Chandra and {Paladi}, Avinash Kumar and {Kolhe}, Neel and {Batra}, Neelam Dhanda and {Agarwal}, Nikita and {Bathula}, Adarsh and {Dandapat}, Subhajit and {Desai}, Shantanu and {Dey}, Lankeswar and {Hisano}, Shinnosuke and {Ingale}, Prathamesh and {Kato}, Ryo and {Kharbanda}, Divyansh and {Kikunaga}, Tomonosuke and {Marmat}, Piyush and {Pandian}, B. Arul and {Prabu}, T. and {Srivastava}, Aman and {Surnis}, Mayuresh and {Susarla}, Sai Chaitanya and {Susobhanan}, Abhimanyu and {Takahashi}, Keitaro and {Arumugam}, P. and {Bagchi}, Manjari and {Banik}, Sarmistha and {De}, Kishalay and {Girgaonkar}, Raghav and {Gopakumar}, A. and {Gupta}, Yashwant and {Maan}, Yogesh and {Manoharan}, P.~K. and {Naidu}, Arun and {Pathak}, Dhruv},
        title = "{The Indian Pulsar Timing Array: First data release}",
      journal = {\pasa},
         year = 2022,
        month = oct,
       volume = {39},
          eid = {e053},
        pages = {e053},
          doi = {10.1017/pasa.2022.46},
archivePrefix = {arXiv},
       eprint = {2206.09289},
 primaryClass = {astro-ph.IM},
       adsurl = {https://ui.adsabs.harvard.edu/abs/2022PASA...39...53T}
}

@ARTICLE{XH+23,
       author = {{Xu}, Heng and {Chen}, Siyuan and {Guo}, Yanjun and others},
        title = "{Searching for the Nano-Hertz Stochastic Gravitational Wave Background with the Chinese Pulsar Timing Array Data Release I}",
      journal = {Research in Astronomy and Astrophysics},
         year = 2023,
        month = jul,
       volume = {23},
       number = {7},
          eid = {075024},
        pages = {075024},
          doi = {10.1088/1674-4527/acdfa5},
archivePrefix = {arXiv},
       eprint = {2306.16216},
 primaryClass = {astro-ph.HE},
       adsurl = {https://ui.adsabs.harvard.edu/abs/2023RAA....23g5024X}
}

@ARTICLE{D79,
       author = {{Detweiler}, S.},
        title = "{Pulsar timing measurements and the search for gravitational waves}",
      journal = {\apj},
         year = 1979,
        month = dec,
       volume = {234},
        pages = {1100-1104},
          doi = {10.1086/157593},
       adsurl = {https://ui.adsabs.harvard.edu/abs/1979ApJ...234.1100D}
}

@ARTICLE{S78,
       author = {{Sazhin}, M.~V.},
        title = "{Opportunities for detecting ultralong gravitational waves}",
      journal = {\sovast},
         year = 1978,
        month = feb,
       volume = {22},
        pages = {36-38},
       adsurl = {https://ui.adsabs.harvard.edu/abs/1978SvA....22...36S}
}

@ARTICLE{HD83,
       author = {{Hellings}, R.~W. and {Downs}, G.~S.},
        title = "{Upper limits on the isotropic gravitational radiation background from pulsar timing analysis.}",
      journal = {\apjl},
         year = 1983,
        month = feb,
       volume = {265},
        pages = {L39-L42},
          doi = {10.1086/183954},
       adsurl = {https://ui.adsabs.harvard.edu/abs/1983ApJ...265L..39H}
}

@ARTICLE{RD+23,
       author = {{Reardon}, Daniel J. and {Zic}, Andrew and {Shannon}, Ryan M. and others},
        title = "{Search for an Isotropic Gravitational-wave Background with the Parkes Pulsar Timing Array}",
      journal = {\apjl},
         year = 2023,
        month = jul,
       volume = {951},
       number = {1},
          eid = {L6},
        pages = {L6},
          doi = {10.3847/2041-8213/acdd02},
archivePrefix = {arXiv},
       eprint = {2306.16215},
 primaryClass = {astro-ph.HE},
       adsurl = {https://ui.adsabs.harvard.edu/abs/2023ApJ...951L...6R}
}

@ARTICLE{AAB+23,
       author = {{EPTA Collaboration} and {InPTA Collaboration} and {Antoniadis}, J. and {Arumugam}, P. and {Arumugam}, S. and {Babak}, S. and {Bagchi}, M. and {Bak Nielsen}, A. -S. and {Bassa}, C.~G. and {Bathula}, A. and {Berthereau}, A. and {Bonetti}, M. and {Bortolas}, E. and {Brook}, P.~R. and {Burgay}, M. and {Caballero}, R.~N. and {Chalumeau}, A. and {Champion}, D.~J. and {Chanlaridis}, S. and {Chen}, S. and {Cognard}, I. and {Dandapat}, S. and {Deb}, D. and {Desai}, S. and {Desvignes}, G. and {Dhanda-Batra}, N. and {Dwivedi}, C. and {Falxa}, M. and {Ferdman}, R.~D. and {Franchini}, A. and {Gair}, J.~R. and {Goncharov}, B. and {Gopakumar}, A. and {Graikou}, E. and {Grie{\ss}meier}, J. -M. and {Guillemot}, L. and {Guo}, Y.~J. and {Gupta}, Y. and {Hisano}, S. and {Hu}, H. and {Iraci}, F. and {Izquierdo-Villalba}, D. and {Jang}, J. and {Jawor}, J. and {Janssen}, G.~H. and {Jessner}, A. and {Joshi}, B.~C. and {Kareem}, F. and {Karuppusamy}, R. and {Keane}, E.~F. and {Keith}, M.~J. and {Kharbanda}, D. and {Kikunaga}, T. and {Kolhe}, N. and {Kramer}, M. and {Krishnakumar}, M.~A. and {Lackeos}, K. and {Lee}, K.~J. and {Liu}, K. and {Liu}, Y. and {Lyne}, A.~G. and {McKee}, J.~W. and {Maan}, Y. and {Main}, R.~A. and {Mickaliger}, M.~B. and {Ni{\c{t}}u}, I.~C. and {Nobleson}, K. and {Paladi}, A.~K. and {Parthasarathy}, A. and {Perera}, B.~B.~P. and {Perrodin}, D. and {Petiteau}, A. and {Porayko}, N.~K. and {Possenti}, A. and {Prabu}, T. and {Quelquejay Leclere}, H. and {Rana}, P. and {Samajdar}, A. and {Sanidas}, S.~A. and {Sesana}, A. and {Shaifullah}, G. and {Singha}, J. and {Speri}, L. and {Spiewak}, R. and {Srivastava}, A. and {Stappers}, B.~W. and {Surnis}, M. and {Susarla}, S.~C. and {Susobhanan}, A. and {Takahashi}, K. and {Tarafdar}, P. and {Theureau}, G. and {Tiburzi}, C. and {van der Wateren}, E. and {Vecchio}, A. and {Venkatraman Krishnan}, V. and {Verbiest}, J.~P.~W. and {Wang}, J. and {Wang}, L. and {Wu}, Z.},
        title = "{The second data release from the European Pulsar Timing Array. III. Search for gravitational wave signals}",
      journal = {\aap},
         year = 2023,
        month = oct,
       volume = {678},
          eid = {A50},
        pages = {A50},
          doi = {10.1051/0004-6361/202346844},
archivePrefix = {arXiv},
       eprint = {2306.16214},
 primaryClass = {astro-ph.HE},
       adsurl = {https://ui.adsabs.harvard.edu/abs/2023A&A...678A..50E}
}

@ARTICLE{AAA+23b,
       author = {{Agazie}, Gabriella and {Anumarlapudi}, Akash and {Archibald}, Anne M. and {Arzoumanian}, Zaven and {Baker}, Paul T. and {B{\'e}csy}, Bence and {Blecha}, Laura and {Brazier}, Adam and {Brook}, Paul R. and {Burke-Spolaor}, Sarah and {Burnette}, Rand and {Case}, Robin and {Charisi}, Maria and {Chatterjee}, Shami and {Chatziioannou}, Katerina and {Cheeseboro}, Belinda D. and {Chen}, Siyuan and {Cohen}, Tyler and {Cordes}, James M. and {Cornish}, Neil J. and {Crawford}, Fronefield and {Cromartie}, H. Thankful and {Crowter}, Kathryn and {Cutler}, Curt J. and {Decesar}, Megan E. and {Degan}, Dallas and {Demorest}, Paul B. and {Deng}, Heling and {Dolch}, Timothy and {Drachler}, Brendan and {Ellis}, Justin A. and {Ferrara}, Elizabeth C. and {Fiore}, William and {Fonseca}, Emmanuel and {Freedman}, Gabriel E. and {Garver-Daniels}, Nate and {Gentile}, Peter A. and {Gersbach}, Kyle A. and {Glaser}, Joseph and {Good}, Deborah C. and {G{\"u}ltekin}, Kayhan and {Hazboun}, Jeffrey S. and {Hourihane}, Sophie and {Islo}, Kristina and {Jennings}, Ross J. and {Johnson}, Aaron D. and {Jones}, Megan L. and {Kaiser}, Andrew R. and {Kaplan}, David L. and {Kelley}, Luke Zoltan and {Kerr}, Matthew and {Key}, Joey S. and {Klein}, Tonia C. and {Laal}, Nima and {Lam}, Michael T. and {Lamb}, William G. and {Lazio}, T. Joseph W. and {Lewandowska}, Natalia and {Littenberg}, Tyson B. and {Liu}, Tingting and {Lommen}, Andrea and {Lorimer}, Duncan R. and {Luo}, Jing and {Lynch}, Ryan S. and {Ma}, Chung-Pei and {Madison}, Dustin R. and {Mattson}, Margaret A. and {McEwen}, Alexander and {McKee}, James W. and {McLaughlin}, Maura A. and {McMann}, Natasha and {Meyers}, Bradley W. and {Meyers}, Patrick M. and {Mingarelli}, Chiara M.~F. and {Mitridate}, Andrea and {Natarajan}, Priyamvada and {Ng}, Cherry and {Nice}, David J. and {Ocker}, Stella Koch and {Olum}, Ken D. and {Pennucci}, Timothy T. and {Perera}, Benetge B.~P. and {Petrov}, Polina and {Pol}, Nihan S. and {Radovan}, Henri A. and {Ransom}, Scott M. and {Ray}, Paul S. and {Romano}, Joseph D. and {Sardesai}, Shashwat C. and {Schmiedekamp}, Ann and {Schmiedekamp}, Carl and {Schmitz}, Kai and {Schult}, Levi and {Shapiro-Albert}, Brent J. and {Siemens}, Xavier and {Simon}, Joseph and {Siwek}, Magdalena S. and {Stairs}, Ingrid H. and {Stinebring}, Daniel R. and {Stovall}, Kevin and {Sun}, Jerry P. and {Susobhanan}, Abhimanyu and {Swiggum}, Joseph K. and {Taylor}, Jacob and {Taylor}, Stephen R. and {Turner}, Jacob E. and {Unal}, Caner and {Vallisneri}, Michele and {van Haasteren}, Rutger and {Vigeland}, Sarah J. and {Wahl}, Haley M. and {Wang}, Qiaohong and {Witt}, Caitlin A. and {Young}, Olivia and {Nanograv Collaboration}},
        title = "{The NANOGrav 15 yr Data Set: Evidence for a Gravitational-wave Background}",
      journal = {\apjl},
         year = 2023,
        month = jul,
       volume = {951},
       number = {1},
          eid = {L8},
        pages = {L8},
          doi = {10.3847/2041-8213/acdac6},
archivePrefix = {arXiv},
       eprint = {2306.16213},
 primaryClass = {astro-ph.HE},
       adsurl = {https://ui.adsabs.harvard.edu/abs/2023ApJ...951L...8A}
}

@ARTICLE{PTV+21,
       author = {{Pol}, Nihan S. and {Taylor}, Stephen R. and {Kelley}, Luke Zoltan and {Vigeland}, Sarah J. and {Simon}, Joseph and {Chen}, Siyuan and others},
        title = "{Astrophysics Milestones for Pulsar Timing Array Gravitational-wave Detection}",
      journal = {\apjl},
         year = 2021,
        month = apr,
       volume = {911},
       number = {2},
          eid = {L34},
        pages = {L34},
          doi = {10.3847/2041-8213/abf2c9},
archivePrefix = {arXiv},
       eprint = {2010.11950},
 primaryClass = {astro-ph.HE},
       adsurl = {https://ui.adsabs.harvard.edu/abs/2021ApJ...911L..34P}
}

@ARTICLE{RTN+25,
       author = {{Rana}, Prerna and {Tarafdar}, Pratik and {Nobleson}, K. and {Dwivedi}, Churchil and {Chandra Joshi}, Bhal and {Deb}, Debabrata and {Mondal}, Sushovan and {Krishnakumar}, M.~A. and {Shukla}, Adya and {Singha}, Jaikhomba and {Grover}, Himanshu and {Tahbildar}, Hemanga and {Susobhanan}, Abhimanyu and {Surnis}, Mayuresh and {Desai}, Shantanu and {Batra}, Neelam Dhanda and {Srivastava}, Aman and {Bharambe}, Vinay and {Jose}, Jibin and {Vyasraj}, Vaishnavi and {Jose Jacob}, Shebin and {Amarnath} and {Singh}, Manpreet and {Zuraiq}, Zenia and {Sengupta}, Sarbartha and {Ogi}, Toki and {Kumar}, Dhruv and {Jagadeesh}, S. and {Kareem}, Fazal and {Maity}, Deep and {Rai}, Kaustubh and {Vara}, Kunjal and {Chowdhury}, Shaswata and {Kato}, Ryo and {Arumugam}, Swetha and {Mamidipaka}, Pragna and {Arul Pandian}, B. and {Shaji}, Kavya and {Thiagaraj}, Prabu and {Arumugam}, Paramasivan and {Bagchi}, Manjari and {Chakraborty}, Manoneeta and {Gopakumar}, Achamveedu and {Gupta}, Yashwant and {Maan}, Yogesh and {Kumar Paladi}, Avinash and {Takahashi}, Keitaro},
        title = "{The Indian Pulsar Timing Array data release 2: I. Dataset and timing analysis}",
      journal = {\pasa},
         year = 2025,
        month = jul,
       volume = {42},
          eid = {e108},
        pages = {e108},
          doi = {10.1017/pasa.2025.10066},
archivePrefix = {arXiv},
       eprint = {2506.16769},
 primaryClass = {astro-ph.IM},
       adsurl = {https://ui.adsabs.harvard.edu/abs/2025PASA...42..108R}
}

@MISC{VM20,
       author = {{Vallisneri}, Michele},
        title = "{libstempo: Python wrapper for Tempo2}",
 howpublished = {Astrophysics Source Code Library, record ascl:2002.017},
         year = 2020,
        month = feb,
          eid = {ascl:2002.017},
archivePrefix = {ascl},
       eprint = {2002.017},
       adsurl = {https://ui.adsabs.harvard.edu/abs/2020ascl.soft02017V}
}

@ARTICLE{VN+09,
       author = {{van Haasteren}, R. and {Levin}, Y. and {McDonald}, P. and {Lu}, T.},
        title = "{On measuring the gravitational-wave background using pulsar timing arrays}",
      journal = {\mnras},
         year = 2009,
        month = may,
       volume = {395},
       number = {2},
        pages = {1005-1014},
          doi = {10.1111/j.1365-2966.2009.14590.x},
archivePrefix = {arXiv},
       eprint = {0809.0791},
 primaryClass = {astro-ph},
       adsurl = {https://ui.adsabs.harvard.edu/abs/2009MNRAS.395.1005V}
}

@ARTICLE{LT+15,
       author = {{Lentati}, L. and {Taylor}, S.~R. and {Mingarelli}, C.~M.~F. and others},
        title = "{European Pulsar Timing Array limits on an isotropic stochastic gravitational-wave background}",
      journal = {\mnras},
         year = 2015,
        month = nov,
       volume = {453},
       number = {3},
        pages = {2576-2598},
          doi = {10.1093/mnras/stv1538},
archivePrefix = {arXiv},
       eprint = {1504.03692},
 primaryClass = {astro-ph.HE},
       adsurl = {https://ui.adsabs.harvard.edu/abs/2015MNRAS.453.2576L}
}

@ARTICLE{TV+16,
       author = {{Taylor}, S.~R. and {Vallisneri}, M. and {Ellis}, J.~A. and {Mingarelli}, C.~M.~F. and {Lazio}, T.~J.~W. and {van Haasteren}, R.},
        title = "{Are We There Yet? Time to Detection of Nanohertz Gravitational Waves Based on Pulsar-timing Array Limits}",
      journal = {\apjl},
         year = 2016,
       volume = {819},
       number = {1},
          eid = {L6},
        pages = {L6},
          doi = {10.3847/2041-8205/819/1/L6},
archivePrefix = {arXiv},
       eprint = {1511.05564},
 primaryClass = {astro-ph.IM},
       adsurl = {https://ui.adsabs.harvard.edu/abs/2016ApJ...819L...6T}
}

@MISC{EVM20,
       author = {{Ellis}, Justin A. and {Vallisneri}, Michele and {Taylor}, Stephen R. and {Baker}, Paul T.},
        title = "{ENTERPRISE: Enhanced Numerical Toolbox Enabling a Robust PulsaR Inference SuitE}",
 howpublished = {Astrophysics Source Code Library, record ascl:1912.015},
         year = 2019,
        month = dec,
          eid = {ascl:1912.015},
        pages = {ascl:1912.015},
archivePrefix = {ascl},
       eprint = {1912.015},
       adsurl = {https://ui.adsabs.harvard.edu/abs/2019ascl.soft12015E}
}

@ARTICLE{VV13,
       author = {{van Haasteren}, Rutger and {Vallisneri}, Michele},
        title = "{New advances in the Gaussian-process approach to pulsar-timing data analysis}",
      journal = {\prd},
         year = 2014,
       volume = {90},
       number = {10},
          eid = {104012},
        pages = {104012},
          doi = {10.1103/PhysRevD.90.104012},
archivePrefix = {arXiv},
       eprint = {1407.1838},
 primaryClass = {gr-qc},
       adsurl = {https://ui.adsabs.harvard.edu/abs/2014PhRvD..90j4012V}
}

@ARTICLE{AB+09,
       author = {{Anholm}, M. and {Ballmer}, S. and {Creighton}, J.~D.~E. and others},
        title = "{Optimal strategies for gravitational wave stochastic background searches in pulsar timing data}",
      journal = {\prd},
         year = 2009,
        month = apr,
       volume = {79},
       number = {8},
          eid = {084030},
        pages = {084030},
          doi = {10.1103/PhysRevD.79.084030},
archivePrefix = {arXiv},
       eprint = {0809.0701},
 primaryClass = {gr-qc},
       adsurl = {https://ui.adsabs.harvard.edu/abs/2009PhRvD..79h4030A}
}

@ARTICLE{DF+13,
       author = {{Demorest}, P.~B. and {Ferdman}, R.~D. and {Gonzalez}, M.~E. and others},
        title = "{Limits on the Stochastic Gravitational Wave Background from the North American Pulsar Timing Array Third Data Release}",
      journal = {\apj},
         year = 2013,
        month = jan,
       volume = {762},
       number = {2},
          eid = {94},
        pages = {94},
          doi = {10.1088/0004-637X/762/2/94},
archivePrefix = {arXiv},
       eprint = {1201.6641},
 primaryClass = {astro-ph.HE},
       adsurl = {https://ui.adsabs.harvard.edu/abs/2013ApJ...762...94D}
}

@ARTICLE{CC+15,
       author = {{Chamberlin}, Sydney J. and {Creighton}, Jolien D.~E. and {Siemens}, Xavier and {Demorest}, Paul and {Ellis}, Justin and {Price}, Larry R. and {Romano}, Joseph D.},
        title = "{Time-domain implementation of the optimal statistic for stochastic gravitational-wave background searches in pulsar timing arrays}",
      journal = {\prd},
         year = 2015,
        month = feb,
       volume = {91},
       number = {4},
          eid = {044048},
        pages = {044048},
          doi = {10.1103/PhysRevD.91.044048},
archivePrefix = {arXiv},
       eprint = {1410.8256},
 primaryClass = {gr-qc},
       adsurl = {https://ui.adsabs.harvard.edu/abs/2015PhRvD..91d4048C}
}

@ARTICLE{VI+18,
       author = {{Vigeland}, Sarah J. and {Islo}, Kristina and {Taylor}, Stephen R. and {Ellis}, Justin A.},
        title = "{Noise-marginalized optimal statistic: A robust hybrid frequentist-Bayesian statistic for the stochastic gravitational-wave background in pulsar timing arrays}",
      journal = {\prd},
         year = 2018,
        month = aug,
       volume = {98},
       number = {4},
          eid = {044003},
        pages = {044003},
          doi = {10.1103/PhysRevD.98.044003},
archivePrefix = {arXiv},
       eprint = {1805.12188},
 primaryClass = {gr-qc},
       adsurl = {https://ui.adsabs.harvard.edu/abs/2018PhRvD..98d4003V}
}

@ARTICLE{AA+15,
       author = {{LIGO Scientific Collaboration} and {Aasi}, J. and {Abbott}, B.~P. and others},
        title = "{Advanced LIGO}",
      journal = {Classical and Quantum Gravity},
         year = 2015,
        month = apr,
       volume = {32},
       number = {7},
          eid = {074001},
        pages = {074001},
          doi = {10.1088/0264-9381/32/7/074001},
archivePrefix = {arXiv},
       eprint = {1411.4547},
 primaryClass = {gr-qc},
       adsurl = {https://ui.adsabs.harvard.edu/abs/2015CQGra..32g4001L}
}

@ARTICLE{AAB+17,
       author = {{Amaro-Seoane}, Pau and {Audley}, Heather and {Babak}, Stanislav and others},
        title = "{Laser Interferometer Space Antenna}",
      journal = {arXiv e-prints},
         year = 2017,
        month = feb,
          eid = {arXiv:1702.00786},
        pages = {arXiv:1702.00786},
          doi = {10.48550/arXiv.1702.00786},
archivePrefix = {arXiv},
       eprint = {1702.00786},
  primaryClass = {astro-ph.GA},
       adsurl = {https://ui.adsabs.harvard.edu/abs/2017arXiv170200786A}
}

@ARTICLE{FB90,
       author = {{Foster}, R.~S. and {Backer}, D.~C.},
        title = "{Constructing a Pulsar Timing Array}",
      journal = {\apj},
         year = 1990,
        month = sep,
       volume = {361},
        pages = {300-309},
          doi = {10.1086/169195},
       adsurl = {https://ui.adsabs.harvard.edu/abs/1990ApJ...361..300F}
}

@ARTICLE{HB+68,
       author = {{Hewish}, A. and {Bell}, S.~J. and {Pilkington}, J.~D.~H. and {Scott}, P.~F. and {Collins}, R.~A.},
        title = "{Observation of a Rapidly Pulsating Radio Source}",
      journal = {\nature},
         year = 1968,
        month = feb,
       volume = {217},
       number = {5130},
        pages = {709-713},
          doi = {10.1038/217709a0},
       adsurl = {https://ui.adsabs.harvard.edu/abs/1968Natur.217..709H}
}

@ARTICLE{BK+82,
       author = {{Backer}, D.~C. and {Kulkarni}, S.~R. and {Heiles}, C. and {Davis}, M.~M. and {Goss}, W.~M.},
        title = "{A millisecond pulsar}",
      journal = {\nature},
         year = 1982,
        month = dec,
       volume = {300},
       number = {5893},
        pages = {615-618},
          doi = {10.1038/300615a0},
       adsurl = {https://ui.adsabs.harvard.edu/abs/1982Natur.300..615B}
}

@ARTICLE{EHM06,
       author = {{Edwards}, R.~T. and {Hobbs}, G.~B. and {Manchester}, R.~N.},
        title = "{TEMPO2, a new pulsar-timing package. II. The timing model and precision estimates}",
      journal = {\mnras},
         year = 2006,
        month = nov,
       volume = {372},
       number = {4},
        pages = {1549-1574},
          doi = {10.1111/j.1365-2966.2006.10870.x},
archivePrefix = {arXiv},
       eprint = {astro-ph/0607664},
 primaryClass = {astro-ph},
       adsurl = {https://ui.adsabs.harvard.edu/abs/2006MNRAS.372.1549E}
}

@ARTICLE{BT+19,
       author = {{Burke-Spolaor}, S. and {Taylor}, S.~R. and {Charisi}, M. and {Dolch}, T. and {Hazboun}, J.~S. and {Holgado}, A.~M. and {Kelley}, L.~Z. and {Lazio}, T.~J.~W. and {Madison}, D.~R. and {McMann}, N. and {Mingarelli}, C.~M.~F. and {Rasskazov}, A. and {Siemens}, X. and {Simon}, J.~J. and {Smith}, T.~L.},
        title = "{The astrophysics of nanohertz gravitational waves}",
      journal = {The Astronomy and Astrophysics Review},
         year = 2019,
        month = aug,
       volume = {27},
       number = {1},
          eid = {5},
        pages = {5},
          doi = {10.1007/s00159-019-0115-7},
archivePrefix = {arXiv},
       eprint = {1811.08826},
 primaryClass = {astro-ph.HE},
       adsurl = {https://ui.adsabs.harvard.edu/abs/2019A&ARv..27....5B}
}

@ARTICLE{DZ+24,
       author = {{Di Marco}, Valentina and {Zic}, Andrew and {Shannon}, Ryan M. and {Thrane}, Eric},
        title = "{Systematic errors in searches for nanohertz gravitational waves}",
      journal = {\mnras},
         year = 2024,
        month = aug,
       volume = {532},
       number = {4},
        pages = {4026-4034},
          doi = {10.1093/mnras/stae1750},
archivePrefix = {arXiv},
       eprint = {2403.13175},
 primaryClass = {astro-ph.HE},
       adsurl = {https://ui.adsabs.harvard.edu/abs/2024MNRAS.532.4026D}
}

@ARTICLE{PS00,
       author = {{Phinney}, E.~S.},
        title = "{A Practical Theorem for Light and Gravitational Radiation from a Cosmic Distribution of Objects - an Application to Massive Black Holes}",
      journal = {arXiv e-prints},
         year = 2001,
        month = aug,
          eid = {astro-ph/0108028},
        pages = {astro-ph/0108028},
archivePrefix = {arXiv},
       eprint = {astro-ph/0108028},
 primaryClass = {astro-ph},
       adsurl = {https://ui.adsabs.harvard.edu/abs/2001astro.ph..8028P}
}

@ARTICLE{LM+16,
       author = {{Lasky}, Paul D. and {Mingarelli}, Chiara M.~F. and {Smith}, Tristan L. and others},
        title = "{Gravitational-Wave Cosmology across 29 Decades in Frequency}",
      journal = {Physical Review X},
         year = 2016,
        month = mar,
       volume = {6},
       number = {1},
          eid = {011035},
        pages = {011035},
          doi = {10.1103/PhysRevX.6.011035},
archivePrefix = {arXiv},
       eprint = {1511.05994},
 primaryClass = {astro-ph.CO},
       adsurl = {https://ui.adsabs.harvard.edu/abs/2016PhRvX...6a1035L}
}

@ARTICLE{BS+22,
       author = {{Bian}, Ligong and {Shu}, Jing sho and {Wang}, Bo and {Yuan}, Qiang and {Zong}, Junchao},
        title = "{Searching for cosmic string induced stochastic gravitational wave background with the Parkes Pulsar Timing Array}",
      journal = {\prd},
         year = 2022,
        month = nov,
       volume = {106},
       number = {10},
          eid = {L101301},
        pages = {L101301},
          doi = {10.1103/PhysRevD.106.L101301},
archivePrefix = {arXiv},
       eprint = {2205.07293},
 primaryClass = {astro-ph.CO},
       adsurl = {https://ui.adsabs.harvard.edu/abs/2022PhRvD.106l1013B}
}

@ARTICLE{MMP+23,
       author = {{Madge}, Eric and {Morgante}, Enrico and {Puchades-Ib{\'a}{\~n}ez}, Cristina and {Ramberg}, Nicklas and {Ratzinger}, Wolfram and {Schenk}, Sebastian and {Schwaller}, Pedro},
        title = "{Primordial gravitational waves in the nano-Hertz regime and PTA data - towards solving the GW inverse problem}",
      journal = {Journal of High Energy Physics},
         year = 2023,
        month = oct,
       volume = {2023},
       number = {10},
          eid = {171},
        pages = {171},
          doi = {10.1007/JHEP10(2023)171},
archivePrefix = {arXiv},
       eprint = {2306.14856},
 primaryClass = {hep-ph},
       adsurl = {https://ui.adsabs.harvard.edu/abs/2023JHEP...10..171M}
}

@ARTICLE{JH06,
       author = {{Jenet}, F.~A. and {Hobbs}, G.~B. and {van Straten}, W. and {Manchester}, R.~N. and {Bailes}, M. and {Verbiest}, J.~P.~W. and {Edwards}, R.~T. and {Hotan}, A.~W. and {Sarkissian}, J.~M. and {Ord}, S.~M.},
        title = "{Upper Bounds on the Low-Frequency Stochastic Gravitational Wave Background from Pulsar Timing Observations: Current Limits and Future Prospects}",
      journal = {\apj},
         year = 2006,
        month = dec,
       volume = {653},
       number = {2},
        pages = {1571-1576},
          doi = {10.1086/508702},
archivePrefix = {arXiv},
       eprint = {astro-ph/0609013},
 primaryClass = {astro-ph},
       adsurl = {https://ui.adsabs.harvard.edu/abs/2006ApJ...653.1571J}
}

@ARTICLE{MS+13,
       author = {{Mingarelli}, C.~M.~F. and {Sidery}, T. and {Mandel}, I. and {Vecchio}, A.},
        title = "{Characterizing gravitational wave stochastic background anisotropy with pulsar timing arrays}",
      journal = {\prd},
         year = 2013,
        month = sep,
       volume = {88},
       number = {6},
          eid = {062005},
        pages = {062005},
          doi = {10.1103/PhysRevD.88.062005},
archivePrefix = {arXiv},
       eprint = {1306.5394},
 primaryClass = {astro-ph.HE},
       adsurl = {https://ui.adsabs.harvard.edu/abs/2013PhRvD..88f2005M}
}

@ARTICLE{GJ+15,
       author = {{Gair}, Jonathan R. and {Romano}, Joseph D. and {Taylor}, Stephen R.},
        title = "{Mapping gravitational-wave backgrounds of arbitrary polarisation using pulsar timing arrays}",
      journal = {\prd},
         year = 2015,
        month = nov,
       volume = {92},
       number = {10},
          eid = {102003},
        pages = {102003},
          doi = {10.1103/PhysRevD.92.102003},
archivePrefix = {arXiv},
       eprint = {1506.08668},
 primaryClass = {gr-qc},
       adsurl = {https://ui.adsabs.harvard.edu/abs/2015PhRvD..92j2003G}
}

@ARTICLE{ABB+20,
       author = {{Arzoumanian}, Zaven and {Baker}, Paul T. and {Blumer}, Harsha and {B{\'e}csy}, Bence and {Brazier}, Adam and {Brook}, Paul R. and {Burke-Spolaor}, Sarah and {Chatterjee}, Shami and {Chen}, Siyuan and {Cordes}, James M. and {Cornish}, Neil J. and {Crawford}, Fronefield and {Cromartie}, H. Thankful and {Decesar}, Megan E. and {Demorest}, Paul B. and {Dolch}, Timothy and {Ellis}, Justin A. and {Ferrara}, Elizabeth C. and {Fiore}, William and {Fonseca}, Emmanuel and {Garver-Daniels}, Nathan and {Gentile}, Peter A. and {Good}, Deborah C. and {Hazboun}, Jeffrey S. and {Holgado}, A. Miguel and {Islo}, Kristina and {Jennings}, Ross J. and {Jones}, Megan L. and {Kaiser}, Andrew R. and {Kaplan}, David L. and {Kelley}, Luke Zoltan and {Key}, Joey Shapiro and {Laal}, Nima and {Lam}, Michael T. and {Lazio}, T. Joseph W. and {Lorimer}, Duncan R. and {Luo}, Jing and {Lynch}, Ryan S. and {Madison}, Dustin R. and {McLaughlin}, Maura A. and {Mingarelli}, Chiara M.~F. and {Ng}, Cherry and {Nice}, David J. and {Pennucci}, Timothy T. and {Pol}, Nihan S. and {Ransom}, Scott M. and {Ray}, Paul S. and {Shapiro-Albert}, Brent J. and {Siemens}, Xavier and {Simon}, Joseph and {Spiewak}, Ren{\'e}e and {Stairs}, Ingrid H. and {Stinebring}, Daniel R. and {Stovall}, Kevin and {Sun}, Jerry P. and {Swiggum}, Joseph K. and {Taylor}, Stephen R. and {Turner}, Jacob E. and {Vallisneri}, Michele and {Vigeland}, Sarah J. and {Witt}, Caitlin A. and {Nanograv Collaboration}},
        title = "{The NANOGrav 12.5 yr Data Set: Search for an Isotropic Stochastic Gravitational-wave Background}",
      journal = {\apjl},
         year = 2020,
        month = dec,
       volume = {905},
       number = {2},
          eid = {L34},
        pages = {L34},
          doi = {10.3847/2041-8213/abd401},
archivePrefix = {arXiv},
       eprint = {2009.04496},
 primaryClass = {astro-ph.HE},
       adsurl = {https://ui.adsabs.harvard.edu/abs/2020ApJ...905L..34A}
}

@ARTICLE{DM71,
       author = {{Dickey}, James M.},
        title = "{The Weighted Likelihood Ratio, Linear Hypotheses on Normal Multi-Variate Distributions}",
      journal = {The Annals of Mathematical Statistics},
         year = 1971,
        month = feb,
       volume = {42},
       number = {1},
        pages = {204-223},
          doi = {10.1214/aoms/1177693507},
       adsurl = {https://ui.adsabs.harvard.edu/abs/1971AnMS...42..204D}
}

@ARTICLE{HM+23,
       author = {{Hourihane}, Sophie and {Meyers}, Patrick and {Johnson}, Aaron and {Chatziioannou}, Katerina and {Vallisneri}, Michele},
        title = "{Accurate characterization of the stochastic gravitational-wave background with pulsar timing arrays by likelihood reweighting}",
      journal = {\prd},
         year = 2023,
        month = apr,
       volume = {107},
       number = {8},
          eid = {084045},
        pages = {084045},
          doi = {10.1103/PhysRevD.107.084045},
archivePrefix = {arXiv},
       eprint = {2212.06276},
 primaryClass = {gr-qc},
       adsurl = {https://ui.adsabs.harvard.edu/abs/2023PhRvD.107h4045H}
}

@ARTICLE{ND+26,
       author = {{Nobleson}, K. and {Dwivedi}, Churchil and {Desai}, Shantanu and {Joshi}, Bhal Chandra and {Grover}, Himanshu and {Deb}, Debabrata and {Vyasraj}, Vaishnavi and {Vara}, Kunjal and {Tahbildar}, Hemanga and {Susobhanan}, Abhimanyu and {Surnis}, Mayuresh and {Srivastava}, Aman and {Sardana}, Shubhit and {Takahashi}, Keitaro and {Amarnath} and {Arumugam}, P. and {Bagchi}, Manjari and {Batra}, Neelam Dhanda and {Chakraborty}, Manoneeta and {Chowdhury}, Shaswata and {Jacob}, Shebin Jose and {Jose}, Jibin and {Kala}, Shubham and {Kato}, Ryo and {Krishnakumar}, M.~A. and {Meena}, Kuldeep and {Paladi}, Avinash Kumar and {Pandian}, Arul and {Rai}, Kaustubh and {Rana}, Prerna and {Singh}, Manpreet and {Singha}, Jaikhomba and {Shukla}, Adya and {Tarafdar}, Pratik and {Thiagraj}, Prabu and {Zuraiq}, Zenia},
        title = "{The Indian Pulsar Timing Array Data Release 2: II. Customised Single-Pulsar Noise Analysis and Noise Budget}",
      journal = {Journal of High Energy Astrophysics},
         year = 2026,
        month = jul,
       volume = {53},
          eid = {100594},
        pages = {100594},
          doi = {10.1016/j.jheap.2026.100594},
archivePrefix = {arXiv},
       eprint = {2512.20455},
 primaryClass = {astro-ph.HE},
       adsurl = {https://ui.adsabs.harvard.edu/abs/2026JHEAp..5300594N}
}

@ARTICLE{JAB+18,
       author = {{Joshi}, Bhal Chandra and {Arumugasamy}, Prakash and {Bagchi}, Manjari and {Bandyopadhyay}, Debades and {Basu}, Avishek and {Dhanda Batra}, Neelam and {Bethapudi}, Suryarao and {Choudhary}, Arpita and {De}, Kishalay and {Dey}, L. and {Gopakumar}, A. and {Gupta}, Y. and {Krishnakumar}, M.~A. and {Maan}, Yogesh and {Manoharan}, P.~K. and {Naidu}, Arun and {Nandi}, Rana and {Pathak}, Dhruv and {Surnis}, Mayuresh and {Susobhanan}, Abhimanyu},
        title = "{Precision pulsar timing with the ORT and the GMRT and its applications in pulsar astrophysics}",
      journal = {Journal of Astrophysics and Astronomy},
         year = 2018,
        month = aug,
       volume = {39},
       number = {4},
          eid = {51},
        pages = {51},
          doi = {10.1007/s12036-018-9549-y},
       adsurl = {https://ui.adsabs.harvard.edu/abs/2018JApA...39...51J}
}

@ARTICLE{SEJ+13,
       author = {{Siemens}, Xavier and {Ellis}, Justin and {Jenet}, Fredrick and {Romano}, Joseph D.},
        title = "{The stochastic background: scaling laws and time to detection for pulsar timing arrays}",
      journal = {Classical and Quantum Gravity},
         year = 2013,
        month = nov,
       volume = {30},
       number = {22},
          eid = {224015},
        pages = {224015},
          doi = {10.1088/0264-9381/30/22/224015},
archivePrefix = {arXiv},
       eprint = {1305.3196},
 primaryClass = {astro-ph.IM},
       adsurl = {https://ui.adsabs.harvard.edu/abs/2013CQGra..30v4015S}
}

@ARTICLE{Joshi2022,
       author = {{Joshi}, Bhal Chandra and {Gopakumar}, Achamveedu and {Pandian}, Arul and {Prabu}, Thiagaraj and {Dey}, Lankeswar and {Bagchi}, Manjari and {Desai}, Shantanu and {Tarafdar}, Pratik and {Rana}, Prerna and {Maan}, Yogesh and {Batra}, Neelam Dhanda and {Girgaonkar}, Raghav and {Agarwal}, Nikita and {Arumugam}, Paramasivan and {Basu}, Avishek and {Bathula}, Adarsh and {Dandapat}, Subhajit and {Gupta}, Yashwant and {Hisano}, Shinnosuke and {Kato}, Ryo and {Kharbanda}, Divyansh and {Kikunaga}, Tomonosuke and {Kolhe}, Neel and {Krishnakumar}, M.~A. and {Manoharan}, P.~K. and {Marmat}, Piyush and {Naidu}, Arun and {Banik}, Sarmistha and {Nobleson}, K. and {Paladi}, Avinash Kumar and {Pathak}, Dhruv and {Singha}, Jaikhomba and {Srivastava}, Aman and {Surnis}, Mayuresh and {Susarla}, Sai Chaitanya and {Susobhanan}, Abhimanyu and {Takahashi}, Keitaro},
        title = "{Nanohertz gravitational wave astronomy during SKA era: An InPTA perspective}",
      journal = {Journal of Astrophysics and Astronomy},
         year = 2022,
        month = dec,
       volume = {43},
       number = {2},
          eid = {98},
        pages = {98},
          doi = {10.1007/s12036-022-09869-w},
archivePrefix = {arXiv},
       eprint = {2207.06461},
 primaryClass = {astro-ph.HE},
       adsurl = {https://ui.adsabs.harvard.edu/abs/2022JApA...43...98J}
}

@ARTICLE{sak+1991,
       author = {{Swarup}, G. and {Ananthakrishnan}, S. and {Kapahi}, V.~K. and {Rao}, A.~P. and {Subrahmanya}, C.~R. and {Kulkarni}, V.~K.},
        title = "{The Giant Metre-Wave Radio Telescope}",
      journal = {Current Science},
         year = 1991,
        month = jan,
       volume = {60},
       number = {2},
        pages = {95-105},
       adsurl = {https://ui.adsabs.harvard.edu/abs/1991CSci...60...95S}
}

@ARTICLE{gak+2017,
       author = {{Gupta}, Y. and {Ajithkumar}, B. and {Kale}, H.~S. and {Nayak}, S. and {Sabhapathy}, S. and {Sureshkumar}, S. and {Swami}, R.~V. and {Chengalur}, J.~N. and {Ghosh}, S.~K. and {Ishwara-Chandra}, C.~H. and {Joshi}, B.~C. and {Kanekar}, N. and {Lal}, D.~V. and {Roy}, S.},
        title = "{The upgraded GMRT: opening new windows on the radio Universe}",
      journal = {Current Science},
         year = 2017,
        month = aug,
       volume = {113},
       number = {4},
        pages = {707-714},
          doi = {10.18520/cs/v113/i04/707-714},
       adsurl = {https://ui.adsabs.harvard.edu/abs/2017CSci..113..707G}
}

@ARTICLE{HGC+20,
       author = {{Hobbs}, G. and {Guo}, L. and {Caballero}, R.~N. and {Coles}, W. and {Lee}, K.~J. and {Manchester}, R.~N. and {Reardon}, D.~J. and {Matsakis}, D. and {Tong}, M.~L. and {Arzoumanian}, Z. and {Bailes}, M. and {Bassa}, C.~G. and {Bhat}, N.~D.~R. and {Brazier}, A. and {Burke-Spolaor}, S. and {Champion}, D.~J. and {Chatterjee}, S. and {Cognard}, I. and {Dai}, S. and {Desvignes}, G. and {Dolch}, T. and {Ferdman}, R.~D. and {Graikou}, E. and {Guillemot}, L. and {Janssen}, G.~H. and {Keith}, M.~J. and {Kerr}, M. and {Kramer}, M. and {Lam}, M.~T. and {Liu}, K. and {Lyne}, A. and {Lazio}, T.~J.~W. and {Lynch}, R. and {McKee}, J.~W. and {McLaughlin}, M.~A. and {Mingarelli}, C.~M.~F. and {Nice}, D.~J. and {Os{\l}owski}, S. and {Pennucci}, T.~T. and {Perera}, B.~B.~P. and {Perrodin}, D. and {Possenti}, A. and {Russell}, C.~J. and {Sanidas}, S. and {Sesana}, A. and {Shaifullah}, G. and {Shannon}, R.~M. and {Simon}, J. and {Spiewak}, R. and {Stairs}, I.~H. and {Stappers}, B.~W. and {Swiggum}, J.~K. and {Taylor}, S.~R. and {Theureau}, G. and {Toomey}, L. and {van Haasteren}, R. and {Wang}, J.~B. and {Wang}, Y. and {Zhu}, X.~J.},
        title = "{A pulsar-based time-scale from the International Pulsar Timing Array}",
      journal = {\mnras},
         year = 2020,
        month = feb,
       volume = {491},
       number = {4},
        pages = {5951-5965},
          doi = {10.1093/mnras/stz3071},
archivePrefix = {arXiv},
       eprint = {1910.13628},
 primaryClass = {astro-ph.IM},
       adsurl = {https://ui.adsabs.harvard.edu/abs/2020MNRAS.491.5951H}
}

@ARTICLE{AAA+23,
       author = {{Agazie}, Gabriella and {Anumarlapudi}, Akash and {Archibald}, Anne M. and {Arzoumanian}, Zaven and {Baker}, Paul T. and {B{\'e}csy}, Bence and {Blecha}, Laura and {Brazier}, Adam and {Brook}, Paul R. and {Burke-Spolaor}, Sarah and {Charisi}, Maria and {Chatterjee}, Shami and {Cohen}, Tyler and {Cordes}, James M. and {Cornish}, Neil J. and {Crawford}, Fronefield and {Cromartie}, H. Thankful and {Crowter}, Kathryn and {Decesar}, Megan E. and {Demorest}, Paul B. and {Dolch}, Timothy and {Drachler}, Brendan and {Ferrara}, Elizabeth C. and {Fiore}, William and {Fonseca}, Emmanuel and {Freedman}, Gabriel E. and {Garver-Daniels}, Nate and {Gentile}, Peter A. and {Glaser}, Joseph and {Good}, Deborah C. and {Guertin}, Lydia and {G{\"u}ltekin}, Kayhan and {Hazboun}, Jeffrey S. and {Jennings}, Ross J. and {Johnson}, Aaron D. and {Jones}, Megan L. and {Kaiser}, Andrew R. and {Kaplan}, David L. and {Kelley}, Luke Zoltan and {Kerr}, Matthew and {Key}, Joey S. and {Laal}, Nima and {Lam}, Michael T. and {Lamb}, William G. and {Lazio}, T. Joseph W. and {Lewandowska}, Natalia and {Liu}, Tingting and {Lorimer}, Duncan R. and {Luo}, Jing and {Lynch}, Ryan S. and {Ma}, Chung-Pei and {Madison}, Dustin R. and {McEwen}, Alexander and {McKee}, James W. and {McLaughlin}, Maura A. and {McMann}, Natasha and {Meyers}, Bradley W. and {Mingarelli}, Chiara M.~F. and {Mitridate}, Andrea and {Ng}, Cherry and {Nice}, David J. and {Ocker}, Stella Koch and {Olum}, Ken D. and {Pennucci}, Timothy T. and {Perera}, Benetge B.~P. and {Pol}, Nihan S. and {Radovan}, Henri A. and {Ransom}, Scott M. and {Ray}, Paul S. and {Romano}, Joseph D. and {Sardesai}, Shashwat C. and {Schmiedekamp}, Ann and {Schmiedekamp}, Carl and {Schmitz}, Kai and {Shapiro-Albert}, Brent J. and {Siemens}, Xavier and {Simon}, Joseph and {Siwek}, Magdalena S. and {Stairs}, Ingrid H. and {Stinebring}, Daniel R. and {Stovall}, Kevin and {Susobhanan}, Abhimanyu and {Swiggum}, Joseph K. and {Taylor}, Stephen R. and {Turner}, Jacob E. and {Unal}, Caner and {Vallisneri}, Michele and {Vigeland}, Sarah J. and {Wahl}, Haley M. and {Witt}, Caitlin A. and {Young}, Olivia and {Nanograv Collaboration}},
        title = "{The NANOGrav 15 yr Data Set: Detector Characterization and Noise Budget}",
      journal = {\apjl},
         year = 2023,
        month = jul,
       volume = {951},
       number = {1},
          eid = {L10},
        pages = {L10},
          doi = {10.3847/2041-8213/acda88},
archivePrefix = {arXiv},
       eprint = {2306.16218},
 primaryClass = {astro-ph.HE},
       adsurl = {https://ui.adsabs.harvard.edu/abs/2023ApJ...951L..10A}
}

@article{CVM+26,
  author        = {Crisostomi, Marco and van Haasteren, Rutger and Meyers, Patrick M. and Vallisneri, Michele},
  title         = {Beyond diagonal approximations: improved covariance modeling for pulsar timing array data analysis},
  journal       = {Physical Review D},
  volume        = {114},
  pages         = {024080},
  year          = {2026},
  doi           = {10.1103/PhysRevD.114.024080},
  eprint        = {2506.13866},
  archivePrefix = {arXiv},
  primaryClass  = {astro-ph.IM},
  adsurl        = {https://ui.adsabs.harvard.edu/abs/2026PhRvD.114b4080C}
}

@article{FFF+26,
  author        = {Ferranti, Irene and Falxa, Mikel and Fantoccoli, Federico and Sesana, Alberto and Shaifullah, Golam Mohiuddin},
  title         = {Data span and frequency coverage requirements for robust detection and inference in {PTAs}: A case study with {EPTA DR2}},
  journal       = {Astronomy \& Astrophysics},
  volume        = {709},
  pages         = {A256},
  year          = {2026},
  doi           = {10.1051/0004-6361/202558270},
  eprint        = {2511.21933},
  archivePrefix = {arXiv},
  primaryClass  = {astro-ph.HE},
  adsurl        = {https://ui.adsabs.harvard.edu/abs/2026A%26A...709A.256F}
}

@ARTICLE{TFM79,
       author = {{Taylor}, J.~H. and {Fowler}, L.~A. and {McCulloch}, P.~M.},
        title = "{Measurements of general relativistic effects in the binary pulsar PSR1913 + 16}",
      journal = {\nature},
         year = 1979,
        month = feb,
       volume = {277},
       number = {5696},
        pages = {437-440},
          doi = {10.1038/277437a0},
       adsurl = {https://ui.adsabs.harvard.edu/abs/1979Natur.277..437T}
}

@ARTICLE{TW82,
       author = {{Taylor}, J.~H. and {Weisberg}, J.~M.},
        title = "{A new test of general relativity - Gravitational radiation and the binary pulsar PSR 1913+16}",
      journal = {\apj},
         year = 1982,
        month = feb,
       volume = {253},
        pages = {908-920},
       adsurl = {https://ui.adsabs.harvard.edu/abs/1982ApJ...253..908T}
}

@ARTICLE{MSR+25,
       author = {{Miles}, Matthew T. and {Shannon}, Ryan M. and {Reardon}, Daniel J. and others},
        title = "{The MeerKAT Pulsar Timing Array: the first search for gravitational waves with the MeerKAT radio telescope}",
      journal = {\mnras},
         year = 2025,
        month = jan,
       volume = {536},
       number = {2},
        pages = {1489-1500},
          doi = {10.1093/mnras/stae2571},
archivePrefix = {arXiv},
       eprint = {2412.01153},
 primaryClass = {astro-ph.HE},
       adsurl = {https://ui.adsabs.harvard.edu/abs/2025MNRAS.536.1489M}
}

\end{document}